\documentclass[aps,prd,10pt,notitlepage,nofootinbib,superscriptaddress]{revtex4-1}

\usepackage[utf8]{inputenc}
\usepackage{newtxtext,newtxmath}

\usepackage{bbold}
\usepackage{bm}
\usepackage{graphicx}
\usepackage[usenames,dvipsnames]{xcolor}
\usepackage{color}
\usepackage[colorlinks=true,linkcolor=blue,urlcolor=blue,citecolor=blue]{hyperref}
\usepackage{slashed}
\usepackage[english]{babel}
\usepackage{dcolumn}
\usepackage{pifont}
\usepackage{dsfont,mathrsfs}
\usepackage{cancel}
\usepackage{bigints}
\usepackage{accents}
\usepackage{soul}
\usepackage{multirow}

\usepackage{lineno}
\newcommand{\nn}{\nonumber}
\newcommand{\ket}[1]{\left|#1\right\rangle}

\newcommand{\expectationvalue}[3]{\langle#1|#2|#3\rangle}
\newcommand{\ExpectationValue}[3]{\left\langle#1\left|#2\right|#3\right\rangle}
\newcommand{\ensembleaverage}[1]{\left\langle#1\right\rangle}

\newcommand{\MB}[1]{\left|#1\right|}
\newcommand{\FB}[1]{\left(#1\right)}
\newcommand{\fb}[1]{(#1)}
\newcommand{\SB}[1]{\left\{#1\right\}}
\newcommand{\TB}[1]{\left[#1\right]}
\newcommand{\AB}[1]{\langle#1\rangle}

\newcommand{\mcT}{\mathcal{T}}

\newcommand{\mcN}{\mathcal{N}}
\newcommand{\mcI}{\mathcal{I}}
\newcommand{\mcF}{\mathcal{F}}
\newcommand{\mcSh}{\hat{\mathcal{S}}}
\newcommand{\mcZ}{\mathcal{Z}}
\newcommand{\mcW}{\mathcal{W}}
\newcommand{\mcL}{\mathcal{L}}

\newcommand{\scrL}{\mathscr{L}}

\newcommand{\scrD}{\mathscr{D}}

\newcommand{\munu}{{\mu\nu}}

\newcommand{\alphabeta}{{\alpha\beta}}
\newcommand{\IM}{\text{Im}}

\newcommand{\Tr}{\text{Tr}}

\newcommand{\Psibar}{\overline{\Psi}}
\newcommand{\psibar}{\overline{\psi}}

\newcommand{\kpll}{k_\parallel}
\newcommand{\qpll}{q_\parallel}
\newcommand{\ppll}{p_\parallel}

\newcommand{\gpll}{g_\parallel}
\newcommand{\gper}{g_\perp}

\newcommand{\del}{\partial}
\newcommand{\identity}{\mathds{1}}

\newcommand{\fsl}{\slashed}

\newcommand{\intzinf}{\int_{0}^{\infty}}

\newcommand{\half}{\dfrac{1}{2}}
\newcommand{\kthreeint}[1]{\int \dfrac{d^3\vec{#1}}{(2\pi)^3}}
\newcommand{\kzint}[1]{\int_{-\infty}^\infty \dfrac{d{#1}_z}{2\pi}}

\newcommand{\potU}{\mathcal{U}\FB{ \Phi,\bar{\Phi} ;T}}
\newcommand{\fnppbT}{\FB{\Phi, \bar{\Phi}, T } }
\newcommand{\F}{\Phi}
\newcommand{\Fb}{\bar{\Phi}}
\newcommand{\paroneder}[2]{\frac{\partial {#1}}{\partial {#2}} }

\begin{document}
\title{Dilepton production from magnetized quark matter with an anomalous magnetic moment of the  quarks using a three-flavor PNJL model}

\author{Nilanjan Chaudhuri}
\email{sovon.nilanjan@gmail.com}
\affiliation{Variable Energy Cyclotron Centre, 1/AF Bidhannagar, Kolkata - 700064, India}
\affiliation{Homi Bhabha National Institute, Training School Complex, Anushaktinagar, Mumbai - 400085, India}
\author{Snigdha Ghosh}
\email{snigdha.physics@gmail.com}
\thanks{Corresponding Author}
\affiliation{Government General Degree College Kharagpur-II, Madpur, Paschim Medinipur - 721149, West Bengal, India}
\author{Sourav Sarkar}
\email{sourav@vecc.gov.in}
\affiliation{Variable Energy Cyclotron Centre, 1/AF Bidhannagar, Kolkata - 700064, India}
\affiliation{Homi Bhabha National Institute, Training School Complex, Anushaktinagar, Mumbai - 400085, India}
\author{ Pradip Roy}
\email{pradipk.roy@saha.ac.in}
\affiliation{Saha Institute of Nuclear Physics, 1/AF Bidhannagar, Kolkata - 700064, India}
\affiliation{Homi Bhabha National Institute, Training School Complex, Anushaktinagar, Mumbai - 400085, India}
%

%
%
%

\begin{abstract}
Dilepton production from hot, dense and magnetized quark matter is studied using the three-flavor Polyakov loop extended Nambu--Jona-Lasinio (PNJL) model in which the anomalous magnetic moment (AMM) of the quarks is also taken into consideration. This is done by first evaluating the thermo-magnetic spectral function of the vector current correlator employing the real time formalism of finite temperature field theory and the Schwinger proper time formalism. The constituent quark mass which goes as an input in the expression of the dilepton production rate (DPR), has been calculated using the three-flavor PNJL model employing Pauli-Villiars (PV) regularization. The obtained constituent quark mass being strongly dependent on the temperature, density, magnetic field and AMM of the quarks, captures the effect of `strong' interactions specifically around the (pseudo) chiral and confinement-deconfinement phase transition regions. The analytic structure of the spectral function in the complex energy plane has been analyzed in detail and a non-trivial Landau cut is found in the physical kinematic domains resulting from the scattering of the Landau quantized quark/antiquark with the photon which is purely a finite magnetic field effect. Due to the emergence of the Landau cut along with the usual unitary cut, the DPR is found to be largely enhanced in the low invariant mass region. Owing to the magnetic field and AMM dependence of the thresholds of these cuts, we find that the kinematically forbidden gap between the Unitary and Landau cuts vanishes at sufficiently high temperature, density and magnetic field leading to the generation of a continuous spectrum of dilepton emission over the whole invariant mass region. In order to see the effects of strangeness and confinement-deconfinement, the rates are compared with the three-flavor NJL and the two-flavor NJL and PNJL models. Finally, an infinite number of quark Landau levels is incorporated in the calculation so that no approximations are made on the strength of the background magnetic field like strong or weak as usually done in the literature.
\end{abstract}

\maketitle
%
\section{INTRODUCTION}\label{sec.intro}

The main objective of heavy ion collision (HIC) experiments in Relativistic Heavy Ion Collider (RHIC) and Large Hadron Collider (LHC) is to study  deconfined state of strongly interacting quarks and gluons in local thermal equilibrium, commonly known as the quark-gluon plasma (QGP). The hot and dense matter created after the HIC goes through different stages while it cools via rapid expansion under its own pressure. However, the whole process is very short lived  (~few fm/c) and direct observations are not possible. So to investigate microscopic as well as bulk properties of QGP, many indirect probes and observables are proposed~\cite{Wong:1995jf} among which the electromagnetic probes i.e. photons and dileptons, have been extensively studied in the literature~\cite{McLerran:1984ay,Kajantie:1986dh,Weldon:1990iw,Alam:1996fd,Alam:1999sc,Rapp:1999ej,Aurenche:2000gf,Arnold:2001ms,Rapp:2009my,Chatterjee:2009rs}. A prime benefit of this probe over hadrons, which are emitted from the freeze out surface after undergoing intense re-scattering, is that photons/dileptons are emitted during all the stages of the expanding fireball. Since they participate only in electromagnetic interaction, their mean free paths are much larger than the typical size and lifetime of this novel state of matter. Consequently once produced, they leave the hot and dense medium without suffering further interactions and reach the detector with unaltered information about the circumstances of their production. 
	
One of the primary theoretical tools to examine various properties of QGP is the study of different $ n $-point current-current correlation functions or in-medium spectral functions of local currents. The electromagnetic spectral function is one such example which is obtained from the vector-vector current correlator which, in turn, is connected to the dilepton prduction rate (DPR)  from  the hot and dense medium~\cite{Mallik:2016anp,Alam:1999sc, Sarkar:2012ty}. In the QGP medium, the asymptotically free quarks can interact with an antiquark to generate a virtual photon, which decays into a dilepton. It should be noted that, there exist several other sources of dileptons in  HIC experiments, like scattering of charged hadrons, hadron resonance decays ($ \pi^0, ~\rho,~\omega,~J/\psi $), Drell-Yan process \textit{etc}~\cite{Sarkar:2012ty}.

Recent studies~\cite{Kharzeev:2007jp,Skokov:2009qp} suggest that, in a non-central or asymmetric HIC experiment, very strong magnetic fields of the order $ \sim 10^{18} $ Gauss or larger are believed  to be generated. 	These fields are, in principle, time dependent and immediately decay after few fm/c.  However, as argued in Refs.~\cite{Tuchin:2013apa,Tuchin:2015oka,Tuchin:2013ie,Gursoy:2014aka}, the presence of large electrical conductivity of the hot and dense 	medium created during HICs can substantially delay the 	decay of these transient magnetic fields. Apart from this, strong magnetic fields can be present in several other physical environments. For example, in the interior of certain astrophysical objects called magnetars~\cite{Duncan:1992hi,Thompson:1993hn}, magnetic field $\sim 10^{15}$ Gauss can be present. Moreover, it is conjectured that, in the early universe during the electroweak phase transition, magnetic fields as high as $ \sim 10^{23}$ Gauss~\cite{Vachaspati:1991nm,Campanelli:2013mea} might have been produced. Since, the magnitude of the magnetic fields are comparable to the typical Quantum Chromodynamics (QCD) energy scale ($eB\sim \Lambda_\text{QCD}^2$), the bulk as well as microscopic properties of the QCD matter could be significantly modified (see ~\cite{Miransky:2015ava,Kharzeev:2013jha} for recent reviews). Furthermore, the presence of a strong background magnetic field results in a large number of interesting physical phenomena~\cite{Kharzeev:2012ph,Kharzeev:2007tn,Chernodub:2010qx, Chernodub:2012tf} owing  to the rich vacuum structure of the underlying QCD, e.g. the Chiral Magnetic Effect (CME)~\cite{Fukushima:2008xe,Kharzeev:2007jp,Kharzeev:2009pj,Bali:2011qj}, 	Magnetic Catalysis (MC)~\cite{Shovkovy:2012zn,Gusynin:1994re,Gusynin:1995nb,Gusynin:1999pq}, Inverse Magnetic Catalysis (IMC)~\cite{Preis:2010cq,Preis:2012fh},  	Chiral Vortical Effect (CVE), vacuum superconductivity and superfluidity~\cite{Chernodub:2011gs,Chernodub:2011mc} {\it etc}.

The modification of the DPR in the presence of a uniform background magnetic field has been studied in the literature using different approaches~\cite{Tuchin:2012mf,Tuchin:2013bda,Sadooghi:2016jyf,Bandyopadhyay:2016fyd,Bandyopadhyay:2017raf,Ghosh:2018xhh,Islam:2018sog,Ghosh:2020xwp}. In~\cite{Tuchin:2012mf,Tuchin:2013bda}, the DPR from hot magnetised quark matter is obtained using phenomenological inputs considering the effects of synchrotron radiation as well as quark-antiquark annihilation.  In~\cite{Sadooghi:2016jyf}, the Ritus formalism is used to evaluate the general structure of photon polarization tensor and DPR at finite temperature and finite magnetic field. The DPR in strong and weak magnetic field approximations using the imaginary and real time formalism of finite temperature field theory have been reported in Refs.~\cite{Bandyopadhyay:2016fyd,Bandyopadhyay:2017raf}. In Ref.~\cite{Ghosh:2018xhh}, the DPR for a hot QGP matter in presence of arbitrary external magnetic field has been estimated using the effective fugacity quasi-particle model (EQPM) in which the effect of strong interactions are captured in the temperature dependent fugacities of the partons and significant enhancement is found in low invariant-mass region.

The imaginary part of the electromagnetic vector current correlator containing the thermo-magnetically modified quark propagators is the most important component in the calculation of DPR which determines the thresholds as well as the intensity of emisson of dileptons~\cite{Alam:1996fd,Alam:1999sc}. These in turn depend crucially on the value of quark mass.   Now, as the system cools down from a (pseudo) chirally symmetric phase to a broken phase, the quark mass acquires large values ($ \sim $ of few hundred MeV) owing to the build up of the quark condensate. Simultaneously the colour deconfined matter converts to hadronic matter through a crossover or phase transition  at smaller values of temperature. The theoretical analysis of these phenomena using first principle calculations is severely restricted by the non-perturbative  nature of QCD at low energies. As an alternative one thus takes recourse to effective theories which	possess some of the essential features of QCD and is	mathematically tractable.
	
The Nambu--Jona-Lasinio (NJL) model~\cite{Nambu1,Nambu2} is one such model, which provides a useful scheme to probe the vacuum structure of QCD at arbitrary values of temperature and density. This model has been extensively used to examine some of the nonperturbative properties of the QCD vacuum as  it respects the global symmetries of QCD, most importantly the chiral symmetry (see~\cite{Klevansky,Vogl,Buballa} for reviews). As mentioned in Ref.~\cite{Klevansky}, the pointlike interaction among the quarks makes the NJL model nonrenormalizable. Thus, a proper regularization scheme has to be chosen to deal with the 	divergent integrals and the parameters associated with the 	model are fixed to reproduce some well-known phenomenological quantities, for example the pion-decay constant, quark condensate \textit{etc}. Moreover, the	NJL model lacks confinement; poles of the massive quark	propagator are present at any temperature and/or chemical potential. But in QCD, both the dynamical chiral symmetry	breaking and the confinement are realized as the global symmetries of the QCD Lagrangian. It is well known that the Polyakov loop can be used as an approximate order parameter for the	deconfinement transition associated with the spontaneous symmetry breaking of the center symmetry~\cite{McLerran:1981pb,Cheng:2007jq}. Thus,	in order to obtain a unified picture of confinement and chiral	symmetry breaking, the Polyakov loop extended Nambu--Jona-Lasinio (PNJL) model was introduced by incorporating a temporal, static and homogeneous gluon-like field~\cite{Ratti:2005jh,Ratti:2006wg}. NJL/PNJL model has been extensively used to study the chiral symmetry restoration in the presence of a background electromagnetic field~\cite{Klevansky:1989vi,Gusynin:1995nb,Gatto:2010qs,Fayazbakhsh:2012vr,Ruggieri:2013cya,Andersen:2014xxa,Ferreira:2013oda,Ferreira:2014exa,Ferreira:2014kpa,Mao:2016fha,Mao:2016lsr,Fayazbakhsh:2014mca,Avancini:2018svs,Chaudhuri:2019lbw,Chaudhuri:2020lga,Ghosh:2020qvg,Mei:2020jzn}.

As discussed above, while calculating DPR from a hot and dense magnetized medium, the thermo-magnetic quark propagator plays a primary role. Using NJL/PNJL model one can further modify the quark propagator by considering the magnetic field, temperature and density dependent constituent quark mass $M=M(B,T,\mu)$ which is dynamically generated. In Ref.~\cite{Ghosh:2020xwp}, two-flavor NJL model in the presence of an arbitrary external magnetic field at finite temperature and baryon density including the anomalous magnetic moment (AMM) of the quarks was used and a continuous spectrum of	the dilepton emission over the whole range of invariant mass was obtained at high magnetic field.

In this work, we calculate the DPR from hot and dense quark matter in the presence of an arbitrary external magnetic field in which the AMM of the quarks is also taken into consideration. For this, we use the three-flavor PNJL model with a gauge invariant regularization scheme namely the Pauli-Villiars scheme. The  constituent quark mass is calculated using the PNJL model by minimizing the thermodynamic potential in the light of mean field approximation. We employ the real time formalism of finite temperature field theory and the Schwinger proper time formalism to evaluate the electromagnetic spectral function of the vector current correlator (which is proportional to the DPR) at finite temperature ($T$), chemical potential ($\mu$), external magnetic field ($B$) and AMM ($\kappa$) of the quarks where the constituent quark mass $M=M(T,\mu,B,\kappa)$ will go as an input. We then analyze the complete analytic structure of the in-medium spectral function to obtain the thresholds of the branch cuts in the complex energy plane in detail. We compare the whole study with the three-flavor NJL and the two-flavor NJL and PNJL models to observe the effects of strangeness and confinement-deconfinement. In all of our calculations, we will consider an infinite number of quark Landau levels so that no approximations will be made on the strength of the background magnetic field like strong or weak.

The paper is organized as follows. In Sec.~\ref{sec.DPR} along with its two subsections, the DPR is obtained at both zero and finite external magnetic field case. Next in Sec.~\ref{sec.NJL}, the evaluation of the constituent quark mass using the two and three flavor NJL/PNJL model is discussed. After that, Sec.~\ref{sec.results} is devoted for showing the numerical results and finally we summarize and conclude in Sec.~\ref{sec.summary}.

%


\section{DILEPTON PRODUCTION RATE} \label{sec.DPR}

The dilepton production rate (DPR) from a hot and dense magnetized medium is already calculated in Refs.~\cite{Mallik:2016anp, Bandyopadhyay:2017raf, Ghosh:2018xhh,Ghosh:2020xwp}. But, for completeness of the presentation, we will recapitulate few important steps here. We consider an initial state  $\ket{i}=\ket{\mcI(p_\mcI)}$ consisting of a quark/antiquark with momentum $p_\mcI$ evolving to a final state $\ket{f}=\ket{\mcF(p_\mcF),l^+(p_+)l^-(p_-)}$ involving a quark/antiquark of momentum $p_\mcF$ plus a pair of leptons of momenta $p_+$ and $p_-$ respectively. The probability amplitude for the transition from the initial to the final state is  $|\expectationvalue{f}{\mcSh}{i}|^2$, where $\mcSh$ is the scattering matrix expressed as
\begin{eqnarray}
\mcSh = \mcT \TB{ \exp \SB{ i\int \scrL_\text{int}(x)d^4x } }
\end{eqnarray}
in which $\mcT$ is the time-ordering operator and 
\begin{eqnarray}
\scrL_\text{int}(x) = j^\mu(x)A_\mu(x)+J^\mu(x)A_\mu(x)
\end{eqnarray}
is the Lagrangian (density) for local interaction. Metric tensor $g^\munu=\texttt{diag}(1,-1,-1,-1)$ will be used through out this paper. In the above equation,  $A^\mu(x)$ is the photon field which is coupled to the conserved vector currents corresponding to the leptons and quarks denoted by $j^\mu(x)$ and $J^\mu(x)$ respectively. After few algebraic steps~\cite{Ghosh:2018xhh}, one can arrive at an expression of the squared amplitude $|\expectationvalue{f}{\mcSh}{i}|^2$ up to the second order given by,
\begin{eqnarray}
|\expectationvalue{f}{\mcSh}{i}|^2 = \int\!\!\!\int d^4x' d^4x e^{i(p_++p_-)\cdot x'}\frac{1}{(p_++p_-)^4}
\ExpectationValue{l^+(p_+)l^-(p_-)}{j^\mu(0)}{0} \ExpectationValue{0}{j^{\nu\dagger}(0)}{l^+(p_+)l^-(p_-)} \nn \\
\ExpectationValue{\mcF(p_\mcF)}{J_\mu(x')}{\mcI(p_\mcI)} \ExpectationValue{\mcI(p_\mcI)}{J_\nu^\dagger(0)}{\mcF(p_\mcF)}. \label{eq.sfi}
\end{eqnarray}
Now the dilepton multiplicity from thermal QGP medium is given by~\cite{Mallik:2016anp}
\begin{eqnarray}
N = \frac{1}{\mcZ} \sum_{\text{spins}}^{}\int\!\!\frac{d^3p_+}{(2\pi)^32p_+^0}\int\!\!\frac{d^3p_-}{(2\pi)^32p_-^0} \sum_{\mcI,\mcF}^{}
\exp\FB{-p_\mcI^0/T} |\expectationvalue{f}{\mcSh}{i}|^2 \label{eq.DM}
\end{eqnarray}
where $\mcZ$ is the partition function of the system and the summation implies  sum over all leptonic spin configurations. Substituting Eq.~\eqref{eq.sfi} into Eq.~\eqref{eq.DM} and simplifying, we get
\begin{eqnarray}
N = \int\!\! d^4x\int\!\!\! \frac{d^4q}{(2\pi)^4} e^{-q^0/T}\frac{1}{q^4} \mcW_{+\munu}(q) \mcL_+^\munu(q) 
\label{eq.dilepton.mult}
\end{eqnarray}
where,
\begin{eqnarray}
\mcW_+^\munu(q) &=& \int\!\! d^4x e^{iq\cdot x} \ensembleaverage{J^\mu(x)J^{\nu\dagger}(0)}, \label{eq.W+}\\
\mcL_+^\munu(q) &=& \int\!\! d^4x e^{iq\cdot x} \ExpectationValue{0}{j^{\nu\dagger}(x)j^{\mu}(0)}{0}. \label{eq.L+}
\end{eqnarray}
Here $ q^0 $ is the total energy of the lepton-pair and $\ensembleaverage{...}$ represents the ensemble average. So the dilepton production rate (DPR) becomes
  \begin{eqnarray}
  \text{DPR} = \frac{dN}{d^4xd^4q} = \frac{1}{(2\pi)^4} \frac{e^{-q^0/T}}{q^4}  \mcW_{+\munu}(q) \mcL_+^\munu(q). \label{eq.DPR.1}
  \end{eqnarray}

Now  it is more useful to express both the $\mcW_+^\munu(q)$ and $\mcL_+^\munu(q)$ in terms of time ordered correlation functions; so that $\mcW_+^\munu(q)$ will be  calculated at finite temperature employing the real time formalism (RTF) of thermal field theory (see ~\cite{Ghosh:2020xwp,Ghosh:2018xhh} for details). Thus we arrive at
\begin{eqnarray}
\text{DPR} = \frac{dN}{d^4xd^4q} = \frac{1}{4\pi^4q^4} \FB{\frac{1}{e^{q^0/T}+1}}\IM \mcW_{11}^\munu(q)\IM \mcL_\munu(q). \label{eq.DPR.2}
\end{eqnarray}

Next we  calculate the quantities $\mcW_{11}^\munu(q)$ and $\mcL^\munu(q)$ for which the explicit form of the currents $J^\mu(x)$ and $j^\mu(x)$ are required. They are given by
\begin{eqnarray}
J^\mu(x) &=& e\Psibar(x)\hat{Q}\gamma^\mu\Psi(x)~, \label{eq.J} \\
j^\mu(x) &=& -e\psibar(x)\gamma^\mu\psi(x) \label{eq.j}
\end{eqnarray}
where, $\Psi=\FB{u~~d~~s}^T$ is the the quark field with three flavors up, down and strange respectively with corresponding charge $\hat{Q}= \texttt{diag}~ \FB{ \frac{2}{3}, -\frac{1}{3}, -\frac{1}{3}  } $, $\psi$ is the lepton field and $e$ is the electric charge of a proton. Using Eqs.~\eqref{eq.J} and \eqref{eq.j} it can be shown that
\begin{eqnarray}
\mcW_{11}^\munu(q) &=& i\int\!\!\!\frac{d^4k}{(2\pi)^4} \Tr_\text{d,f,c} 
\TB{\gamma^\mu \hat{Q} S_{11}(p=q+k)\gamma^\nu \hat{Q} S_{11}(k)}~, \label{eq.W11.2}\\
\mcL^\munu(q) &=& i e^2\int\!\!\!\frac{d^4k}{(2\pi)^4} \Tr_\text{d} \TB{\gamma^\nu S(p=q+k)\gamma^\mu S(k)} \label{eq.L.2}
\end{eqnarray}
where the subscript `d', `f', and `c' in the trace correspond to trace over Dirac, flavor and color spaces respectively; $S_{11}(p)$ is the 11-component of the real time quark propagator and $S(k)$ is the Feynman propagator for leptons given by 
\begin{eqnarray}
S(p) = \frac{-(\cancel{p}+m_L)}{p^2-m_L^2+i\varepsilon} \label{eq.lepton.propagator}
\end{eqnarray}
with $m_L$ being the mass of the lepton. Notice that, as a consequence of the conservation of the currents $J^\mu(x)$ and $j^\mu(x)$: $\del_\mu J^\mu(x)=\del_\mu j^\mu(x)=0$, both the matter tensor $\mcW_{11}^\munu(q)$ and the leptonic tensor $\mcL^\munu(q)$ are transverse to the momentum $q^\mu$ i.e.
\begin{eqnarray}
q_\mu \mcW_{11}^\munu(q) = q_\mu \mcL^\munu(q) = 0. \label{eq.transversality}
\end{eqnarray}
 Now we will consider the following two cases separately: (i) zero external magnetic field $(B=0)$ and (ii) non-zero external magnetic field $(B\ne 0)$ in order to calculate the DPR in the following subsections.
\subsection{DPR AT $B=0$} \label{subsec.DPR.1}
\begin{figure}[h]
	\begin{center}
		\includegraphics[angle=0,scale=0.24]{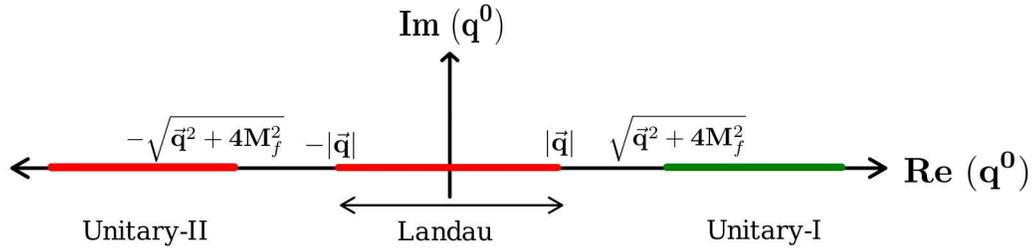}
	\end{center}
	\caption{(Color Online) The branch cuts of $\mcW_{11}^\munu(q)$ in the complex $q^0$ plane for a given $\vec{q}$. Kinematic domain for the physical dileptons production defined in terms of $q^0>0$ and $q^2>0$ corresponds to the green region. }
	\label{fig.analytic0}
\end{figure}

At $B=0$, the transversility condition $q_\mu \mcL^\munu(q) = 0$ of Eq.~\eqref{eq.transversality} suggests that the Lorentz structure of $\mcL^\munu(q)$ has to be of the following form:
\begin{eqnarray}
\mcL^\munu(q) = \FB{g^\munu-\frac{q^\mu q^\nu}{q^2}}\FB{\frac{1}{3}g_\alphabeta \mcL^\alphabeta}. \label{eq.L.0}
\end{eqnarray}
Substituting the above equation in Eq.~\eqref{eq.DPR.2}, and making use of $q_\mu \mcW_{11}^\munu(q)=0$ we get the DPR at $B=0$ as
\begin{eqnarray}
\text{DPR}_{B=0} = \FB{\frac{dN}{d^4xd^4q}}_{B=0} = \frac{1}{12\pi^4q^4} \FB{\frac{1}{e^{q^0/T}+1}}
g_\munu\IM \mcW_{11}^\munu(q) g_\alphabeta \IM \mcL^\alphabeta(q). \label{eq.DPR.3}
\end{eqnarray}
$g_\alphabeta \IM \mcL^\alphabeta(q)$ can be calculated in a straightforward way by  substituting Eq.~\eqref{eq.lepton.propagator} into Eq.~\eqref{eq.L.2} and performing some algebraic steps we get
\begin{eqnarray}
g_\alphabeta \IM \mcL^\alphabeta(q) = \frac{-e^2}{4\pi}q^2\FB{1+\frac{2m_L^2}{q^2}}\sqrt{1-\frac{4m_L^2}{q^2}}\Theta\FB{q^2-4m_L^2}.
\label{eq.L.3}
\end{eqnarray}
In order to calculate $g_\munu\IM \mcW_{11}^\munu(q)$, we note that the 11-component of the real time thermal quark propagator is 
\begin{eqnarray}
S_{11}(p,M_f) = \FB{\cancel{p}+M_f}\TB{\frac{-1}{p^2-M_f^2+i\varepsilon}-2\pi i \eta(p\cdot u)\delta(p^2-M_f^2)}\label{eq.S11.T}
\end{eqnarray}
where, $M_f$ is the \textit{constituent quark mass} of flavor $ f $, $u^\mu$ is the four-velocity of the thermal bath, $\eta(x)=\Theta(x)f^+(x)+\Theta(-x)f^-(-x)$, 
\begin{eqnarray}
f^\pm(x) = \TB{\exp\FB{\frac{x\mp\mu_B/3}{T}}+1}^{-1}
\end{eqnarray}
and $\mu_B$ is the baryon chemical potential. In the local rest frame (LRF) of the medium, $u^\mu_\text{LRF}\equiv(1,\vec{0})$. Substitution of Eq.~\eqref{eq.S11.T} into Eq.~\eqref{eq.W11.2} yields, after some simplifications,
\begin{eqnarray}
g_\munu\IM \mcW_{11}^\munu(q) = N_c\sum_{f} e_f^2 \pi \int\frac{d^3k}{(2\pi)^3}&&\frac{1}{4\omega_k\omega_p}\big[
\SB{1-f^-(\omega_k)-f^+(\omega_p)+2f^-(\omega_k)f^+(\omega_p)}\mcN(k^0=-\omega_k)\delta(q^0-\omega_k-\omega_p) \nn \\
&&+ \SB{1-f^+(\omega_k)-f^-(\omega_p)+2f^+(\omega_k)f^-(\omega_p)}\mcN(k^0=\omega_k)\delta(q^0+\omega_k+\omega_p) \nn \\
&&+ \SB{-f^-(\omega_k)-f^-(\omega_p)+2f^-(\omega_k)f^-(\omega_p)}\mcN(k^0=-\omega_k)\delta(q^0-\omega_k+\omega_p) \nn \\
&&+ \SB{-f^+(\omega_k)-f^+(\omega_p)+2f^+(\omega_k)f^+(\omega_p)}\mcN(k^0=\omega_k)\delta(q^0+\omega_k-\omega_p) \big]
\label{eq.W11.7}
\end{eqnarray}
where, $e_u=2e/3$, $e_d=-e/3$, $e_s=-e/3$, $\omega_k=\sqrt{\vec{k}^2+M_f^2}$, $\omega_p=\sqrt{\vec{p}^2+M_f^2}=\sqrt{(\vec{q}+\vec{k})^2+M_f^2}$ and $\mcN (q,k)= 8(k^2+q\cdot k - 2M_f^2)$. Now, let us briefly discuss the analytic structure of $ \IM \mcW_{11}^\munu(q)$ in the complex $q^0$ plane following~\cite{Ghosh:2018xhh, Ghosh:2020xwp}.  Note that, the four different Dirac delta functions appearing in the Eq.~\eqref{eq.W11.7} are responsible for different physical processes such as decay or scattering. The first delta function, termed as the unitary-I cut, corresponds to the contribution from quark-antiquark  annihilation to a positive energy time-like virtual photon (and the time reversed process as the decay of the photon to the quark-antiquark pair). The delta function appearing in the second term (the unitary-II cut) corresponds to the quark-antiquark  annihilation to a negative energy time-like virtual photon (and the time reversed process as the vacuum to quark-antiquark-photon transition).  Similarly, the last two delta functions (called the Landau cuts, are purely a medium effect) correspond to the scattering processes where  a quark/antiquark absorbs a space-like virtual photon (and the corresponding time reversed process involves the emission of the photon by the quark/antiquark). All these terms are non-vanishing in their respective kinematic domain controlled by the delta functions~\cite{Ghosh:2018xhh}. It can be shown that,  the kinematic domains for the unitary-I and unitary-II cuts are respectively $\sqrt{\vec{q}^2+4M_f^2}<q^0<\infty$ and $-\infty<q^0<-\sqrt{\vec{q}^2+4M_f^2}$. On the other hand, nonzero contributions from the Landau cuts is  expected when $|q^0|<|\vec{q}|$. Now, we will restrict ourselves to the case of physical dileptons with positive energy and time-like four momentum i.e $ q^0 >0 $ and $q^2>0$ respectively. Then, it follows that, only unitary-I cut contributes as shown by the green region in Fig.~\ref{fig.analytic0}. So, in the kinematically allowed region considered here, processes like photon decay and formation are permitted via the unitary-I cut. However, the scattering and emission processes cease to occur.  Notice that, in the above analysis, the kinematic domains can be different for different flavors and it is directly related to constituent quark mass $ M_f $.

Now with the physical restrictions $q^0>0$ and $q^2>0$, the angular integrals of Eq.~\eqref{eq.W11.7} can be performed  using the first Dirac delta function. As a result we get:
\begin{eqnarray}
g_\munu\IM \mcW_{11}^\munu(q) = \sum_{f} \Theta\FB{q^2-4M_f^2} \frac{N_c e_f^2}{16\pi|\vec{q}|} \int_{\omega_-}^{\omega_+}d\omega_k
\SB{1-f^-(\omega_k)-f^+(\omega_p)+2f^-(\omega_k)f^+(\omega_p)}\mcN(k^0=-\omega_k)\Big|_{\theta=\theta_0} \label{eq.W11.3}
\end{eqnarray}
where, $\omega_\pm = \frac{1}{2q^2}\TB{q^0q^2\pm|\vec{q}|\lambda^{1/2}(q^2,M_f^2,M_f^2)}$, $\theta$ is the angle between 
$\vec{q}$ and $\vec{k}$ and $\theta_0 = \cos^{-1}\FB{\frac{q^2-2q^0\omega_k}{2|\vec{q}||\vec{k}|}}$ with $\lambda(x,y,z)$ being the 
K\"all\'en function. Simplifying further we arrive at
\begin{eqnarray}
g_\munu\IM \mcW_{11}^\munu(q) = -\sum_f \Theta\FB{q^2-4M_f^2}N_c e_f^2 \frac{Tq^2}{4\pi|\vec{q}|} \coth\FB{\frac{q^0}{2T}}
\FB{1+\frac{2M_f^2}{q^2}} \ln \TB{\SB{\frac{e^{(q^0+q_-)/T}+1}{e^{(q^0+q_+)/T}+1}}\FB{\frac{e^{q_+/T}+1}{e^{q_-/T}+1}}} \label{eq.W11.4}
\end{eqnarray}
where, $q_\pm = -\frac{1}{2}\TB{q^0 \pm |\vec{q}|\sqrt{1-\frac{4M_f^2}{q^2}}} + \mu_B/3$. Finally substituting Eqs.~\eqref{eq.L.3} and 
\eqref{eq.W11.4} into Eq.~\eqref{eq.DPR.3} yields the following analytical expression for the DPR
\begin{eqnarray}
\text{DPR}_{B=0} = \FB{\frac{dN}{d^4xd^4q}}_{B=0} =  \sum_f \Theta\FB{q^2-4m_L^2} \Theta\FB{q^2-4M_f^2}  N_c\frac{ e^2 e_f^2}{ 192\pi^6} \frac{T}{|\vec{q}|} \FB{\frac{1}{e^{q^0/T}-1}} \nn \\ \FB{1+\frac{2m_L^2}{q^2}}\sqrt{1-\frac{4m_L^2}{q^2}}
\FB{1+\frac{2M_f^2}{q^2}} \ln \TB{\SB{\frac{e^{(q^0+q_-)/T}+1}{e^{(q^0+q_+)/T}+1}}\FB{\frac{e^{q_+/T}+1}{e^{q_-/T}+1}}}
 \label{eq.DPR.4}
\end{eqnarray}
which is consistent with the results previously obtained in~\cite{Greiner:2010zg}. The step functions appearing in the above expression indicates that the production of dileptons  with invariant mass $ q^2<4m_L^2 $ and $ q^2<4M_f^2  $ are kinematically restricted.

\subsection{DPR AT $B\ne0$ INCLUDING AMM OF THE QUARKS} \label{subsec.DPR.2}

Let us consider the same situation in presence of a finite background magnetic field along the positive z-direction. Since, the presence of external magnetic field leads to breaking of the rotational symmetry, any four vector $a^\mu$ can be decomposed into its parallel and perpendicular component in the following manner, $a^\mu=(a_\parallel^\mu+a_\perp^\mu)$ where $a_\parallel^\mu = \gpll^\munu a_\nu$ and $a_\perp^\mu = \gper^\munu a_\nu$. To achieve this the metric tensor should be decomposed accordingly as $g^\munu=(\gpll^\munu+\gper^\munu)$ satisfying $\gpll^\munu=\text{diag}(1,0,0,-1)$ and $\gper^\munu=\text{diag}(0,-1,-1,0)$.

\begin{figure}[h]
	\begin{center}
		\includegraphics[angle=0,scale=0.23]{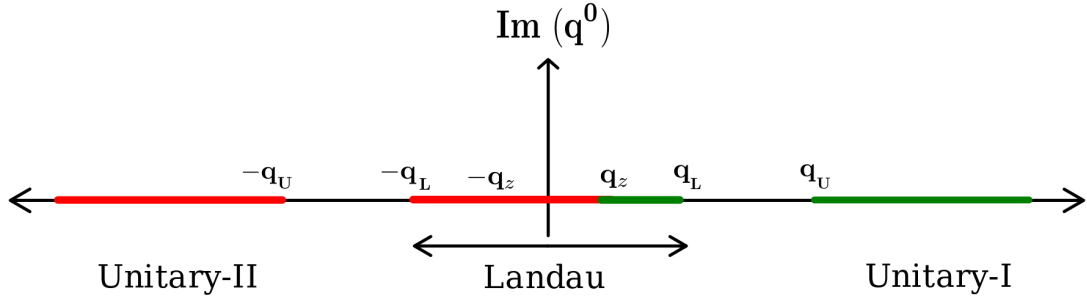}
	\end{center}
	\caption{(Color Online) The branch cuts of $\mcW_{11}^\munu(\qpll)$ in the complex $q^0$ plane for a given $q_z$ and $B$. The points  $q_\text{U}$ and $q_\text{L}$ refer to 
		$q_\text{U} = \sqrt{q_z^2+4\FB{M_f-\kappa_fe_fB}^2}$ and $q_\text{L} = \sqrt{q_z^2+\FB{\sqrt{M_f^2+2|e_fB|}+\kappa_fe_fB-\MB{M_f-\kappa_fe_fB}}^2}$. 
		The green region corresponds to the physical dileptons defined in terms of $q^0>0$ and $\qpll^2>0$.}
	\label{fig.analytic1}
\end{figure}

To ease our analytical calculations, it is convenient to take the resultant transverse momentum of the dileptons to be zero i.e. $q_\perp=0$. Then, from the transversility condition $\qpll^\mu \mcL_\munu(\qpll)=0$ of Eq.~\eqref{eq.transversality}, one can immediately write down the  Lorentz structure of $\mcL^\munu(\qpll)$ as:
\begin{eqnarray}\label{eq.L.01}
\mcL^\munu(\qpll) = \FB{\gpll^\munu-\frac{\qpll^\mu \qpll^\nu}{\qpll^2}}\FB{\gpll^\alphabeta \mcL_\alphabeta}
+ \gper^\munu\FB{\frac{1}{2}\gper^\alphabeta \mcL_\alphabeta}.
\end{eqnarray}
Putting the above result in Eq.~\eqref{eq.DPR.2}, and simplifying by using the condition $q_{\parallel\mu} \mcW_{11}^\munu(\qpll)=0$, we obtain the DPR in presence of external magnetic field as 
\begin{eqnarray}
\text{DPR}_{B\ne0} = \FB{\frac{dN}{d^4xd^4q}}_{B\ne0} = \frac{1}{4\pi^4\qpll^4} \FB{\frac{1}{e^{q^0/T}+1}}
\TB{ g_\parallel^\munu\IM \mcW^{11}_\munu(\qpll) ~ g_\parallel^\alphabeta \IM \mcL_\alphabeta(\qpll)
	+ \frac{1}{2}g_\perp^\munu\IM \mcW^{11}_\munu(\qpll) ~ g_\perp^\alphabeta \IM \mcL_\alphabeta(\qpll)}. \nn \\
\label{eq.DPR.5}
\end{eqnarray}
The quantities $g_{\parallel,\perp}^\munu\IM \mcL_\munu(\qpll)$ appearing within the third bracket of the above equation can be easily calculated by comparing Eqs.~\eqref{eq.L.0} and \eqref{eq.L.01}. The result is
\begin{equation}\label{IMLeB}
g_{\parallel,\perp}^\munu\IM \mcL_\munu(\qpll) =  \zeta_{\parallel,\perp} ~ g_\alphabeta \IM \mcL^\alphabeta(q)
\end{equation}
where $ \zeta_{\parallel,\perp} = (1,2) $. Now in order to  calculate  $g_{\parallel,\perp}^\munu\IM \mcW^{11}_\munu(\qpll)$, first we note that, the 11-component of the thermo-magnetic quark propagator becomes

\begin{eqnarray}
S_{11}(p) =  \begin{pmatrix}
S_{11}^u(p) & 0 & 0 \\ 
0 & S_{11}^d(p) & 0 \\
0 & 0  & S_{11}^s(p)
\end{pmatrix} 
\end{eqnarray}
in which,  the diagonal elements corresponding to up, down and strange quarks are modified differently due to the presence of the external magnetic field. In the above equation, 
\begin{eqnarray}
S_{11}^f(p) = \sum_{s\in\{\pm1\}}^{}\sum_{n=0}^{\infty} \scrD^f_{ns}(p)\TB{\frac{-1}{\ppll^2-(M_n^f-s\kappa_f e_fB)^2+i\varepsilon}
	-2\pi i \eta(p\cdot u)\delta\FB{\ppll^2-(M_n^f-s\kappa_f e_fB)^2}} \label{eq.S11.TB}
\end{eqnarray}
where $M_n^f = \sqrt{M_f^2+2n|e_fB|}$, $\kappa_f$ is the anomalous magnetic moment of quark flavor $f\in \{u,d,s\}$ and
\begin{eqnarray}
\scrD^f_{ns}(p) = (-1)^ne^{-\alpha_p^f}\frac{1}{2M_n^f}(1-\delta_n^0\delta_s^{-1})
\Big[(M_n^f+sM_f)(\cancel{p}_\parallel-\kappa_f e_f B+sM_n^f) \FB{\identity+\text{sign}(e_f)i\gamma^1\gamma^2}L_n(2\alpha_p^f) \nn \\
-(M_n^f-sM_f)(\cancel{p}_\parallel-\kappa_f e_fB-sM_n^f)\FB{\identity-\text{sign}(e_f)i\gamma^1\gamma^2}L_{n-1}(2\alpha_p^f) \nn \\
-4s \FB{ \cancel{p}_\parallel - \text{sign}(e_f)i\gamma^1\gamma^2 (\kappa_f e_fB-sM_n^f)} \text{sign}(e_f)i\gamma^1\gamma^2 \cancel{p}_\perp
L_{n-1}^1(2\alpha_p^f)   \Big]
\end{eqnarray}
where $\alpha_p^f=-p_\perp^2/|e_fB|$. Substituting Eq.~\eqref{eq.S11.TB} into Eq.~\eqref{eq.W11.2} and simplifying, we get
\begin{eqnarray}
g^\munu_{\parallel,\perp}\IM \mcW^{11}_\munu(\qpll) = N_c\sum_{f\in\{u,d,s\}} e_f^2 &&\sum_{s_k\in\{\pm1\}} \sum_{s_p\in\{\pm1\}}
\sum_{l=0}^{\infty}\sum_{n=0}^{\infty} \pi \int\frac{d^3k}{(2\pi)^3}\frac{1}{4\omega_k^{lf}\omega_p^{nf}} \nn \\
&&\times \Big[ \SB{1-f^-(\omega_k^{lf})-f^+(\omega_p^{nf})+2f^-(\omega_k^{lf})f^+(\omega_p^{nf})}
\mcN_{\parallel,\perp}^{fln}(k^0=-\omega_k^{lf})\delta(q^0-\omega_k^{lf}-\omega_p^{nf}) \nn \\
&&+ \SB{1-f^+(\omega_k^{lf})-f^-(\omega_p^{nf})+2f^+(\omega_k^{lf})f^-(\omega_p^{nf})}
\mcN_{\parallel,\perp}^{fln}(k^0=\omega_k^{lf})\delta(q^0+\omega_k^{lf}+\omega_p^{nf}) \nn \\
&&+\SB{-f^-(\omega_k^{lf})-f^-(\omega_p^{nf})+2f^-(\omega_k^{lf})f^-(\omega_p^{nf})}
\mcN_{\parallel,\perp}^{fln}(k^0=-\omega_k^{lf})\delta(q^0-\omega_k^{lf}+\omega_p^{nf}) \nn \\
&&+\SB{-f^+(\omega_k^{lf})-f^+(\omega_p^{nf})+2f^+(\omega_k^{lf})f^+(\omega_p^{nf})}
\mcN_{\parallel,\perp}^{fln}(k^0=\omega_k^{lf})\delta(q^0+\omega_k^{lf}-\omega_p^{nf}) \Big] \label{eq.W11.5}
\end{eqnarray}
where, $\omega_k^{lf} = \sqrt{k_z^2+(M_l^f-s_k\kappa_f e_fB)^2}$, $\omega_p^{nf} = \sqrt{p_z^2+(M_n^f-s_p\kappa_f e_fB)^2}$ and
\begin{eqnarray}
\mcN_{\parallel,\perp}^{fln}(\qpll,\kpll) = g^\munu_{\parallel,\perp}\Tr_\text{d}\TB{\gamma^\mu\scrD_{ns_p}^f(p=q+k)\gamma^\nu\scrD_{ls_k}^f(k)}. \label{eq.N.1}
\end{eqnarray}
It can be seen that, Eq.~\eqref{eq.W11.5} also contains four Dirac delta functions corresponding to the unitary and Landau cuts similar to  Eq.~\eqref{eq.W11.7}, although, the kinematic regions for the different cuts are modified as a consequence of the non zero background field as well as the finite values of the AMM of the quarks~\cite{Ghosh:2017rjo,Ghosh:2019fet,Ghosh:2020xwp}. Analyzing Eq.~\eqref{eq.W11.5},one can find that, the new kinematic domains for the unitary-I and unitary-II cuts are respectively $\sqrt{q_z^2+4\FB{M_f-\kappa_fe_f B}^2}<q^0<\infty$ and $-\infty<q^0<-\sqrt{q_z^2+4\FB{M_f-\kappa_fe_fB}^2}$ whereas the same for the Landau cuts comes out to be 
\begin{eqnarray}
|q^0|<\sqrt{q_z^2+\FB{\sqrt{M_f^2+2|e_fB|}+\kappa_fe_fB-\MB{M_f-\kappa_fe_fB}}^2}. \label{eq.LandauCut}
\end{eqnarray}
The modified analytic structure of $\mcW_{11}^\munu(\qpll)$ in presence of external magnetic field with AMM of quarks is shown in Fig.~\ref{fig.analytic1}. Again restricting ourselves to the situation where  the physical dileptons are produced with positive total energy and time-like resultant momentum quantified as $q^0>0$ and $\qpll^2>0$  (shown as green region in Fig.~\ref{fig.analytic1}), we find that the unitary-II cut does not contribute.  However, contributions from  a portion of the Landau cut is added to the physical region $q^0>0$ and $\qpll^2>0$ which is purely a magnetic field dependent effect. The physical processes associated with this portion of the Landau cut correspond to the emission/absorption of time-like virtual photon with positive energy by  a quark/antiquark which changes its Landau level by unity i.e. the quark in the Landau level $l$ goes to the Landau level $l\pm1$ after absorbing/emitting the photon. This will lead to an enhancement of dilepton yield in the low invariant mass region. 

Next, we perform the $d^2k_\perp$ integral of Eq.~\eqref{eq.W11.5} by using the orthogonality of the  Laguerre polynomials and evaluate  the remaining $dk_z$ integral using the Dirac delta functions which will be converted to theta functions to ensure kinematic restrictions discussed above. This leads to the following  expression
\begin{eqnarray}
g^\munu_{\parallel,\perp}\IM \mcW^{11}_\munu(\qpll) &=& N_c\sum_{f\in\{u,d,s\}} e_f^2 \sum_{s_k\in\{\pm1\}} \sum_{s_p\in\{\pm1\}}
\sum_{l=0}^{\infty}\sum_{n=(l-1)}^{(l+1)} \sum_{k_z\in\{k_z^\pm\}} 
\frac{1}{4}\lambda^{-\frac{1}{2}}\FB{\qpll^2,M_{lfs_k}^2,M_{nfs_p}^2} \nn \\
&& \times  \Big[ \SB{1-f^-(\omega_k^{lf})-f^+(\omega_p^{nf})+2f^-(\omega_k^{lf})f^+(\omega_p^{nf})}
\tilde{\mcN}_{\parallel,\perp}^{fln}(k^0=-\omega_k^{lf}) \Theta \FB{q^0-\sqrt{q_z^2+(M_{lfs_k}+M_{nfs_p})^2}} \nn \\
&& + \SB{1-f^+(\omega_k^{lf})-f^-(\omega_p^{nf})+2f^+(\omega_k^{lf})f^-(\omega_p^{nf})}
\tilde{\mcN}_{\parallel,\perp}^{fln}(k^0=\omega_k^{lf})\Theta \FB{-q^0-\sqrt{q_z^2+(M_{lfs_k}+M_{nfs_p})^2}} \nn \\
&&+\SB{-f^-(\omega_k^{lf})-f^-(\omega_p^{nf})+2f^-(\omega_k^{lf})f^-(\omega_p^{nf})}
\tilde{\mcN}_{\parallel,\perp}^{fln}(k^0=-\omega_k^{lf})\Theta\FB{-q^0-q_\text{min}} \Theta\FB{q^0+q_\text{max}}\nn \\
&& +\SB{-f^+(\omega_k^{lf})-f^+(\omega_p^{nf})+2f^+(\omega_k^{lf})f^+(\omega_p^{nf})}
\tilde{\mcN}_{\parallel,\perp}^{fln}(k^0=\omega_k^{lf})\Theta\FB{q^0-q_\text{min}} \Theta\FB{-q^0+q_\text{max}} \Big]
\label{eq.W11.6}
\end{eqnarray}
where 
\begin{eqnarray}
\tilde{\mcN}_{\parallel}^{fln}(\qpll,\kpll) &=& \frac{|e_fB|}{2\pi}\frac{1}{M_l^fM_n^f}
(1-\delta_l^0\delta_{s_k}^{-1})(1-\delta_n^0\delta_{s_p}^{-1})(M_l^f-s_k\kappa_f e_fB)(M_n^f-s_p\kappa_f e_fB) \nn \\
&& \times \TB{-4|e_fB|n\delta_{l-1}^{n-1} -\delta_{l-1}^{n-1}(M-s_kM_l^f)(M-s_pM_n^f)-\delta_l^n(M+s_kM_l^f)(M+s_pM_n^f)}, \\
\tilde{\mcN}_{\perp}^{fln}(\qpll,\kpll) &=& -\frac{|e_fB|}{2\pi}\frac{1}{M_l^fM_n^f}
(1-\delta_l^0\delta_{s_k}^{-1})(1-\delta_n^0\delta_{s_p}^{-1}) 
\TB{ s_ks_p(\kpll^2+\qpll\cdot\kpll)+ (M_l^f-s_k\kappa_f e_fB)(M_n^f-s_p\kappa_f e_fB)} \nn \\
&& \times \TB{ \delta_{l-1}^{n}(M-s_kM_l^f)(M+s_pM_n^f)-\delta_{l}^{n-1}(M+s_kM_l^f)(M-s_pM_n^f)}
\end{eqnarray}
in which 
\begin{eqnarray}
M_{lfs_k}=|M_l^f-s_k\kappa_f e_fB| 
\end{eqnarray}
is  the effective quark mass when finite values of AMM of the quarks are considered and 
\begin{eqnarray}
k_z^\pm &=& \frac{1}{2\qpll^2}\TB{-q_z(\qpll^2+M_{lfs_k}^2-M_{nfs_p}^2) \pm |q^0|\lambda^{\frac{1}{2}}\fb{\qpll^2,M_{lfs_k}^2,M_{nfs_p}^2} }, \\
q_\text{min} &=& \text{min}\FB{q_z,\frac{M_{lfs_k}-M_{nfs_p}}{|M_{lfs_k}\pm M_{nfs_p}|}\sqrt{q_z^2+(M_{lfs_k}\pm M_{nfs_p})^2}},\\
q_\text{max} &=& \text{max}\FB{q_z,\frac{M_{lfs_k}-M_{nfs_p}}{|M_{lfs_k}\pm M_{nfs_p}|}\sqrt{q_z^2+(M_{lfs_k}\pm M_{nfs_p})^2}}.
\end{eqnarray}
The presence of the step functions in Eq.~\eqref{eq.W11.6} dictates the kinematic domains in which the quantity 
$g^\munu_{\parallel,\perp}\IM \mcW^{11}_\munu(\qpll)$ is non-zero. It can now clearly be seen that, for $q^0>0$ and $\qpll^2>0$, 
the second term within the third bracket (the unitary-II cut) of Eq.~\eqref{eq.W11.6} does not contribute.

Now DPR in presence of arbitrary external magnetic field can be obtained by substituting Eqs.~\eqref{IMLeB} and \eqref{eq.W11.6} into Eq.~\eqref{eq.DPR.5}. Note that, while calculating DPR one needs the constituent quark mass which depends on the external parameters, such as, temperature $ T $, chemical potential $ \mu_B $, external magnetic field $ eB $ and $ \kappa_f $. So we will use NJL model and its modifications to estimate the dependence of constituent quark mass of different flavors on external parameters, i.e. $ M_f = M_f(T, \mu_B, eB, \kappa_f) $.


\section{THE CONSTITUENT QUARK MASS IN THE (P)NJL MODEL} \label{sec.NJL}
In this section we briefly outline few important steps to calculate constituent quark mass using NJL model. Since we want to study the effect of both Polyakov loop and the strange quark in the DPR  from a hot and dense magnetized medium we will consider both two and three-flavor PNJL model. One can always obtain NJL model results starting from PNJL model by modifying the distribution functions~\cite{Hansen:2006ee}.
 \subsection{TWO-FLAVOR MODEL}
The Lagrangian for the two-flavor PNJL model considering
the AMM of free quarks in presence of constant background
magnetic field is given by~\cite{Chaudhuri:2019lbw,Chaudhuri:2020lga}
\begin{equation}
\scrL = \overline{\Psi}(x)\FB{i\fsl{D}-\hat{m} + \gamma_0 \mu_q +\half e\hat{a} \sigma^\munu F_\munu^{\rm ext}  }\Psi(x)  
+ G\SB{ \FB{\overline{\Psi}(x) \Psi (x)}^2 + \FB{ \overline{\Psi}(x) i\gamma_5\tau \Psi(x)}^2} - \mathcal{U} \FB{\Phi , \bar{\Phi } ; T} \label{PNJL_lagrangian}
\end{equation}
where the flavor ($ f=u,d $) and color ($ c=r,g,b $) indices are omitted from the Dirac field $ \FB{ \Psi^{fc}} $ for a convenient representation. In Eq.~\eqref{PNJL_lagrangian}, $\hat{m} = \texttt{diag} \FB{m_u, m_d}$ is current quark mass representing the explicit chiral  symmetry breaking. We will take $ m_u= m_d = m_0 $ to ensure isospin symmetry of the theory at vanishing magnetic field and  $ \mu_q $ is the chemical potential of the quark. The constituent quarks interact with the Abelian gauge fields $ A_\mu  $, the external electromagnetic field $ A_\mu^{\rm ext} $ and the $ {\rm SU_c(3)} $ gauge field $ \mathcal{A}_\mu  $ via the covariant derivative
\begin{eqnarray}
D_\mu = \partial_\mu  -ie\hat{Q}\FB{A_\mu + A_\mu^{\rm ext}}- i \mathcal{A}_\mu^a .
\end{eqnarray}
The factor $\hat{a}= \hat{Q}\hat{\kappa }$,  where $ \hat{Q} = \texttt{diag}(2/3,-1/3) $ and  $\hat{\kappa}=\texttt{diag}(\kappa_u,\kappa_d) $, are  $2\times 2 $ matrices in the flavor space, $ F_\munu^{\rm ext} = (\partial_\mu A_\nu^{\rm ext}- \partial_\nu A_\mu^{\rm ext}) $ and  $ \sigma^\munu=i[\gamma^\mu,\gamma^\nu]/2$.  All the other details can be found in Ref.~\cite{Chaudhuri:2020lga}.  The potential $ \mathcal{U}\FB{\Phi,\bar{\Phi};T} $ in the Lagrangian (Eq.~\eqref{PNJL_lagrangian}) governs the dynamics of the traced Polyakov loop and its conjugate and is given by~\cite{Roessner:2006xn}
\begin{eqnarray}\label{polyakov_potential}
\frac{\mathcal{U}\FB{ \Phi,\bar{\Phi} ;T}}{T^4} = -\frac{a(T) }{2}  \bar{\Phi} \Phi + b(T) \ln \TB{1 - 6 \bar{\Phi} \Phi   + 4 \FB {\bar{\Phi }^3 - \Phi^3   } - 3 \FB{\bar{\Phi} \Phi}^2   }
\end{eqnarray} 
where 
\begin{eqnarray}
a(T) = a_0 + a_1 \FB{\frac{T_0}{T}}+ a_2 \FB{\frac{T_0}{T}}^2 ~~~,~~~
b(T) = b_3 \FB{\frac{T_0}{T}}^3.
\end{eqnarray}
Values of different co-efficients~\cite{Roessner:2006xn} are tabulated in Table~\ref{Table_parameters}.
\begin{center}
	\begin{table}[h]
		\caption{Parameter set for Polyakov potential}
		\begin{tabular} { p{2cm}p{2cm}p{2cm}p{2cm} }
			\hline \hline
			\hspace{0.02IN}$ a_0 $ & \hspace{0.1IN}$ a_1 $ &  \hspace{0.07IN}$ a_2 $   &\hspace{0.082IN} $ b_3 $ \\ 
			\vspace{0.02IN} $ 3.51 $ \vspace{0.02IN}& \vspace{0.02IN}$ -2.47$ & \vspace{0.02IN} $ 15.2 $ &\vspace{0.02IN}  $ -1.75 $  \\
			\hline
		\end{tabular} \label{Table_parameters}
	\end{table}
\end{center}
Following the argument in Ref.~\cite{Gatto:2010qs} we have chosen $ T_0 = 210 $ MeV.  Now under the mean field approximation one can show that, the thermodynamic potential for a  two-flavor Polyakov NJL model considering the AMM of the quarks at finite temperature ( $ T $) and chemical potential ($ \mu_q $) in presence of a uniform background magnetic field can be expressed as
\begin{eqnarray}\label{Omega_PNJL}
\Omega &=&   \frac{(M- m_0)^2}{4 G}+ \potU - 3 \sum_{n,f,s} \frac{\MB{e_f B}}{2\pi } \kzint{p}\omega_{nfs}  \nn \\&& - \frac{1}{\beta} \sum_{n,f,s} \frac{\MB{e_f B}}{2\pi } \kzint{p}\TB{\ln g^{(+)}\FB{\Phi, \bar{\Phi}, T } + \ln g^{(-)}\FB{\Phi, \bar{\Phi}, T } }
\end{eqnarray}
where $ \omega_{nfs} $ are the energy eigenvalues of the  quarks in the presence of external magnetic field as a consequence of the Landau quantization of the transverse momenta of the quarks and is  given by
\begin{equation} 
\omega_{nfs}  = \TB{p_z^2 + \SB{ \FB{\sqrt{\MB{e_f B} (2n+1-s) +M^2} - s\kappa_fe_fB }^2 }}^{\half}
\label{energy}
\end{equation}
with $n$ and $ s $ being the Landau level and the spin indices respectively. The quantities $ g^{(+)}\fnppbT $ and $ g^{(-)}\fnppbT $  are defined as
\begin{eqnarray}
g^{(+)}\fnppbT &=& 1 + 3 \FB{  \Phi + \Fb e^ { - \beta ( \omega_{nfs} -\mu_q )} } e^ { - \beta ( \omega_{nfs} -\mu_q )} + e^ { - 3\beta ( \omega_{nfs} -\mu_q )} ,\\
g^{(-)}\fnppbT &=& 1 + 3 \FB{  \Fb + \F e^ { - \beta ( \omega_{nfs} + \mu_q )} } e^ { - \beta ( \omega_{nfs} + \mu_q )} + e^ { - 3\beta ( \omega_{nfs}+ \mu_q )} .
\end{eqnarray}
Now from Eq.~\eqref{Omega_PNJL}  one can obtain the expressions for the constituent quark mass ($ M $) and the expectation values of the Polyakov loops $ \F $ and $ \Fb $ using the following stationary conditions: 
\begin{equation}
\paroneder{\Omega}{M} = 0 ~~~,~~~ \paroneder{\Omega}{\F} = 0 ~~~\text{and}~~~ \paroneder{\Omega}{\Fb} =0
\end{equation} 
which leads to the following sets of coupled equations
\begin{eqnarray}
&& M = m_0 + 6G \sum_{n,f,s} \frac{\MB{e_f B }}{2\pi^2 } \intzinf dp_z \frac{M}{\omega_{nfs}} \FB{1 -  \frac{s\kappa_f q_f B}{\tilde{M}_{nfs}}}
-6G \sum_{n,f,s} \frac{\MB{e_f B}}{2\pi^2 } \intzinf dp_z \frac{M}{\omega_{nfs} } \FB{ 1 - \frac{ s\kappa_f e_f B}{\tilde{M}_{nfs}}  } 
\nn \\ && \hspace{10cm}
\times \TB{ \frac{}{}  f^+ \fnppbT + f^- \fnppbT} \label{Gap_M}~, \\
&& \SB{- \frac{a ( T )}{2}\Fb - 6b(T)\frac{\Fb  - 2 \F^2 + \FB{ \Fb \F } \F}{  1 - 6 \Fb \F + 4 \FB{\F^3 + \Fb^3} - 3\FB{\Fb \F}^2 }   } = \frac{3}{T^3} \sum_{n,f,s} \frac{\MB{e_f B}}{2 \pi^2 } \intzinf dp_z \TB{\frac{e^{- \beta( \omega_{nfs} -\mu_q) }}{g^{(+)}}  + \frac{e^{- 2\beta( \omega_{nfs} +\mu_q) }}{g^{(-)}}        }  \label{Gap_p}, \\
&& \SB{- \frac{a ( T )}{2}\F - 6b(T)\frac{ \F - 2 \Fb^2 + \FB{ \Fb \F } \Fb}{  1 - 6 \Fb \F + 4 \FB{\F^3 + \Fb^3} - 3\FB{\Fb \F}^2 }   } = \frac{3}{T^3} \sum_{n,f,s} \frac{\MB{e_f B}}{2 \pi^2 } \intzinf dp_z \TB{\frac{e^{- 2\beta( \omega_{nfs} -\mu_q) }}{g^{(+)}}  + \frac{e^{- \beta( \omega_{nfs} +\mu_q) }}{g^{(-)}}        }  \label{Gap_pb}
\end{eqnarray}
where 
\begin{eqnarray}
\tilde{M}_{nfs} &=& \sqrt{ \MB{e_f B} \FB{2n + 1 -s   }  + M^2  }~, \\
f^+ \fnppbT &=& \frac{	\FB{\F + 2\Fb e^{-\beta( \omega_{nfs}  -\mu_q ) }} e^{-\beta( \omega_{nfs}  -\mu_q ) } + e^{-3\beta( \omega_{nfs}  -\mu_q ) }  	}{ 1 + 3 \FB{  \Phi + \Fb e^ { - \beta ( \omega_{nfs} -\mu_q )} } e^ { - \beta ( \omega_{nfs} -\mu_q )} + e^ { - 3\beta ( \omega_{nfs} -\mu_q )}  }~,\label{eq.fplus} \\
f^- \fnppbT &=& \frac{	\FB{\Fb + 2\F e^{-\beta( \omega_{nfs}  +\mu_q ) }} e^{-\beta( \omega_{nfs}  +\mu_q ) } + e^{-3\beta( \omega_{nfs} + \mu_q ) }  	}{ 1 + 3 \FB{  \Fb + \F e^ { - \beta ( \omega_{nfs} +\mu_q )} } e^ { - \beta ( \omega_{nfs} +\mu_q )} + e^ { - 3\beta ( \omega_{nfs} +\mu_q )}  }~. \label{eq.fminus} 
\end{eqnarray}
One has to solve Eqs.~\eqref{Gap_M},~\eqref{Gap_p} and \eqref{Gap_pb} self-consistently of obtain $ T $ and/or $ \mu_q $ dependence of $ M, \Phi  $ and $ \bar{\Phi} $ for different values of background magnetic field.   Note that in Eq.~\eqref{Gap_M}, the medium independent integral is ultraviolet divergent. Since the theory is known to be non-renormalizable owing to the point-like interaction between the quarks, a proper regularization scheme is necessary. Our preferred regularization scheme throughout this article is the Pauli-Villars (PV) regularization which we apply only to the diverging vacuum terms. One of the most important features of the PV scheme is that, it preserves the Gauge invariance~\cite{Klevansky}. With PV regularization, the calculations are more reliable at larger external parameters like temperature, density and electromagnetic fields as it allows for a larger energy scale~\cite{Wang:2017vtn}. It is shown in Refs.~\cite{Mao:2016fha, Mao:2016lsr} that, when a cut-off scheme is used in the chiral limit, the transverse velocity of the Goldstone mode is larger than the speed of light violating the causality. However with the PV scheme, causality is always guaranteed. In Refs.~\cite{Carignano:2011gr,Carignano:2011gr}, it is discussed that, a sharp three-momentum cutoff introduces large regularization artifacts when dealing with inhomogeneous phases. 

Let us now outline few important steps describing the implementation of PV scheme. In case of vanishing magnetic field, under the PV regularization, the vacuum integrals can be regularized by making the following replacements~\cite{Klevansky}
\begin{equation}\label{PVrep}
\kthreeint{p}f(\omega) \rightarrow \sum_{i=0}^N C_i\kthreeint{p} f(\omega_i) 
\end{equation}
 where $ \omega_i^2 = \vec{p}^2 + M^2 + b_i \Lambda^2 $ with the regularized mass $ M^2 \rightarrow M_i^2 =  M^2 + b_i \Lambda^2 $. The coefficients $ C_i  $ and $ b_i $ are determined by following constrain relations
 \begin{equation}\label{Constrain_PVrep}
 \sum_{i = 0}^{N} C_i = 0 ~~~~\text{and}~~~~~ \sum_{i = 0}^{N} C_i M_i^{2} = 0
  \end{equation}
  with $ C_0 = 1 $ and $ b_0 = 0 $.

  In the case of finite magnetic field considering the nonzero values of the AMM of the quarks, the energy of quarks are given by $ \omega_{nfs}^2 = {p_z^2 + M_{\rm eff}^2}$, where $ M_{\rm eff}^2 = \FB{\sqrt{\MB{e_f B} (2n+1-s) +M^2} - s\kappa_fe_fB }^2 = \FB{\tilde{M}_{nfs}^2 - s  \kappa_f e_f B }^2 $. Here, under PV scheme we make following replacements~\cite{Mao:2016fha,Mao:2014hga,Mao:2016lsr,Mei:2020jzn}
  \begin{equation}\label{PVrepeB}
  \sum_n \kzint{p} f ( \omega_{nfs}) \rightarrow \sum_{i=0}^N C_i \sum_n \kzint{p} f ( \omega_{nfs,i}) 
  \end{equation}
  where $  \omega_{nfs,i} = p_z^2 + M_{\rm eff}^2 + b_i \Lambda^2 $.  The co-efficients are determined using 
  \begin{equation}\label{constrain_PVeB}
  \sum_{i = 0}^{N} C_i M_i^{2j} = 0 ~~~~{\rm for~} j= 0,1,2,\dots, N - 1
  \end{equation}
  with $ C_0 = 1, b_0 =0 $ and $ M_i^2 =  m_{\rm eff}^2 + b_i \Lambda^2  $. In all the treatments in this article, we have used $ N = 3 $  with $ b_i=\SB{0,1,2,3 }$ and $ C_i = \SB{1,-3,3,-1}$ for convergence.

\subsection{THREE-FLAVOR MODEL}
The Lagrangian for the three-flavor PNJL model considering the AMM of free quarks in presence of constant background magnetic field is given by
\begin{eqnarray}
\scrL &=& \overline{\Psi}(x)\FB{i\fsl{D}-\hat{m} + \gamma_0 \mu_q +\half e\hat{a} \sigma^\munu F_\munu^{\rm ext}  }\Psi(x)  
+ G_S \sum_{a=0}^{8}\SB{ \FB{\overline{\Psi}(x) \lambda^a \ \Psi (x)}^2 + \FB{ \overline{\Psi}(x) i\gamma_5\lambda^a \Psi(x)}^2} \nn \\&& \hspace{0.92IN}-~ K \TB{\det \overline{\Psi}\FB{1 + \gamma_5} \Psi + \det \overline{\Psi}\FB{1 - \gamma_5} \Psi} + \mathcal{U} \FB{\Phi , \bar{\Phi } ; T}. \label{3PNJL_lagrangian}
\end{eqnarray}
In the above, $ \Psi = (u~ d~ s)^T $ is the quark field with three flavors ($ N_f = 3 $) and three colors ($ N_c = 3 $), $ \hat{m}= \texttt{diag}(m_u , m_d , m_s ) $ is the current quark mass matrix, and $ \lambda^a $ are the flavor $ SU_f(3) $ Gell-Mann matrices. The isospin symmetry on the Lagrangian level is assumed, i.e., $ m_u = m_d = m_0 $ , while $ SU_f(3) $-symmetry is explicitly broken, so that $ m_s \ne m_0 $.  $ G_S $ represents the strength of scalar  coupling and the $ K $ term represents the six-point Kobayashi-Maskawa-t’Hooft (KMT) interaction that breaks the axial $ U(1)_A $ symmetry (see~\cite{Klevansky} for a general review on three-flavor model). The factor $\hat{a}= \hat{Q}\hat{\kappa }$,  where $ \hat{Q} = \texttt{diag}(2/3,-1/3, -1/3) $ and  $\hat{\kappa}=\texttt{diag}(\kappa_u,\kappa_d, \kappa_s) $, are  $3\times 3 $ matrix in the flavor space. All the other terms are defined in same manner as in case of two-flavor model. Again using mean field approximation one can show that
\begin{eqnarray}
\Omega &=&\potU + G_S \sum_{ f\in~ \{u,d,s\} } \AB{\overline{\Psi}\Psi }_f^2 + 4K \AB{\overline{\Psi}\Psi }_u\AB{\overline{\Psi}\Psi }_d\AB{\overline{\Psi}\Psi }_s - 3 \sum_{n,f,s} \frac{\MB{e_f B}}{2\pi } \kzint{p}\omega_{nfs}  \nn \\&& - \frac{1}{\beta} \sum_{n,f,s} \frac{\MB{e_f B}}{2\pi } \kzint{p}\TB{\ln g^{(+)}\FB{\Phi, \bar{\Phi}, T } + \ln g^{(-)}\FB{\Phi, \bar{\Phi}, T } } \label{pot3fPNJL}
\end{eqnarray}
where
\begin{eqnarray}
 \AB{\overline{\Psi}\Psi }_f =- N_c \frac{\MB{e_f B}}{2 \pi } \sum_{n,s} \kzint{p} \frac{M_f}{\omega_{nfs}}  \FB{1 - \frac{s\kappa_f e_f B}{\tilde{M}_{nfs}} } \TB{ \frac{}{} 1 -  f^+ \fnppbT + f^- \fnppbT }.
\end{eqnarray}
We obtain the gap equations by minimizing the thermodynamic potential given by Eq.~\eqref{pot3fPNJL}  with respect to the order parameters $ (M_u , M_d , M_s , \F, \Fb) $ which yields a set of five coupled equations
\begin{eqnarray}
 && \hspace{2IN}M_u = m_0 - 4 G_s \AB{\overline{\Psi}\Psi }_u + 2K \AB{\overline{\Psi}\Psi }_d \AB{\overline{\Psi}\Psi }_s  \label{Gap_M31}~, \\
&& \hspace{2IN}M_d = m_0 - 4G_s \AB{\overline{\Psi}\Psi }_d + 2K \AB{\overline{\Psi}\Psi }_s \AB{\overline{\Psi}\Psi }_u  \label{Gap_M32}~, \\
&&\hspace{2IN} M_s = m_s - 4G_s \AB{\overline{\Psi}\Psi }_s + 2K \AB{\overline{\Psi}\Psi }_u \AB{\overline{\Psi}\Psi }_d  \label{Gap_M33}~, \\
&& \SB{- \frac{a ( T )}{2}\Fb - 6b(T)\frac{\Fb  - 2 \F^2 + \FB{ \Fb \F } \F}{  1 - 6 \Fb \F + 4 \FB{\F^3 + \Fb^3} - 3\FB{\Fb \F}^2 }   } = \frac{3}{T^3} \sum_{n,f,s} \frac{\MB{e_f B}}{2 \pi^2 } \intzinf dp_z \TB{\frac{e^{- \beta( \omega_{nfs} -\mu_q) }}{g^{(+)}}  + \frac{e^{- 2\beta( \omega_{nfs} +\mu_q) }}{g^{(-)}}        }  \label{Gap_p3}, \\
&& \SB{- \frac{a ( T )}{2}\F - 6b(T)\frac{ \F - 2 \Fb^2 + \FB{ \Fb \F } \Fb}{  1 - 6 \Fb \F + 4 \FB{\F^3 + \Fb^3} - 3\FB{\Fb \F}^2 }   } = \frac{3}{T^3} \sum_{n,f,s} \frac{\MB{e_f B}}{2 \pi^2 } \intzinf dp_z \TB{\frac{e^{- 2\beta( \omega_{nfs} -\mu_q) }}{g^{(+)}}  + \frac{e^{- \beta( \omega_{nfs} +\mu_q) }}{g^{(-)}}  }.  \label{Gap_pb3}
\end{eqnarray}
Again the self-consistent solution of Eqs.~\eqref{Gap_M31}--\eqref{Gap_pb3} results in $ T $ and/or $ \mu_q $ dependence of $ M_u, M_d, M_s, \F $ and  $ \Fb  $ for different background magnetic field. We would like to emphasise that, in all the calculations presented in this paper, no approximation is  made while considering the strength of the magnetic field.

 \subsection{PARAMETERS OF THE (P)NJL MODEL}
 For the two-flavor model , we take a parameter set of Table~\ref{Table_2flv} which has been fitted to the vacuum values of the pion mass and the pion decay constant decay constant ($f_\pi$). 
 	\begin{table}[h]
 		 \begin{center}
 		\caption{Parameter set for two-flavor model}
 		\begin{tabular} { p{2cm}p{2cm}p{2cm} }
 			\hline \hline
 			\hspace{0.02IN}$ m_0 $ (MeV) & \hspace{0.1IN}$ \Lambda $ (MeV) &  \hspace{0.07IN}$ G\Lambda^2  $  \\ 
 			\vspace{0.032IN} \hspace{0.109IN}$ 10.3 $ \vspace{0.032IN}& \vspace{0.032IN} \hspace{0.109IN}$744.2$ & \vspace{0.032IN} \hspace{0.09IN}$6.21$  \\
 			\hline
 		\end{tabular}\label{Table_2flv}
 	 \end{center}
 	\end{table}
 	\begin{table}[h]
 	\begin{center}
 		\caption{Parameter set for three-flavor model}
 		\begin{tabular} { p{3cm}p{2cm}p{2cm}p{1.5cm}p{1.5cm} }
 			\hline \hline
 			\hspace{0.02IN}$ m_u = m_d $ (MeV) & $ m_s  $ (MeV)&  \hspace{0.1IN}$ \Lambda $ (MeV) &  \hspace{0.07IN}$ G_s\Lambda^2  $  & \hspace{0.08IN}$ K\Lambda^5 $ \\ 
 			\vspace{0.032IN} \hspace{0.2IN}$ 10.3 $ \vspace{0.032IN}& \vspace{0.032IN} \hspace{0.109IN}$236.9$ & \vspace{0.032IN} \hspace{0.15IN}$781.2$ & \vspace{0.032IN} \hspace{0.07IN}$ 4.90 $ &\vspace{0.032IN} \hspace{0.0IN} $ 129.8 $ \\
 			\hline
 		\end{tabular} \label{Table_3flv}
 	\end{center}
 \end{table}

For the three-flavor model we have chosen the parameters following Ref.~\cite{Carignano:2019ivp} and they are tabulated in Table~\ref{Table_3flv}. The following values of AMM of the quarks are used $ \kappa_u = 0.101,~\kappa_d = 0.175  $ and $ \kappa_s = 0.049$ in units of  $ ~\rm GeV^{-1}$~\cite{Aguirre:2020tiy} through out this article.
\begin{figure}[h]
	\begin{center}
		\includegraphics[angle=-90,scale=0.32]{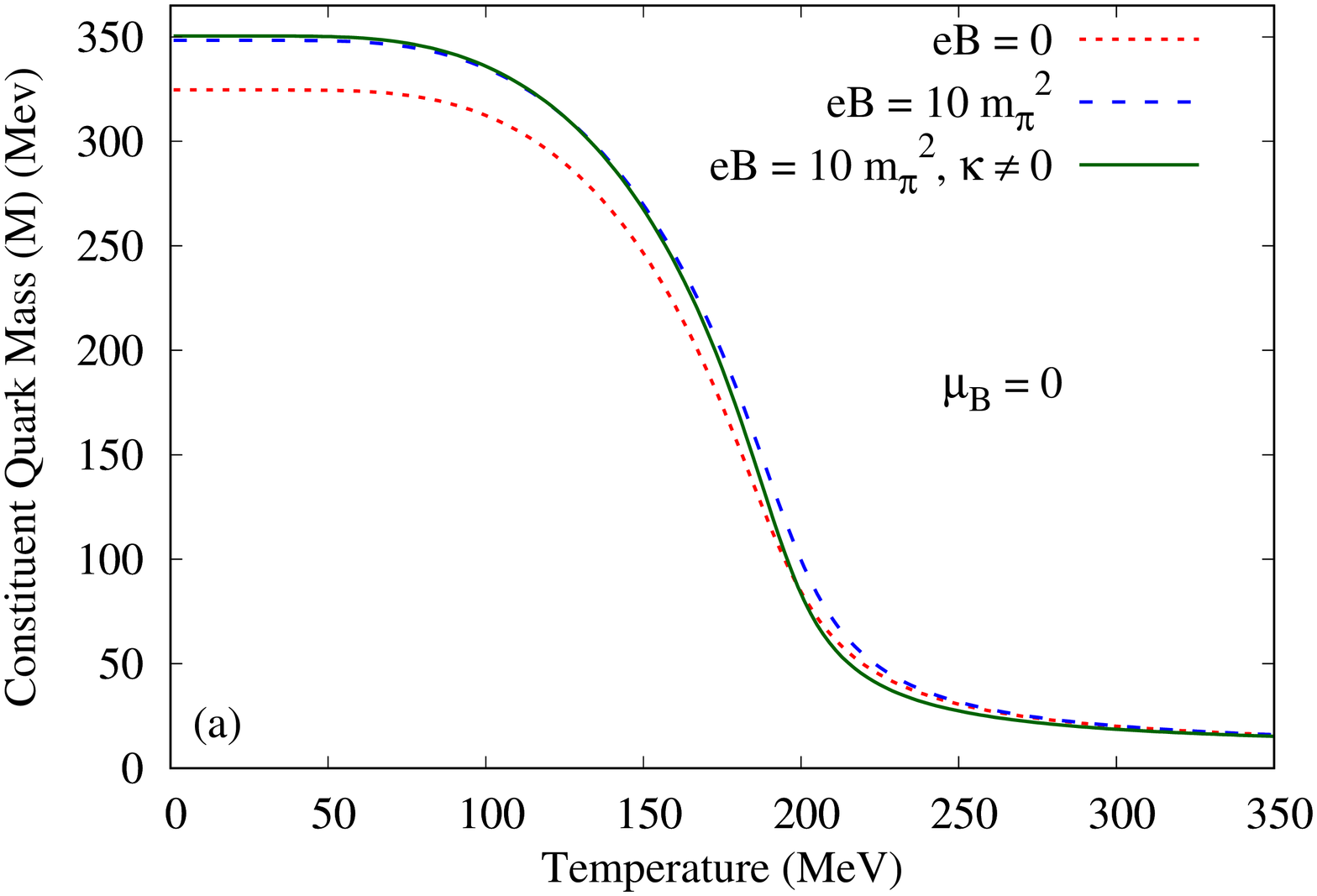}  \includegraphics[angle=-90,scale=0.32]{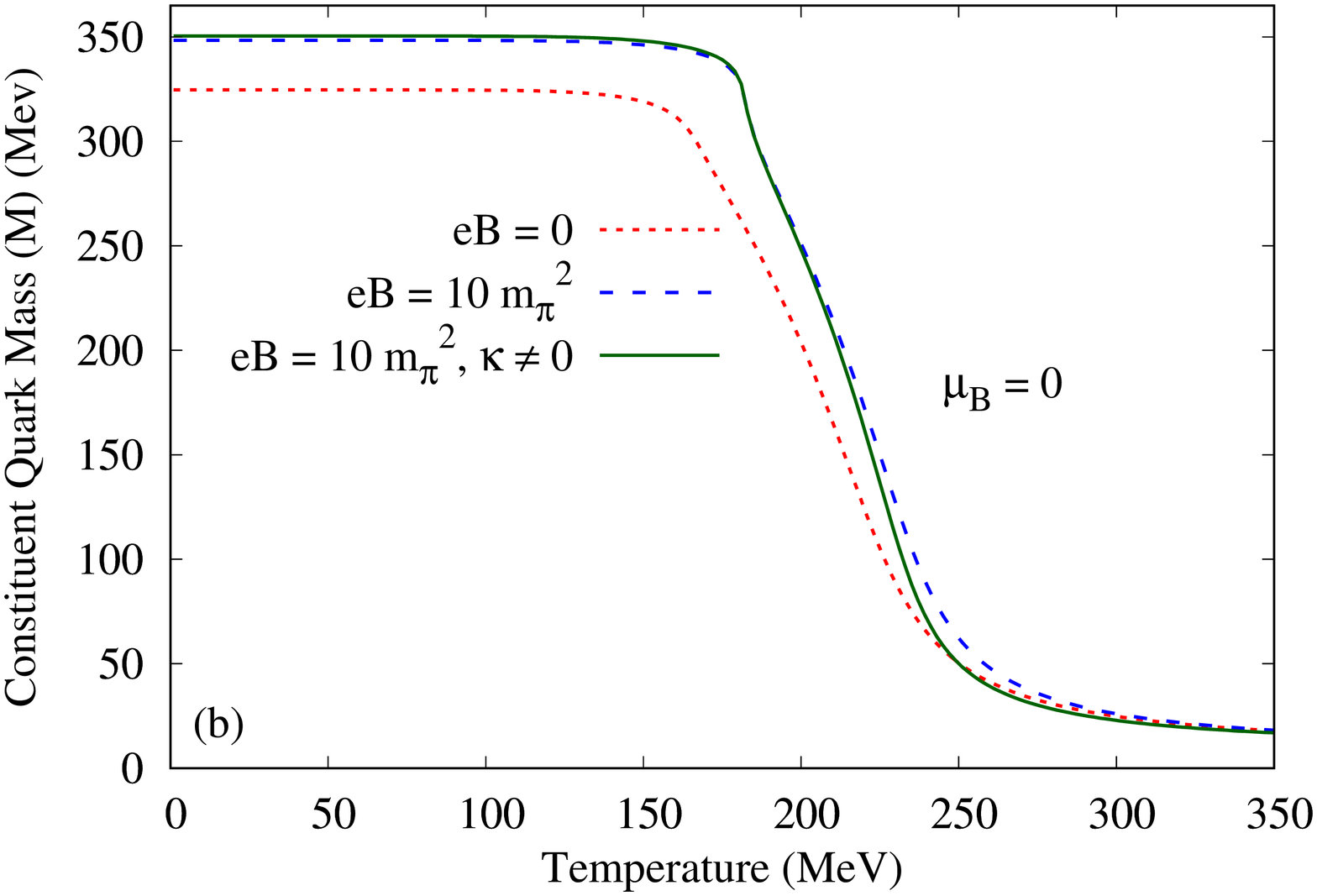} 
		\includegraphics[angle=-90,scale=0.32]{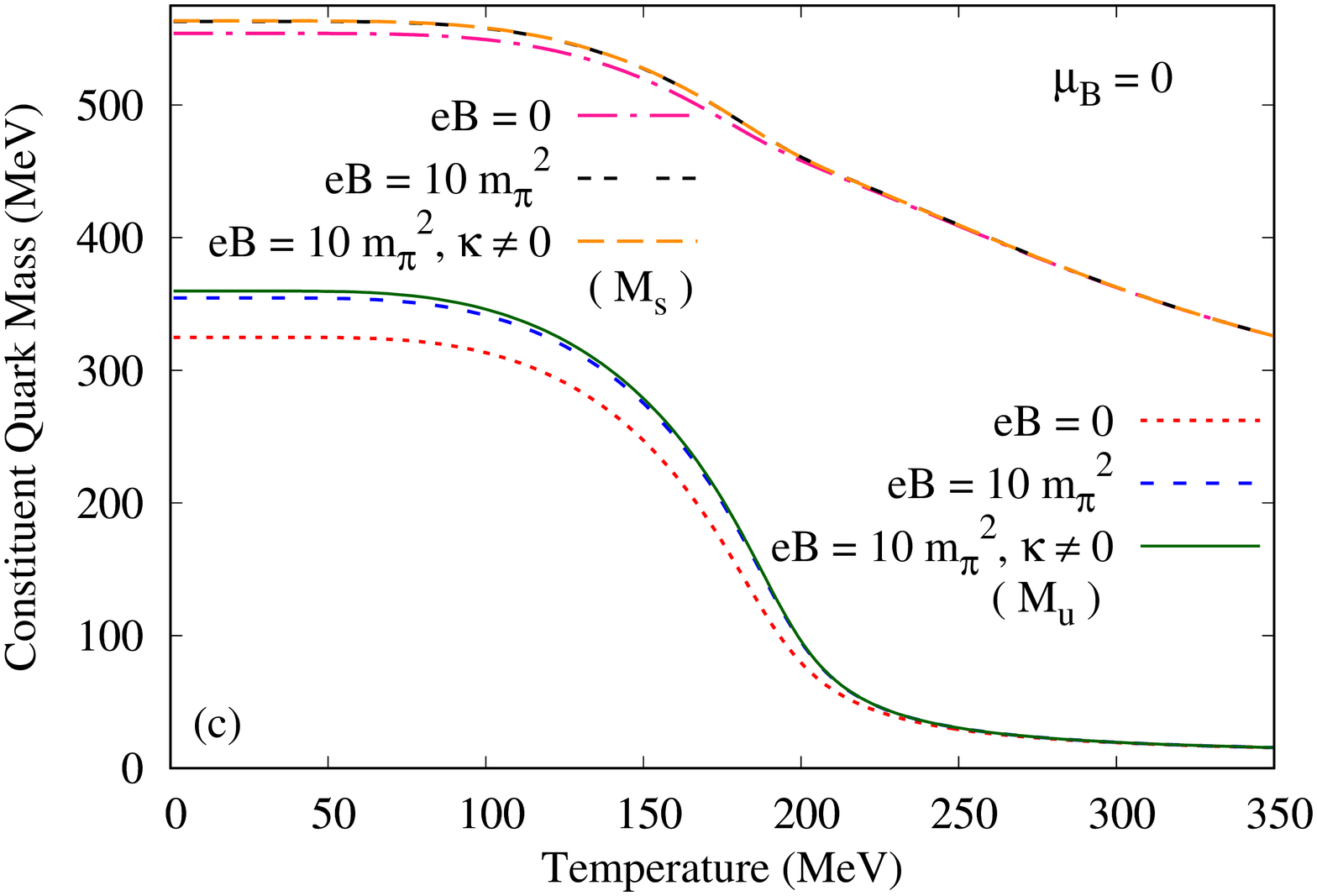}  \includegraphics[angle=-90,scale=0.32]{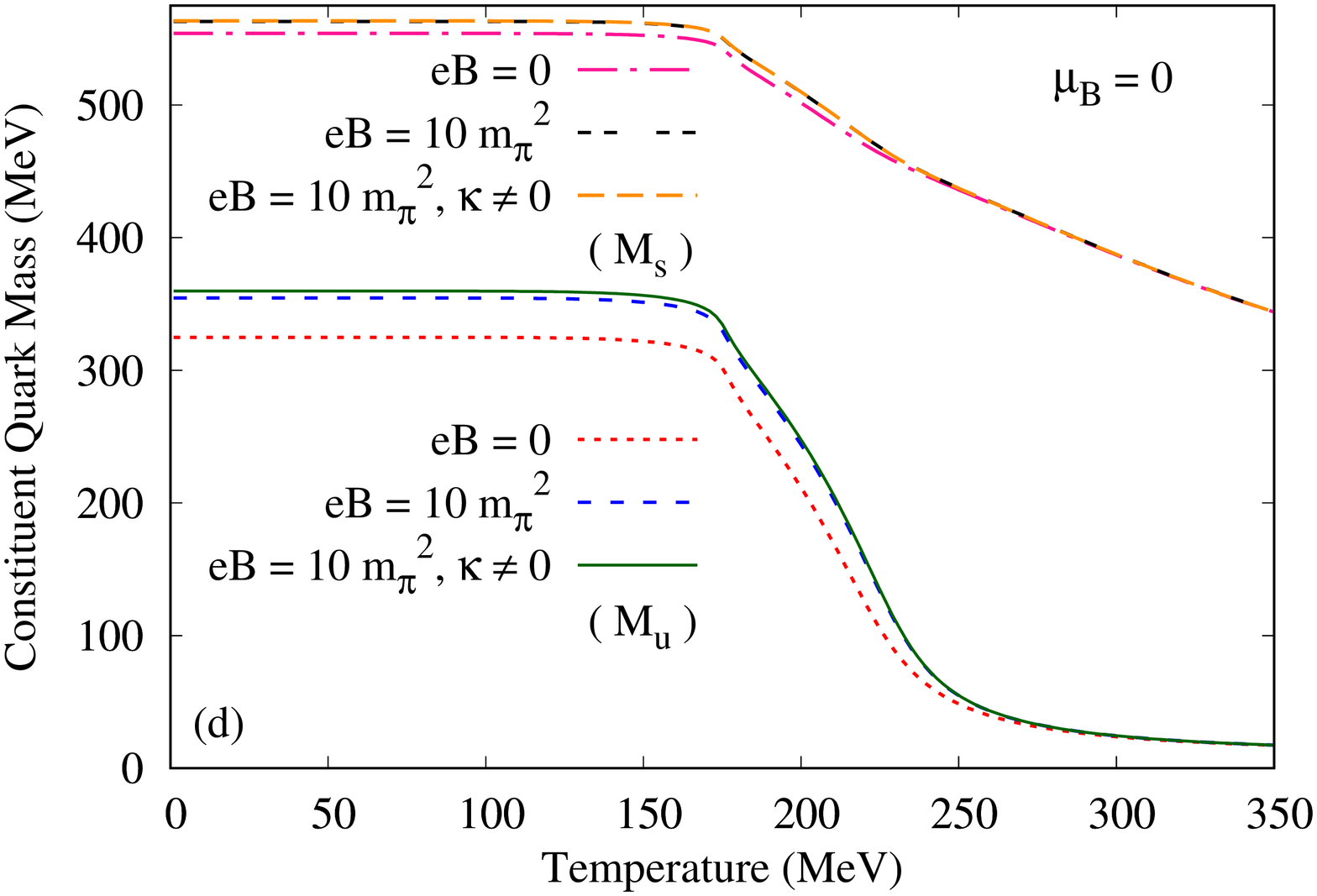}
	\end{center}
	\caption{(Color Online) Variation of the constituent quark mass ($M$) as a function of temperature for different values of external magnetic field and AMM of the quarks at $ \mu_B = 0 $ using (a) two-flavor NJL model  (b) two-flavor PNJL model (c) three-flavor NJL model and (d) three-flavor PNJL model.}
	\label{fig.M1}
\end{figure}
\section{NUMERICAL RESULTS}\label{sec.results}
We first briefly discuss the numerical results for constituent quark masses of different flavors which are the main inputs while calculating the  DPR in the presence of a background magnetic field. In Fig.~\ref{fig.M1} we have shown the variation of the constituent quark masses ($ M $) as a function of temperature $ (T) $ at zero baryon chemical potential $ (\mu_B) $ for both zero and nonzero values of the background magnetic field with and without considering the finite values of AMM of the quarks using  (a) two-flavor NJL model, (b) two-flavor PNJL model, (c) three-flavor NJL model and (d) three-flavor PNJL model respectively. The overall behaviour of constituent masses of the low lying quarks (up and down) are qualitatively similar in all the cases as it starts from a high value at low $ T  $, remains almost constant in the smaller values of $ T $, falls off sharply in a small range of temperature and finally become nearly equal to the bare masses of the quarks at high $ T $ values representing the (partial) restoration of the chiral symmetry. However, the constituent mass for the strange quark decreases smoothly when compared to that of the up/down quark, as shown in Figs.~\ref{fig.M1} (c) and (d).  It is also worth mentioning that even at $ T= 350  $~MeV, the $ s $ quark mass is  still higher than its current mass. Now, from Fig.~\ref{fig.M1}~(a), it can be seen that in presence of background magnetic field, the constituent quark mass increases at $T\rightarrow 0$ limit and the transitions to the symmetry restored phase take place at the larger values of temperature. This phenomenon is known as magnetic catalysis (MC)~\cite{Shovkovy:2012zn,Gusynin:1994re,Gusynin:1995nb,Gusynin:1999pq}, which explains that the magnetic field has a strong tendency to enhance (or catalyze) spin-zero fermion-antifermion condensates $ \langle\overline{\Psi}\Psi\rangle $. Moreover, when finite values of the AMM of the quarks are considered at low $ T $ the effect of MC slightly strengthen which leads to small increase in the values of constituent quark masses. From Fig.~\ref{fig.M1}~(b), one can observe that, as a result of including the Polyakov loop in the two-flavor model constituent mass goes to bare quark mass limit at higher values of $ T $ compare to the simple two-flavor NJL model. In three-flavor case the effects of magnetic field and the Polyakov loop are similar as described above. However, it is important to  note that,  the magnitude of electric charge of each quark  plays a dominant role in MC effect~\cite{Ferreira:2013oda}. This is manifested in the fact that, increase in the constituent mass of strange quarks are smaller compared to that of up quarks.  A similar effect can also be seen in constituent mass of down quarks which are not plotted here. 

\begin{figure}[h]
	\begin{center}
		\includegraphics[angle=-90,scale=0.30]{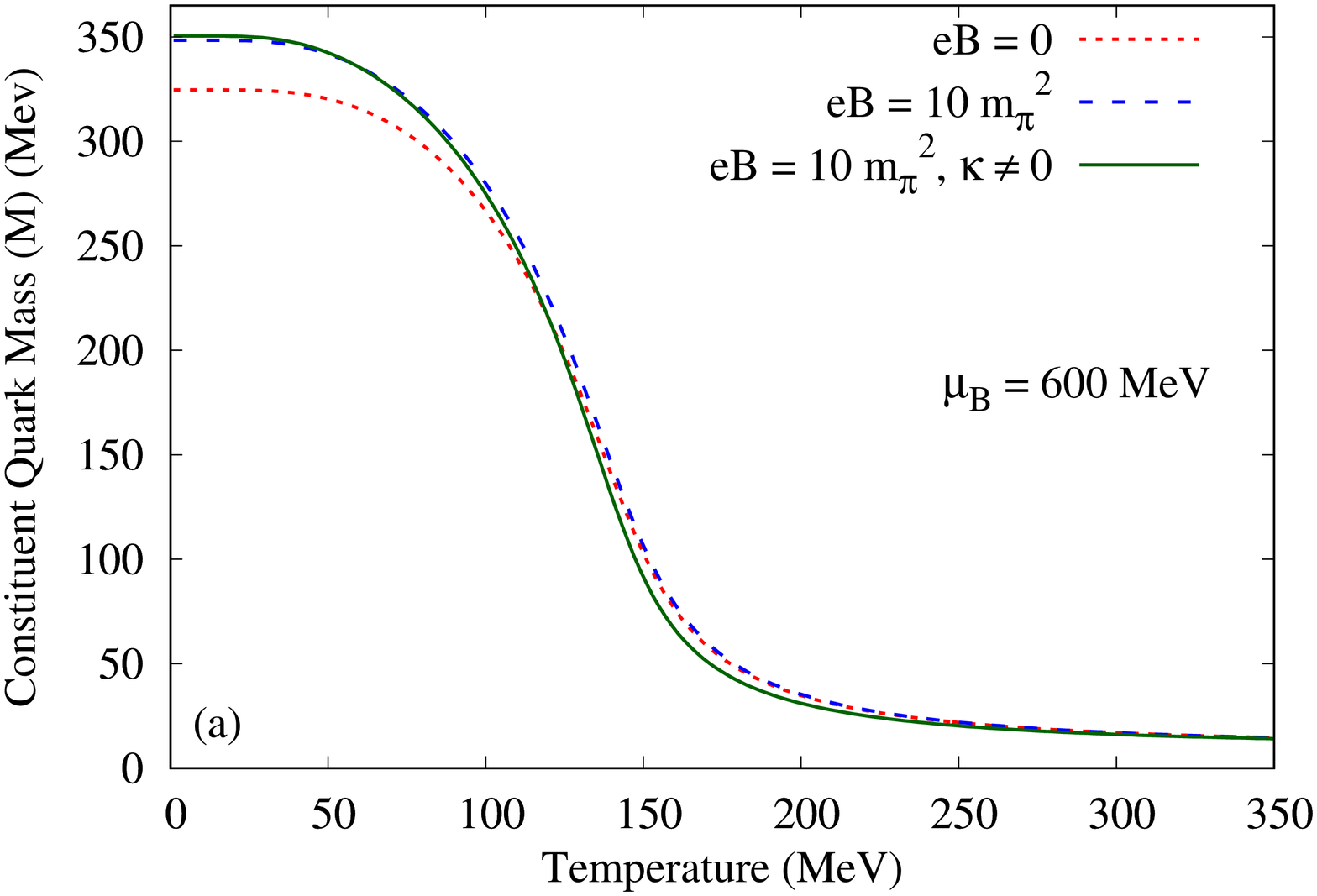}  \includegraphics[angle=-90,scale=0.30]{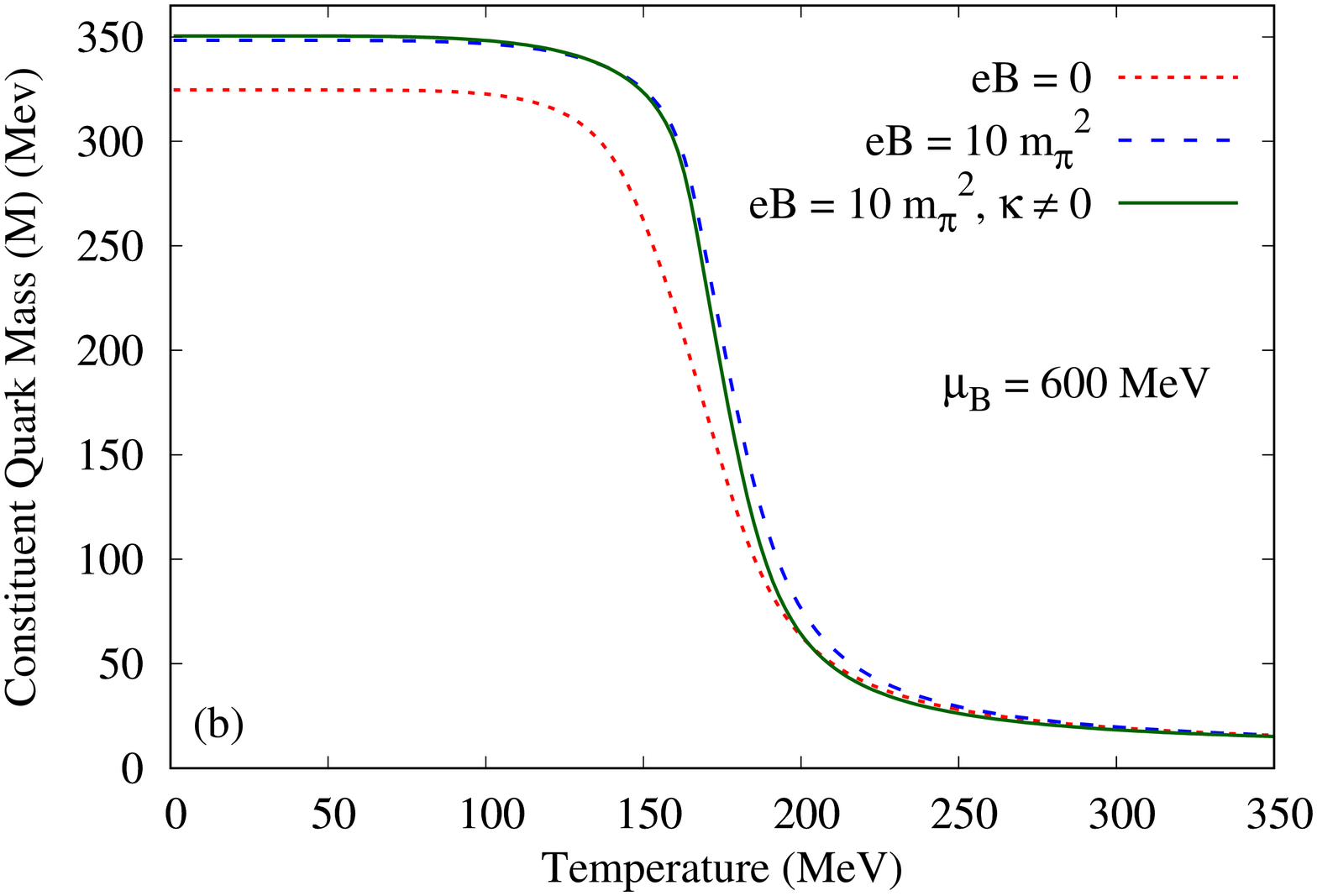} \\
		\includegraphics[angle=-90,scale=0.30]{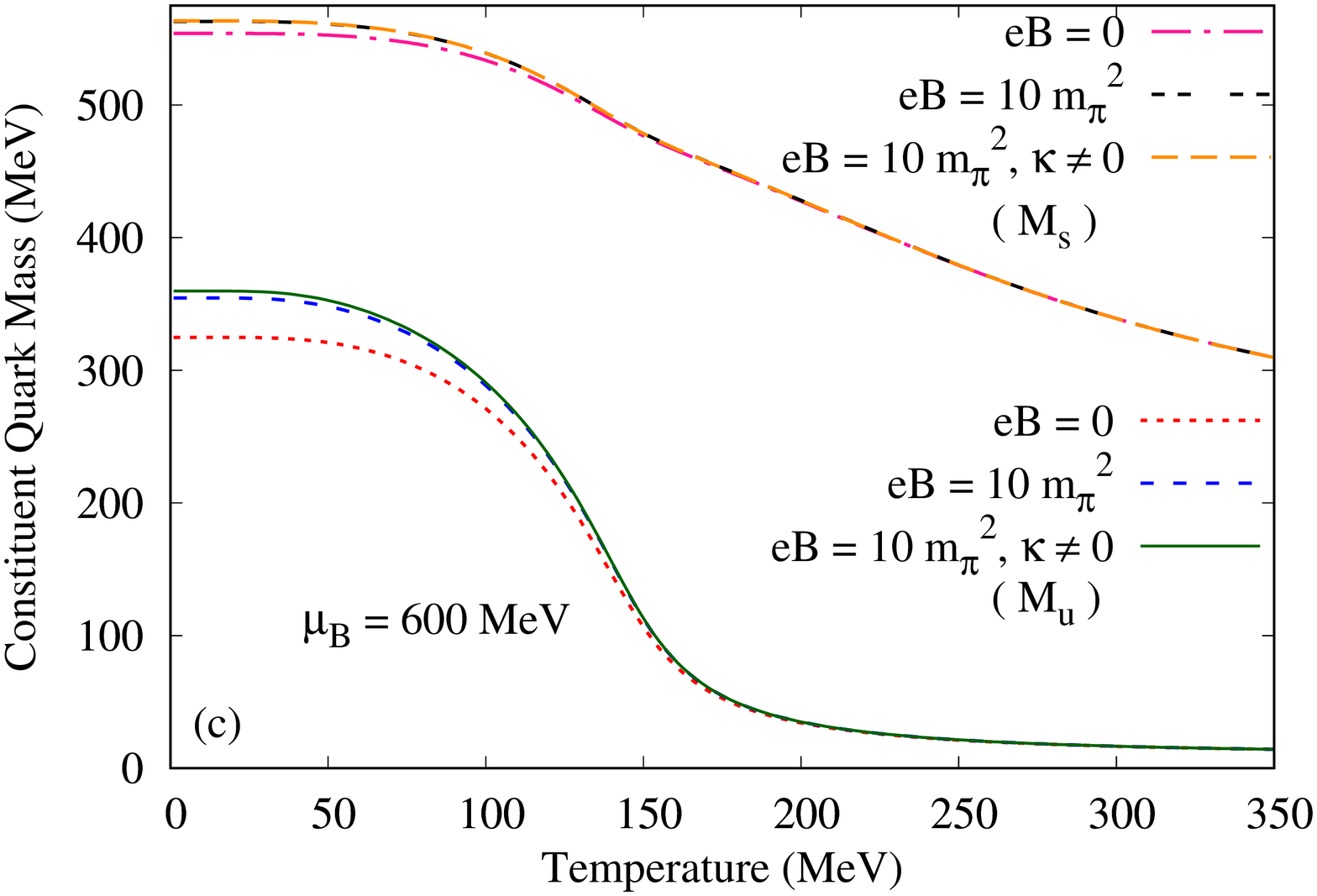}  \includegraphics[angle=-90,scale=0.30]{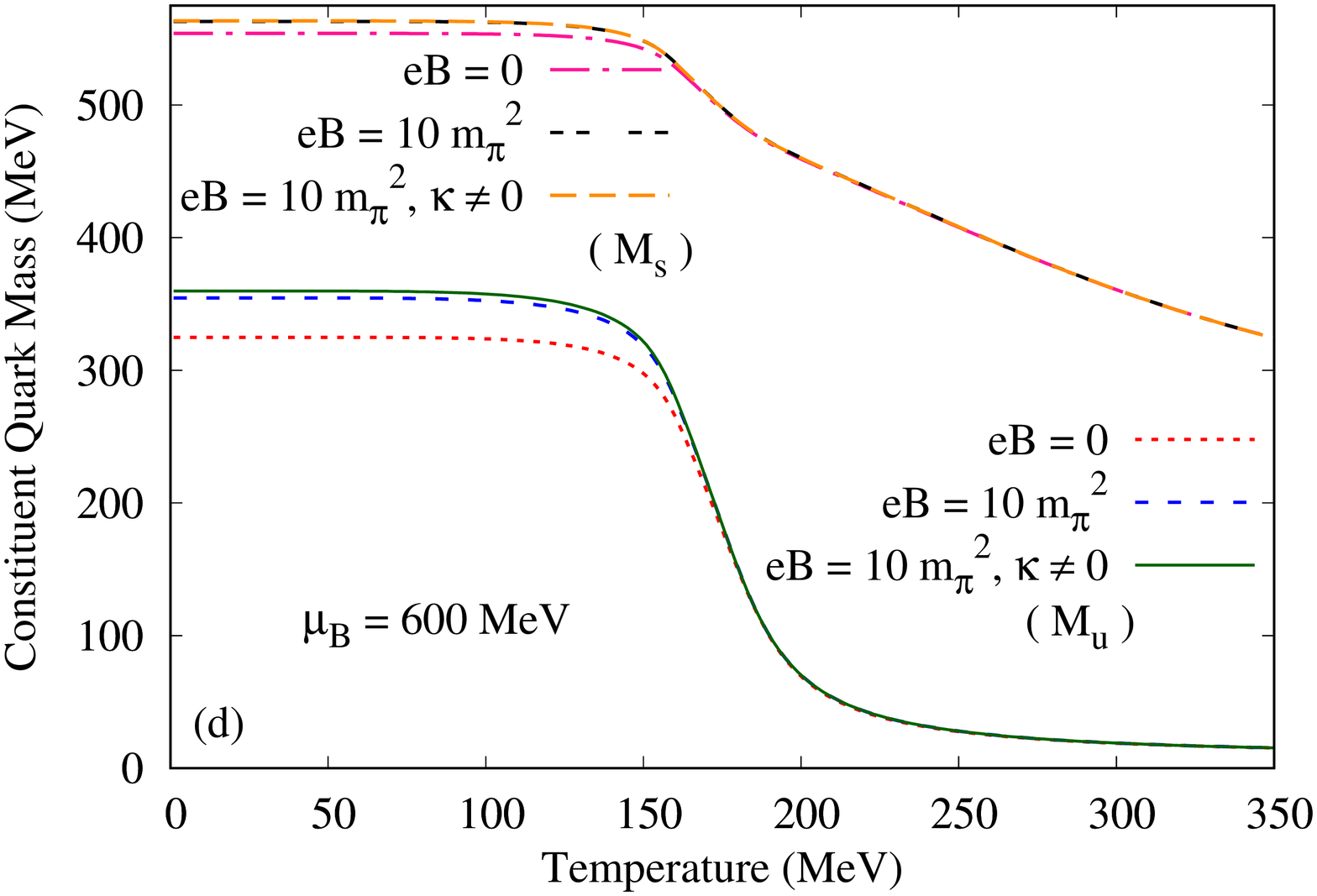}
	\end{center}
	\caption{(Color Online) Variation of the constituent quark mass ($M$) as a function of temperature for different values of external magnetic field and AMM of the quarks at $ \mu_B = 600  $  MeV using (a) two-flavor NJL model  (b) two-flavor PNJL model (c) three-flavor NJL model and (d) three-flavor PNJL model.}
	\label{fig.M2}
\end{figure}

In Fig.~\ref{fig.M2}, the temperature dependence of constituent quark masses are plotted at finite baryon chemical potential considering the same physical scenario as previously mentioned.  One can observe that, there are no major modifications  at lower temperatures but the system goes to (pseudo) chiral symmetric phase at smaller values of temperature in contrast to the $ \mu_B  = 0$ case.
\begin{figure}[h]
	\begin{center}
		\includegraphics[angle=-90,scale=0.32]{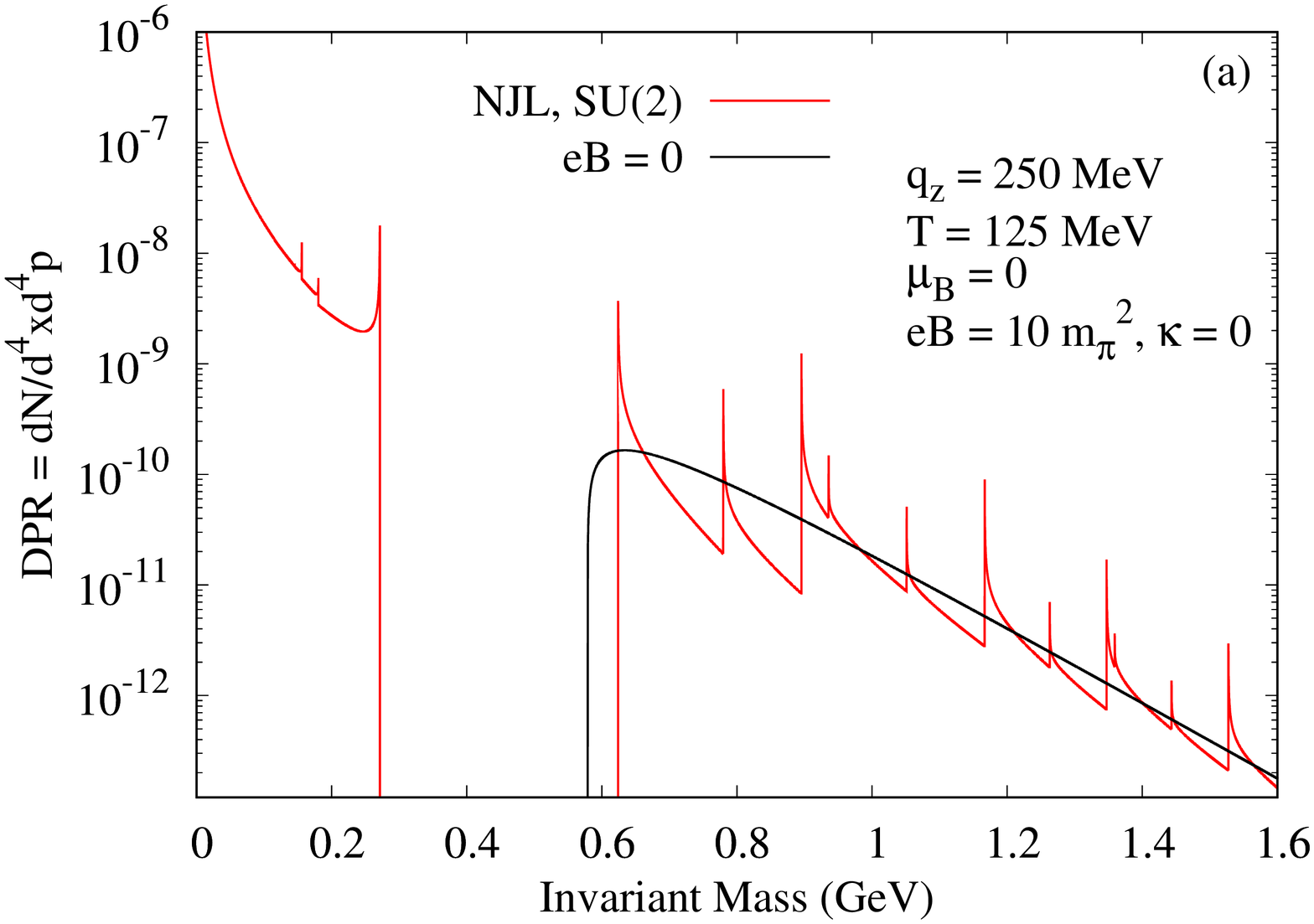}  \includegraphics[angle=-90,scale=0.32]{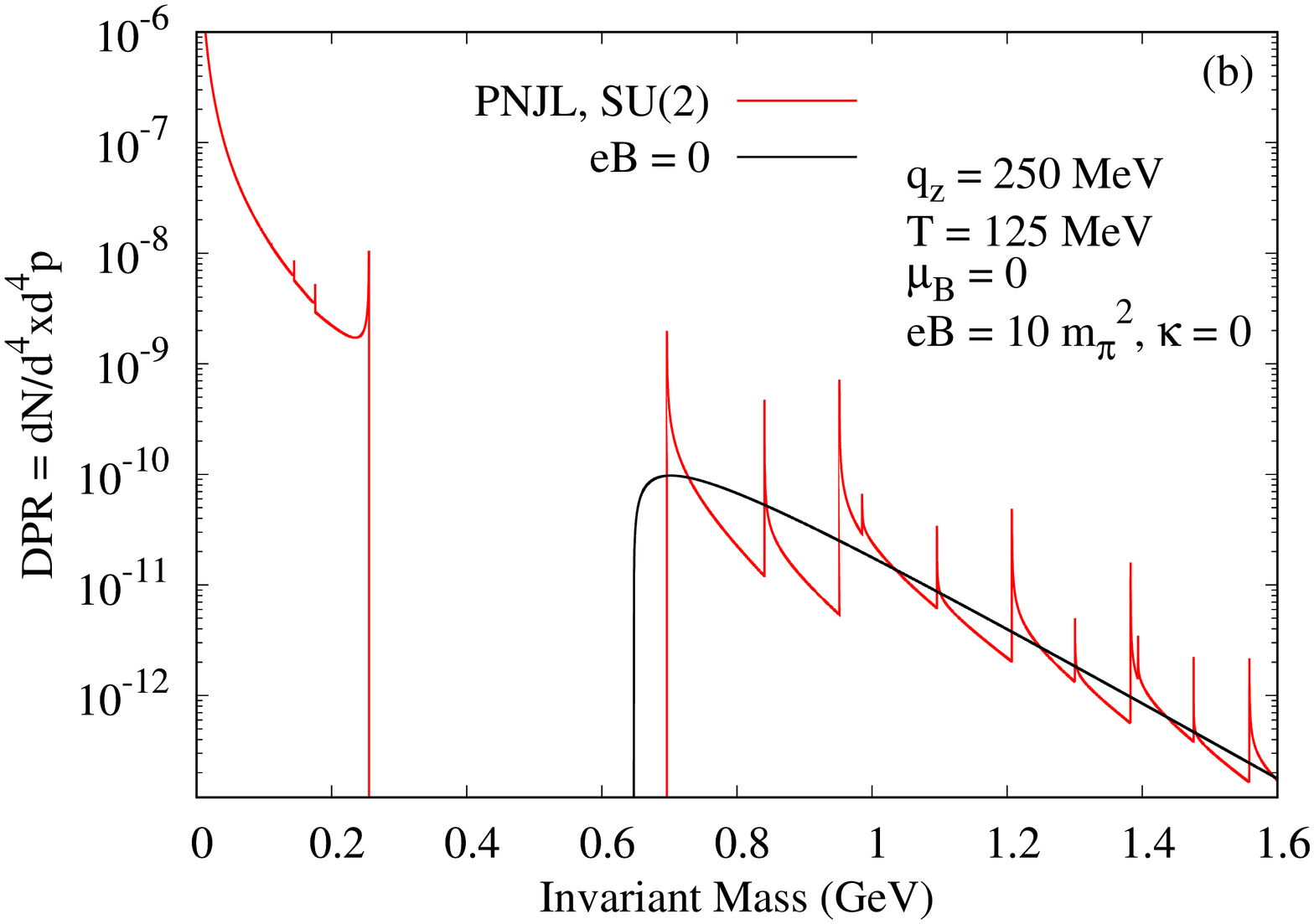}\\
		\includegraphics[angle=-90,scale=0.32]{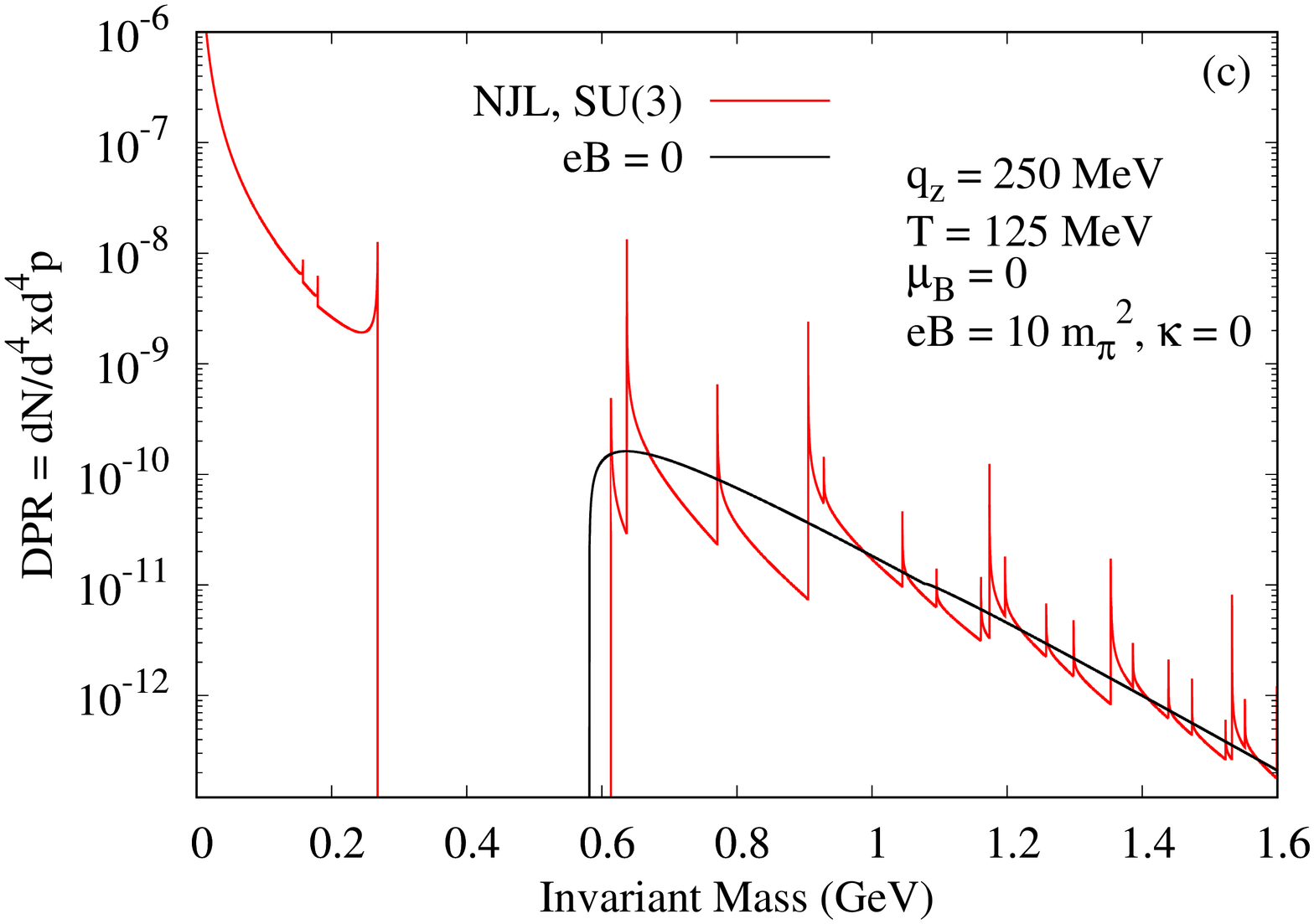}  \includegraphics[angle=-90,scale=0.32]{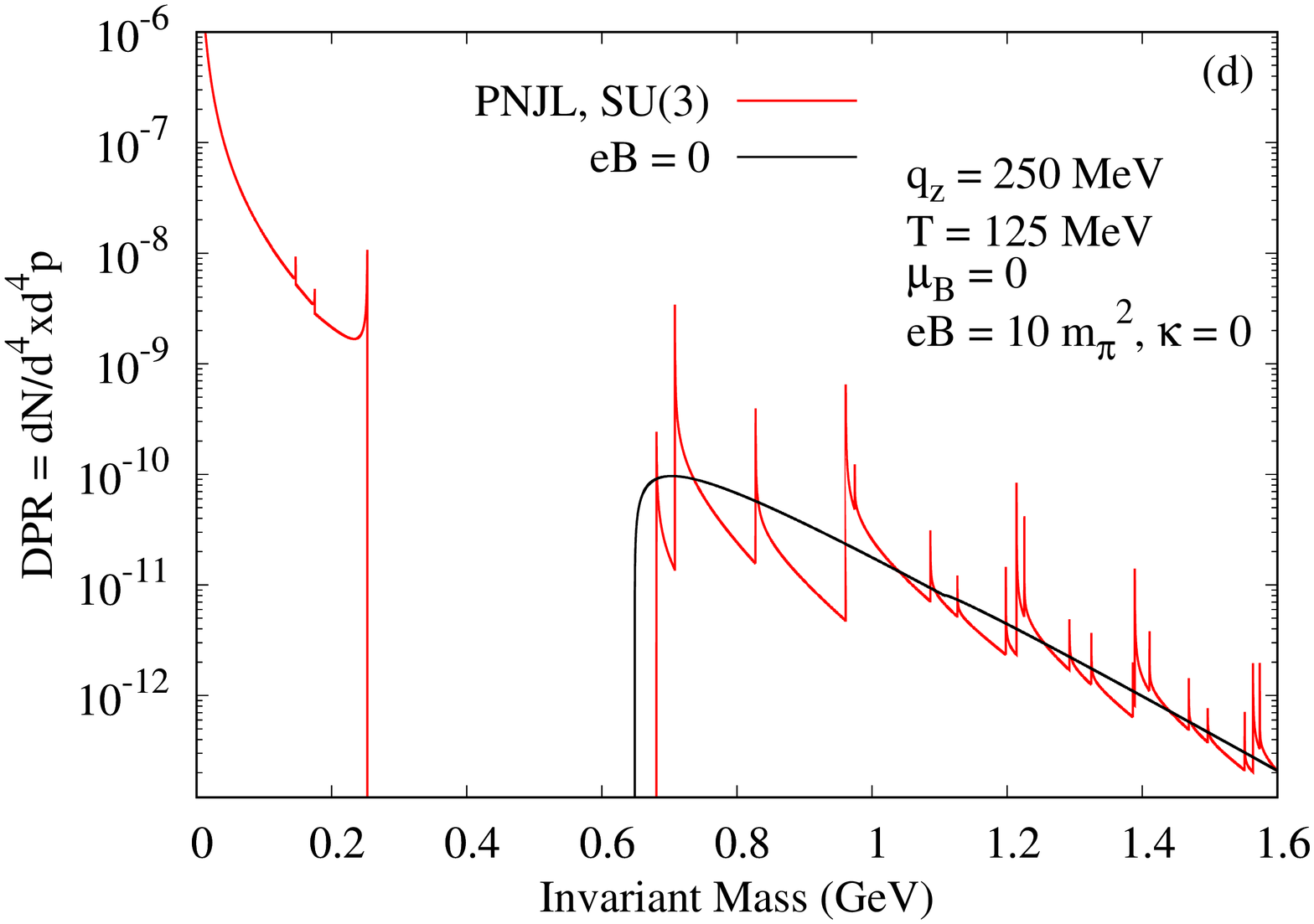}
	\end{center}
	\caption{(Color Online) Dilepton production rate at $q_z=250$ MeV, $T=125$ MeV, $\mu_B=0$ at $ eB = 10~m_\pi^2 $ using (a) two-flavor NJL model  (b) two-flavor PNJL model (c) three-flavor NJL model and (d) three-flavor PNJL model without considering the finite values of AMM of the quarks. The  DPR at $ eB = 0 $ (black solid line) is also shown for comparison. }
	\label{fig.dpr1}
\end{figure}

Next we present the numerical results for DPR from a hot and dense magnetized medium.  Note that, the results are evaluated ignoring the lepton mass i.e. $ m_L = 0 $ and considering the longitudinal momentum $ q_z = 250  $~MeV. In Fig.~\ref{fig.dpr1} we have shown DPRs at $ T = 125 $~MeV and $ \mu_B = 0 $ in presence of a uniform background magnetic field i.e. $ eB = 10~ m_\pi^2  $ as a function of the dilepton invariant mass $ \sqrt{q^2} $ without considering finite values of the AMM of the quarks using (a) two-flavor NJL model,  (b) two-flavor PNJL model, (c) three-flavor NJL model and (d) three-flavor PNJL model respectively. For all the cases mentioned, the  DPRs at $ eB = 0 $ (black solid line) are also shown for comparison. Since we have taken $ m_L = 0 $, the only nontrivial theta function in Eq.~\eqref{eq.DPR.4} is $ \Theta \FB{q^2 - M_f^2} $. As a result the threshold invariant mass for dilepton production at $ eB = 0 $ coincides with unitary-I cut threshold of $ \IM \mcW_{11}^\munu(q) $ which is $ \sqrt{q^2} > 2M_f \FB{T, \mu_B} $ where $ M_f $ is the constituent quark mass of flavor $ f $.  This feature is observed in all the plots shown in Fig.~\ref{fig.dpr1}. Considering Fig.~\ref{fig.dpr1}~(a), one can observe that, the threshold for dilepton production starts just below $ 0.6  $~GeV which can be understood from Fig.~\ref{fig.M1}~(a) as at $ T = 125  $~MeV, $ \mu_B = 0$ in two-flavor NJL model where the constituent quark mass is around $ \approx 0.290 $~GeV in absence of background field. However, when the magnetic field is turned on, DPR receives contributions from both Landau cut as well as unitary-I cut which is understandable  from the discussions below Eq.~\eqref{eq.N.1}. In presence of the background magnetic field without considering the finite values of AMM of the quarks the thresholds for unitary-I and Landau cuts are given by
\begin{equation}\label{eq.posi_UL_cuts}
{\text{unitary-I cut :}}~~\sqrt{q_\parallel^2}> 2M_f~~~~\text{and}~~~~{\text{Landau cut :}}~~\sqrt{q_\parallel^2} < \FB{\sqrt{M^2_f + 2\MB{e_f B}} -M_f}.
\end{equation}
 Now, at $ eB = 10 m_\pi^2,~ T =125 $~MeV and $ \mu_B  = 0 $ for two-flavor NJL model without Polyakov loop, $ M_u = M_d \approx 0.310 $~GeV. This explains, when put into the inequalities previously mentioned, the shifting of unitary-I cut towards higher values of invariant mass as well as termination point of the contribution from the Landau cut which is around $ \approx 0.27 $ GeV. In all the cases with finite background field, the spikelike structures can be observed  over whole range of allowed invariant mass for dilepton production. The appearance of spikes at nonzero magnetic field is a well-known phenomena and can be explained in terms of `threshold singularities' which occurs at each Landau level. These threshold singularities are basically manifestations of the fact that, in Eq.~\eqref{eq.W11.6}, the K\"all\'en function appearing in the denominator vanishes at each threshold defined in terms of step functions. In Fig.~\ref{fig.dpr1}~(b) we have shown the variation of DPRs with the invariant mass for same external parameters using Polyakov extended two-flavor NJL model. 
 While describing Fig.~\ref{fig.M1}, we have already discussed that the inclusion of Polyakov loop results in an increase of transition temperature from chiral symmetry broken to the restored phase and as a consequence the constituent quark mass remains almost constant for larger values of temperature (upto $ \sim 150 $~MeV). This explains the shift of the unitary-I cuts for both zero and nonzero values of $ eB $ towards higher values of invariant mass compared to simple NJL model. As a result the forbidden gap between unitary-I and Landau cut contributions also increases due to the inclusion of Polyakov loop. In case of three-flavor NJL model (Fig.~\ref{fig.dpr1}~(c)) the position of the unitary-I and Landau cuts can be explained in a similar fashion as described earlier. Moreover,  a close observation , reveals that, DPRs at higher invariant mass are slightly higher in case of three-flavor models. Later, we will discuss this feature in more detail. Inclusion of Polyakov loop in the three-flavor NJL model, depicted in Fig.~\ref{fig.dpr1}~(d), leads to the same qualitative behaviour as seen in case of two-flavor model.  
 
 It may be pointed out that, our results for $ SU(3) $ PNJL model at finite magnetic field are in good agreement with the results presented in Ref.~\cite{Ghosh:2018xhh} where the authors have used the Effective fugacity Quasi-Particle Model (EQPM) to incorporate the effect of `strong' interaction and no approximations were made on the strength of the magnetic field as in the present work. Moreover, we have also checked that our results in the strong field approximation (also known as Lowest Landau Level (LLL) approximation) are in good qualitative agreement with Refs.~\cite{Bandyopadhyay:2016fyd,Islam:2018sog} in which LLL approximation is used. The spike-like structures as seen in the invariant mass dependence of the dilepton production rate (DPR) in presence of magnetic field are also in good agreement with the earlier works cited in Refs.~\cite{Hattori:2020htm,Ghosh:2020xwp,Ghosh:2018xhh,Sadooghi:2016jyf}.
\begin{figure}[h]
	\begin{center}
		\includegraphics[angle=-90,scale=0.32]{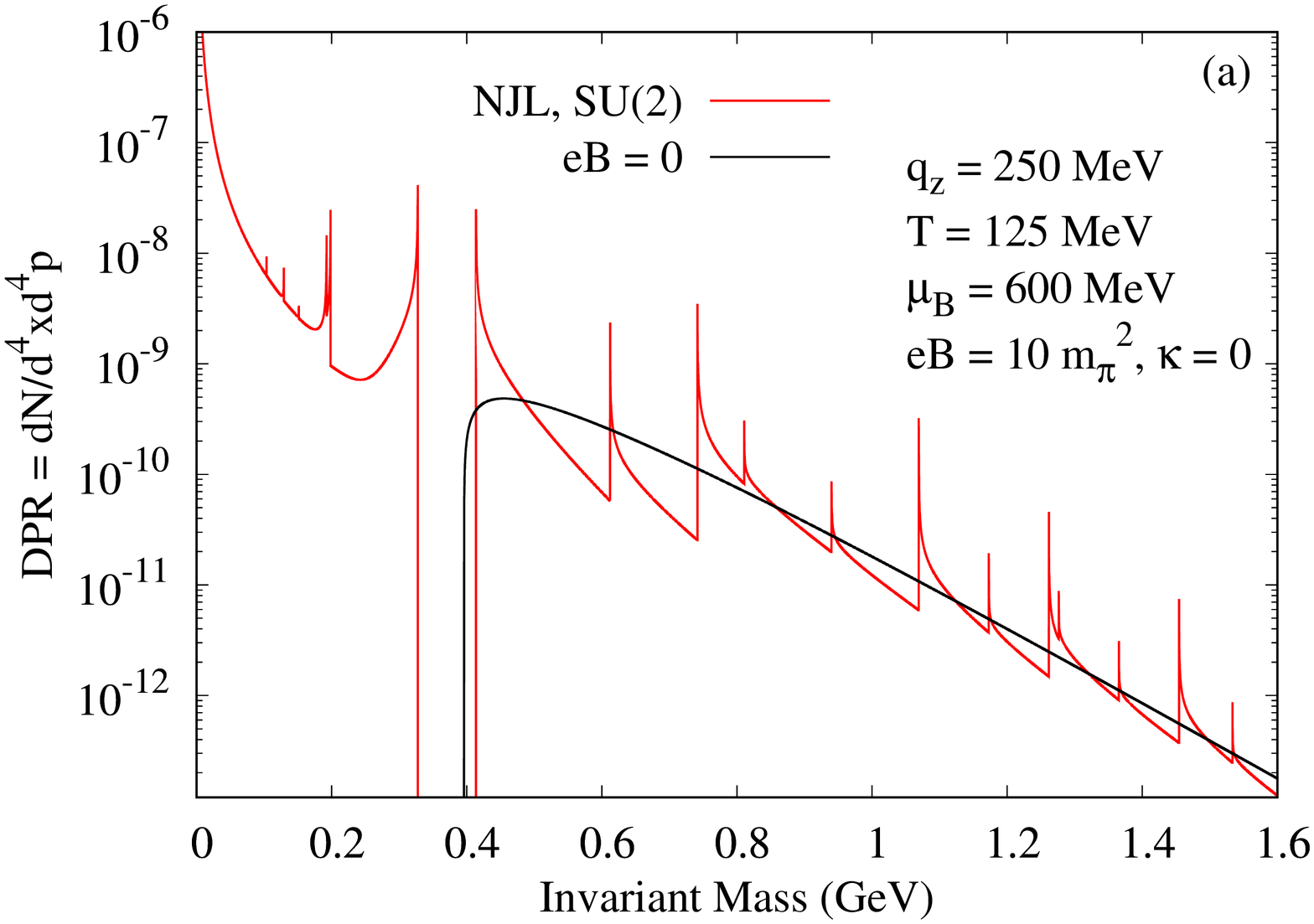}  \includegraphics[angle=-90,scale=0.32]{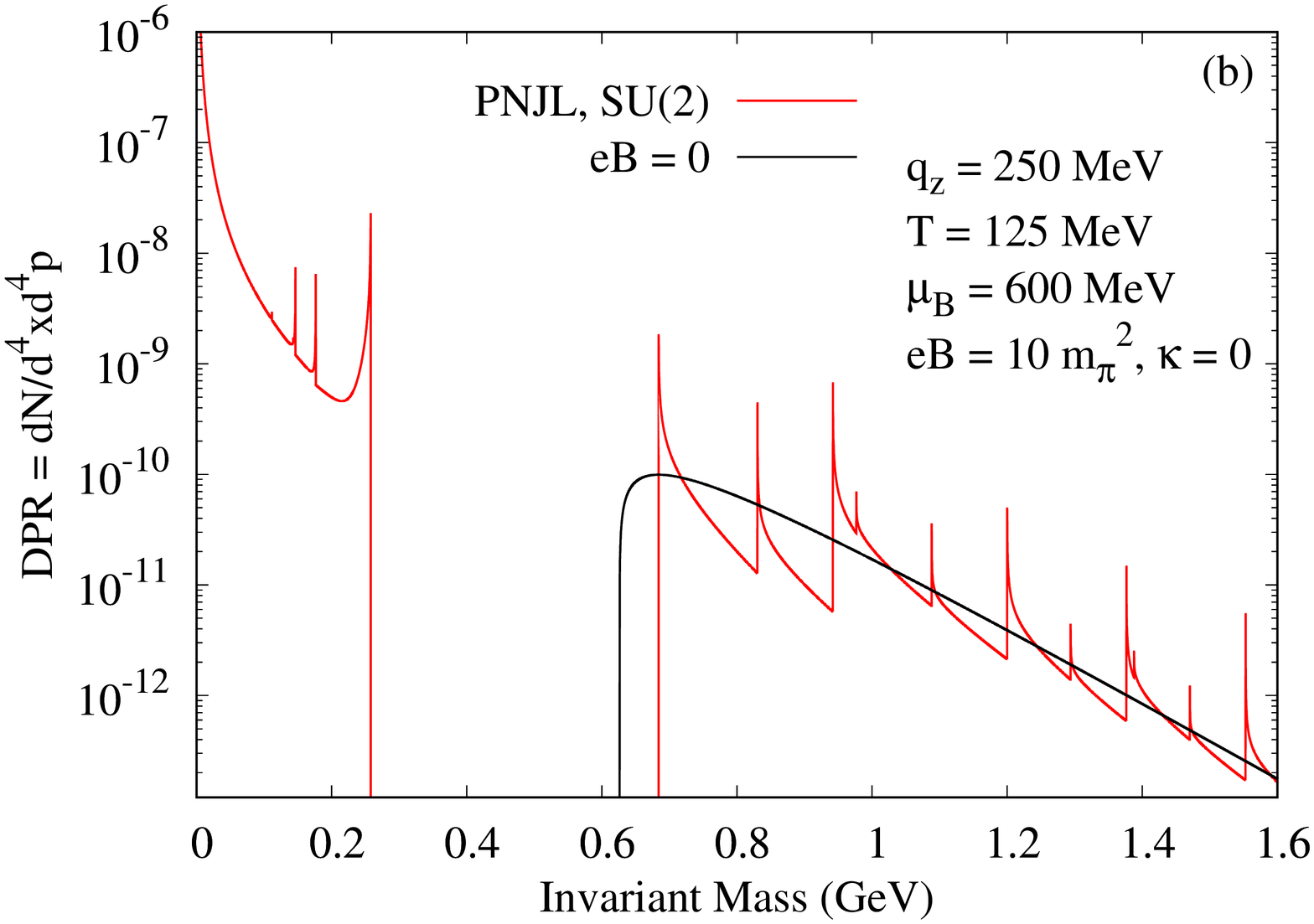}\\
		\includegraphics[angle=-90,scale=0.32]{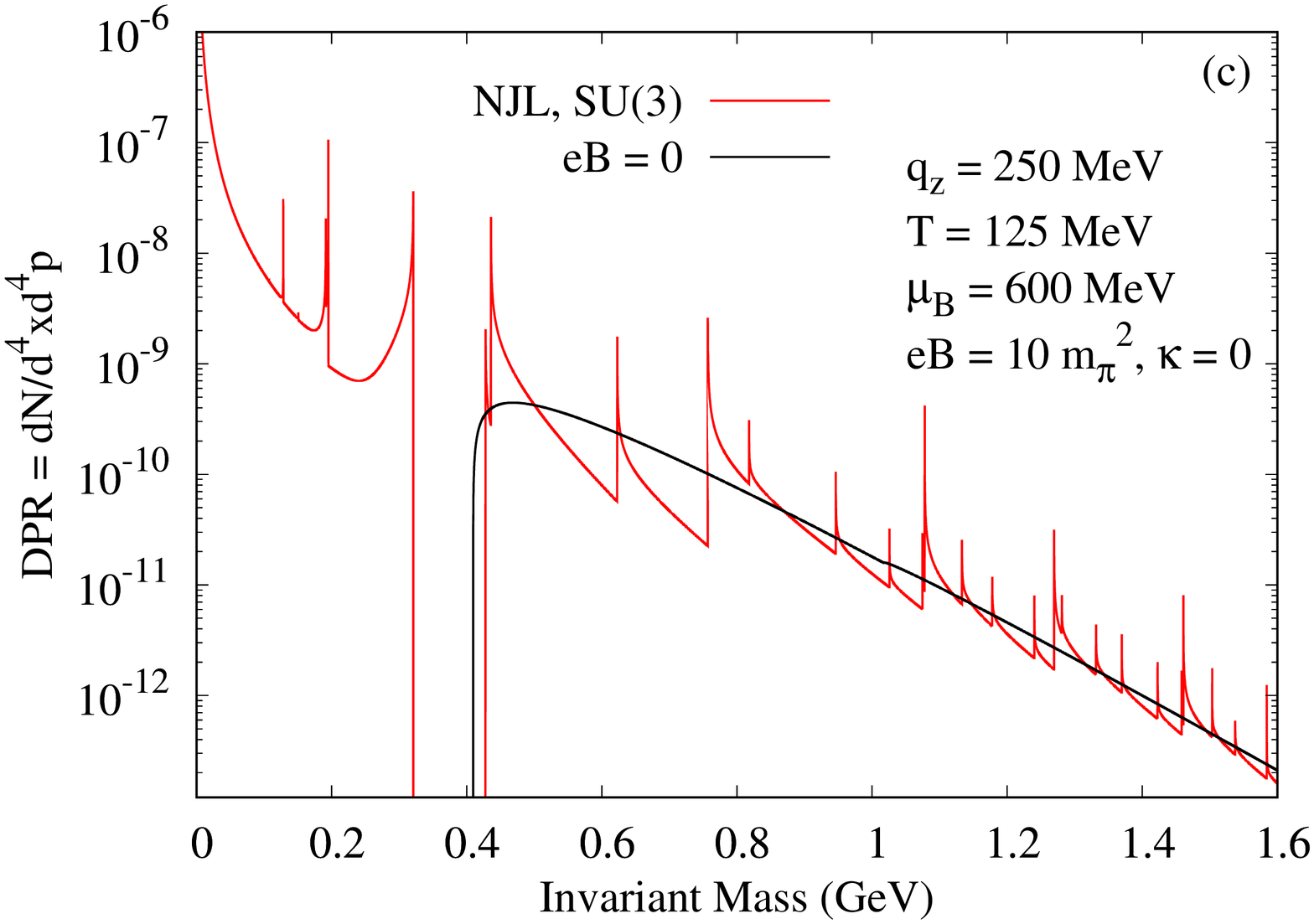}  \includegraphics[angle=-90,scale=0.32]{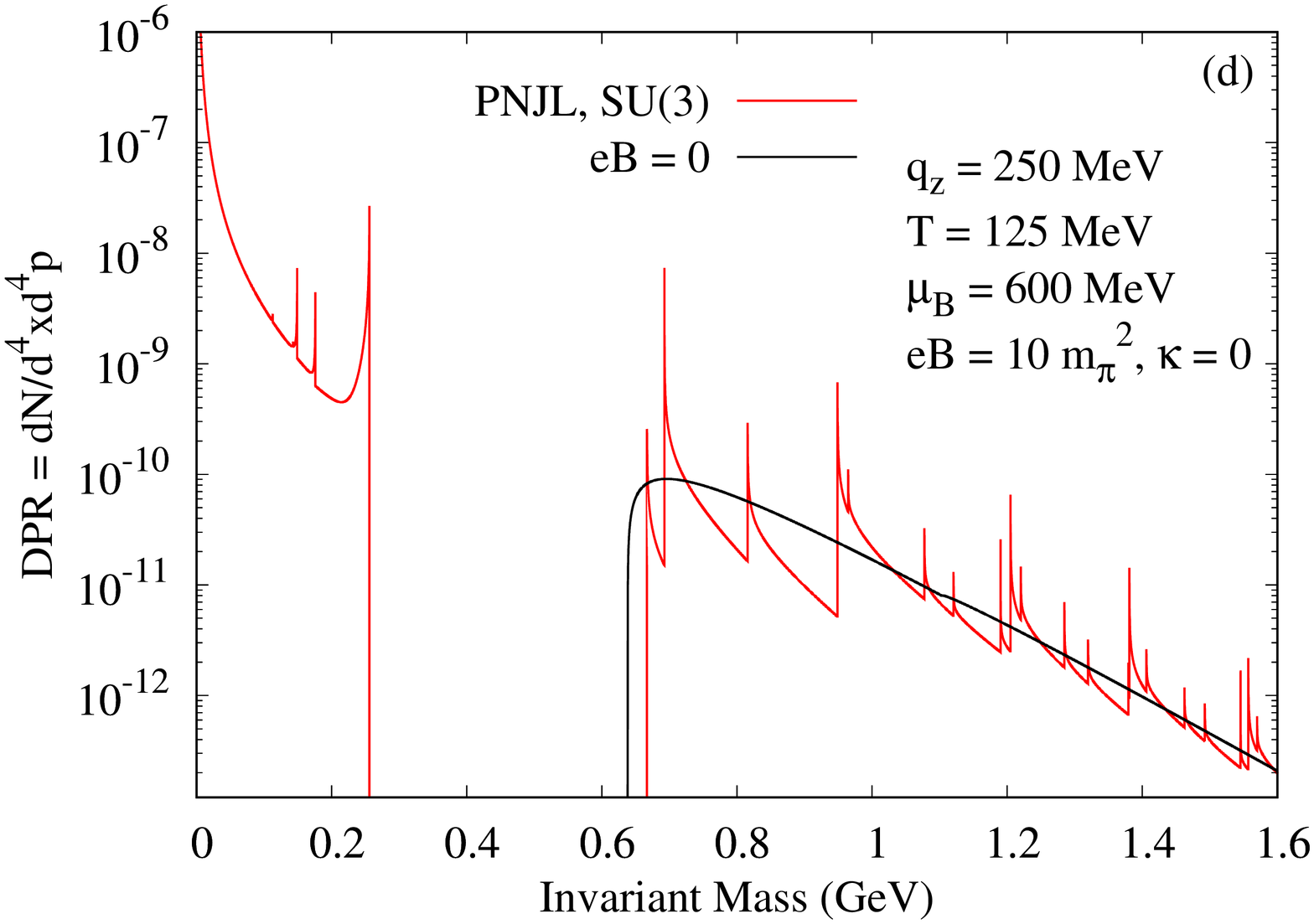}
	\end{center}
	\caption{(Color Online) Dilepton production rate at $q_z=250$ MeV, $T=125$ MeV, $\mu_B=600$ MeV at $ eB = 10~m_\pi^2 $ using (a) two-flavor NJL model  (b) two-flavor PNJL model (c) three-flavor NJL model and (d) three-flavor PNJL model without considering the finite values of AMM of the quarks. The  DPR at $ eB = 0 $ (black solid line) is also shown for comparison. }
	\label{fig.dpr2}
\end{figure}

In Fig.~\ref{fig.dpr2} we have taken the same parameters described in Fig.~\ref{fig.dpr1} except  finite baryon chemical potential which is  $ 600 $~MeV in this case. From Fig.~\ref{fig.dpr2}~(a), where we have used a simple two-flavor NJL model to calculate DPR from a hot and dense magnetized medium, one can observe that, finite values of baryon chemical potential results in a substantial change in the DPR when compared to the $ \mu_B = 0 $ case (see Fig.~\ref{fig.dpr1}~(a)). This is a manifestation of the fact that, at $ T=125  $ MeV, the constituent quark mass in two-flavor NJL model significantly modified at $ \mu_B=600 $~MeV in contrast to $ \mu_B = 0 $ case as can be seen from Fig~\ref{fig.M2}~(a). The positions of the unitary and Landau cuts again can be precisely determined using Eq.~\eqref{eq.posi_UL_cuts}. Note that, the threshold for unitary-I cut moves towards the lower invariant mass as the constituent mass decreases. However, the contributions from Landau cut survives for higher values of invariant mass when constituent quark mass reduces. This explains the decrease in the gap between thresholds of unitary-I and Landau cuts. Moreover, the overall magnitude of the DPRs for both $B=0$ and $B\ne0$ cases increase compared to the $ \mu_B = 0 $ case shown in Fig~\ref{fig.dpr1}~(a) because of the enhancement of the thermal phase space. The similar effect of inclusion of nonzero baryon chemical potential is also visible in case of three-flavor NJl model. However, for Polyakov extended two and three-flavor NJL model, at $ \mu_B = 600 $~MeV (see Figs.~\ref{fig.dpr2}~(b) and (d)), there are no noticeable modifications in DPRs for zero and finite $ eB $ case compared to the $ \mu_B = 0 $ scenario as depicted in Figs.~\ref{fig.dpr1}~(b) and (d). This is understandable from Figs.~\ref{fig.M2}~(b) and (d) where it can be seen that, at $ T=125 $~MeV and $ \mu_B = 600 $~MeV, constituent mass of quarks in Polyakov loop extended models are nearly equal to $ \mu_B =0 $ case.  
\begin{figure}[h]
	\begin{center}
		\includegraphics[angle=-90,scale=0.32]{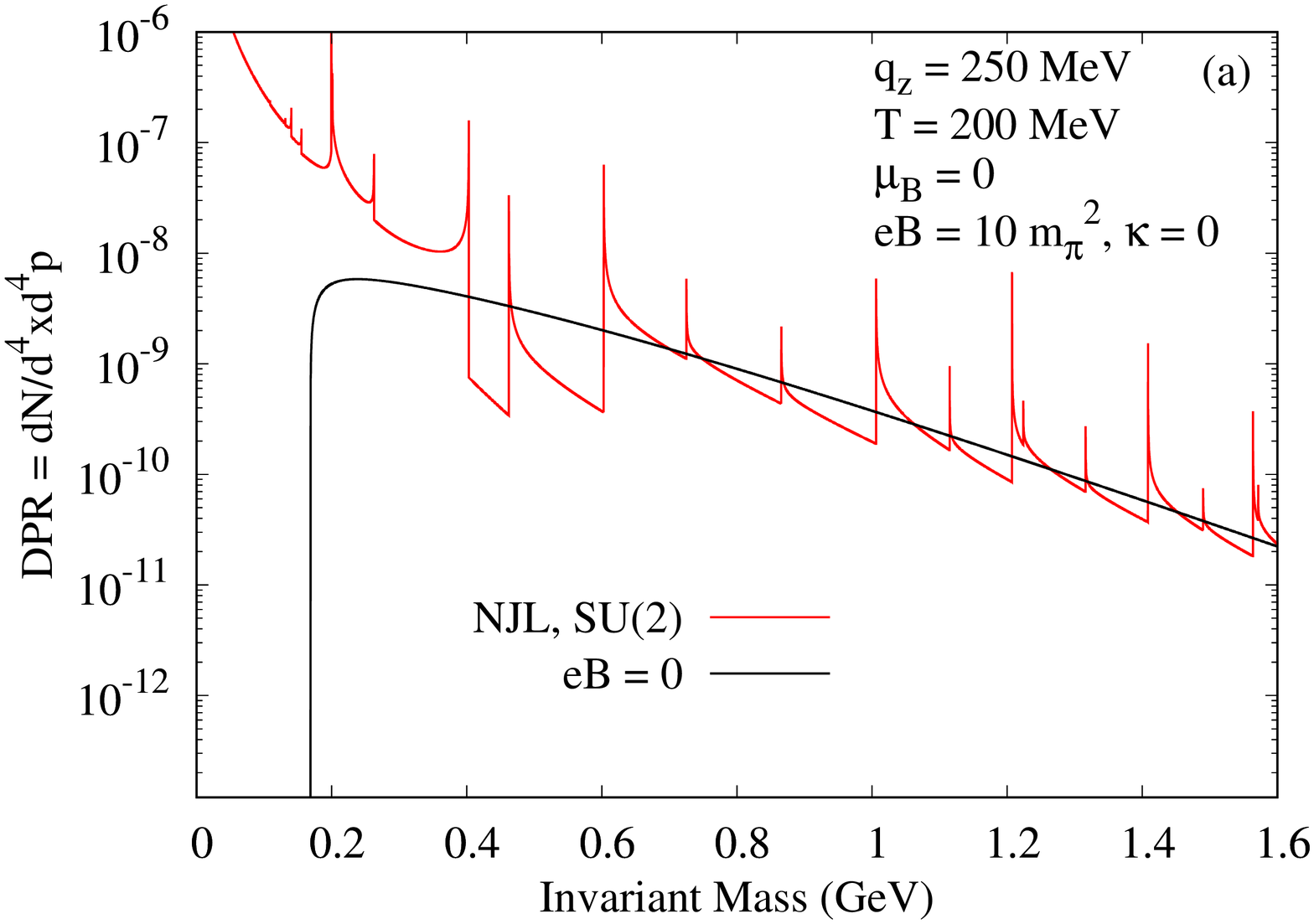}  \includegraphics[angle=-90,scale=0.32]{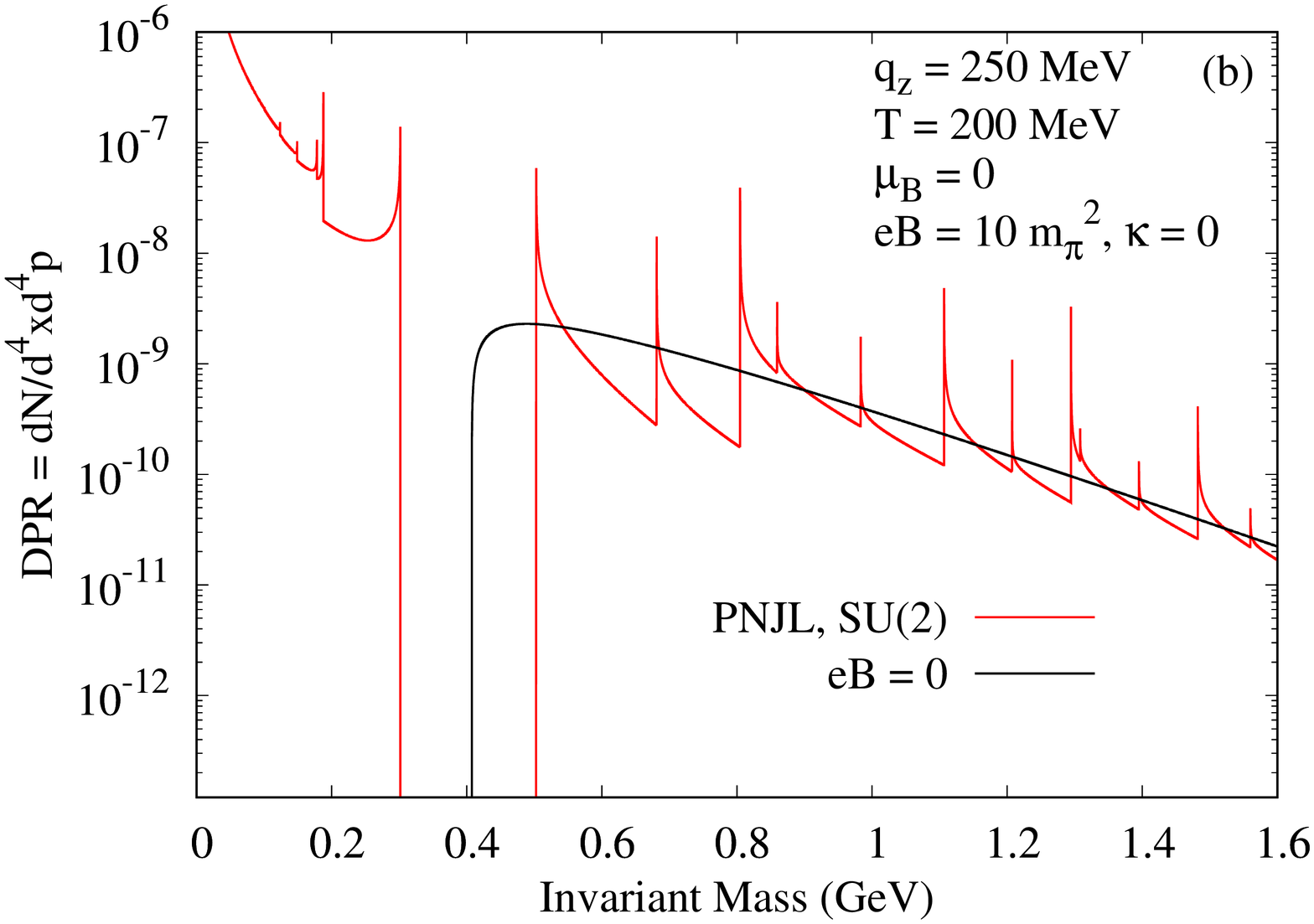}\\
		\includegraphics[angle=-90,scale=0.32]{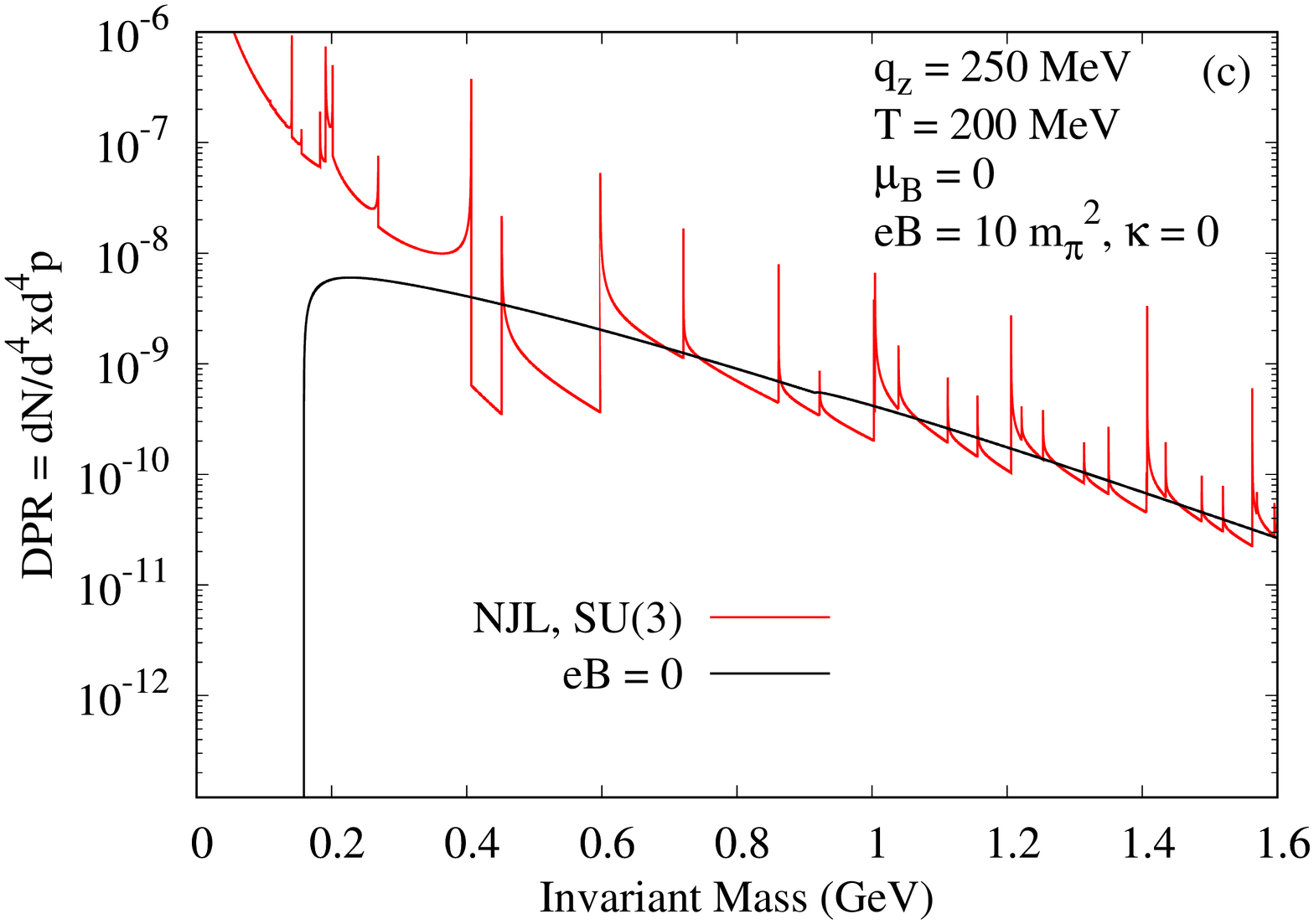}  \includegraphics[angle=-90,scale=0.32]{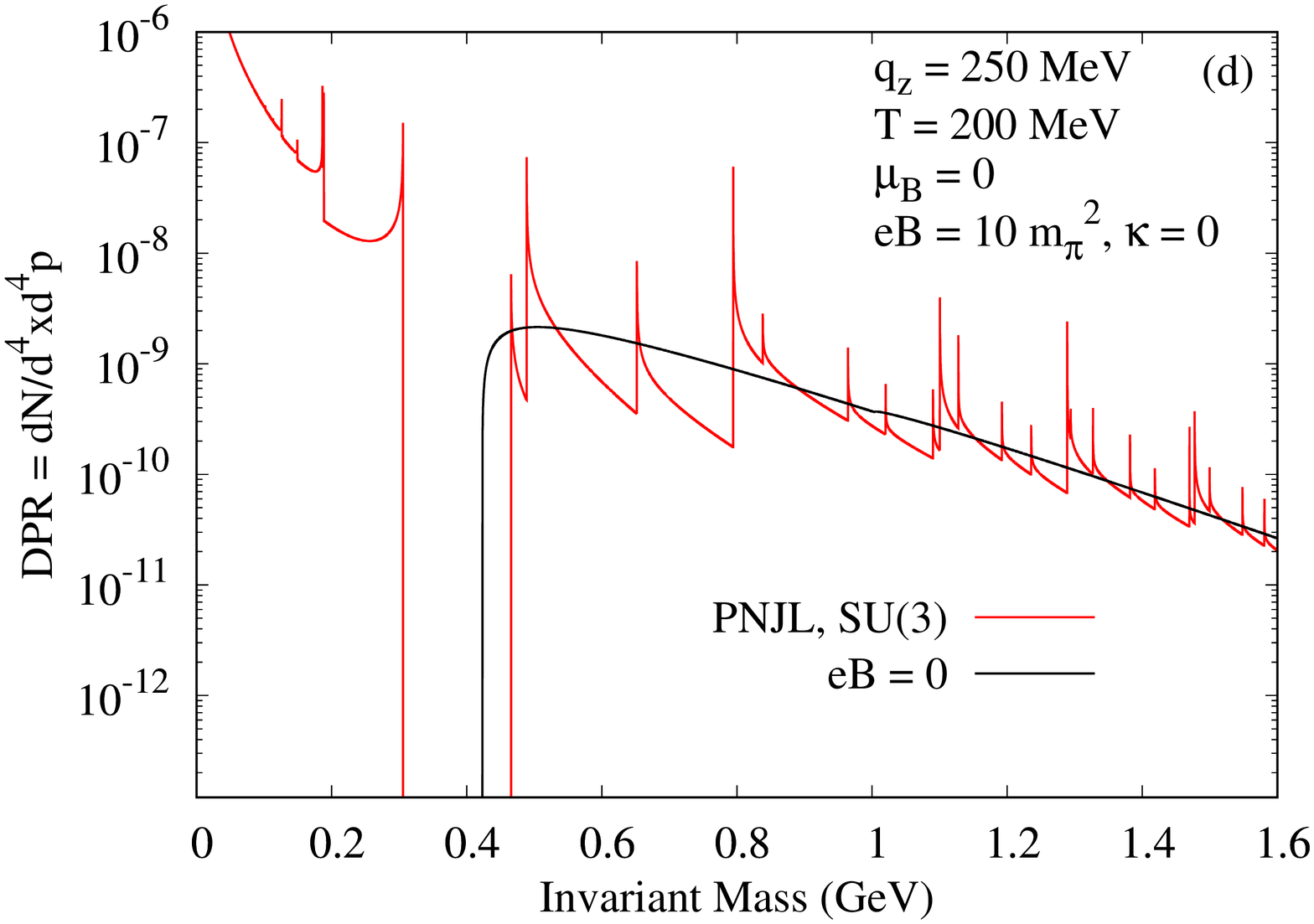}
	\end{center}
	\caption{(Color Online) Dilepton production rate at $q_z=250$ MeV, $T=200$ MeV, $\mu_B=0$ at $ eB = 10~m_\pi^2 $ using (a) two-flavor NJL model  (b) two-flavor PNJL model (c) three-flavor NJL model and (d) three-flavor PNJL model without considering the finite values of AMM of the quarks. The  DPR at $ eB = 0 $ (black solid line) is also shown for comparison. }
	\label{fig.dpr3}
\end{figure}

From the discussions of Figs.~\ref{fig.dpr1} and ~\ref{fig.dpr2}, we observe that different thresholds for DPR at zero and nonzero background fields are strongly related to the constituent masses of quarks which in turn depends highly on $ T, \mu_B $ and $ eB $. From Eq.~\eqref{eq.posi_UL_cuts} we have seen that unitary-I (Landau) cut shifts towards lower (higher) values of invariant mass as constituent mass decreases. Moreover, it is well-known that in NJL-type models with increasing temperature and/or baryon chemical potential the constituent mass goes to bare mass limit. So it is expected that for sufficiently high values $ T $ and/or $ \mu_B $ the contributions from unitary-I and Landau cuts merge with each other and dilepton production can occur for the whole range of invariant mass. In Fig.~\ref{fig.dpr3} we have plotted DPRs using the four different models for both zero and nonzero $ eB $ at $ T = 200  $~MeV and $ \mu_B = 0 $. Now in case of two and three-flavor NJL models at zero $ \mu_B $, as can be seen from Fig.~\ref{fig.M1}~(a) and (c), at $ T = 200 $~MeV the constituent mass of low lying quarks are less than $ 100 $~MeV. As a result,  in Figs.~\ref{fig.dpr3}~(a) and (c), the threshold for unitary-I cut contributions in zero $ eB $ case starts at invariant mass less than $ 200 $~MeV and when finite background magnetic field is present  unitary-I and Landau cut contributions have merged with each other resulting in production of dileptons throughout the whole range of invariant mass. One should also observe that, due to large temperature value considered in this plot, the availability of the thermal phase space increases  resulting a substantial change in the overall magnitude of DPRs in Figs.~\ref{fig.dpr3}~(a) and (c) (almost two orders of magnitude when compared to Figs.~\ref{fig.dpr1}~(a) and (c) where $ T = 125 $~MeV). In Figs.\ref{fig.dpr3}~(b) and (d), we have used Polyakov loop extended models to evaluate DPRs. In this case there are appreciable changes in the positions of unitary-I and Landau cuts  as well as the in the magnitude of DPRs. This is because, from Figs.~\ref{fig.M1}~(b) and (d) one can observe that, at $ T = 200  $~MeV, constituent masses of low lying quarks are smaller than the same at $ T= 125 $~MeV. So, one can infer that, for a particular value of $ \mu_B $, the presence of the Polyakov loop can shift the merging of unitary-I and Landau cut contributions towards higher values of temperature. 
\begin{figure}[h]
	\begin{center}
		\includegraphics[angle=-90,scale=0.32]{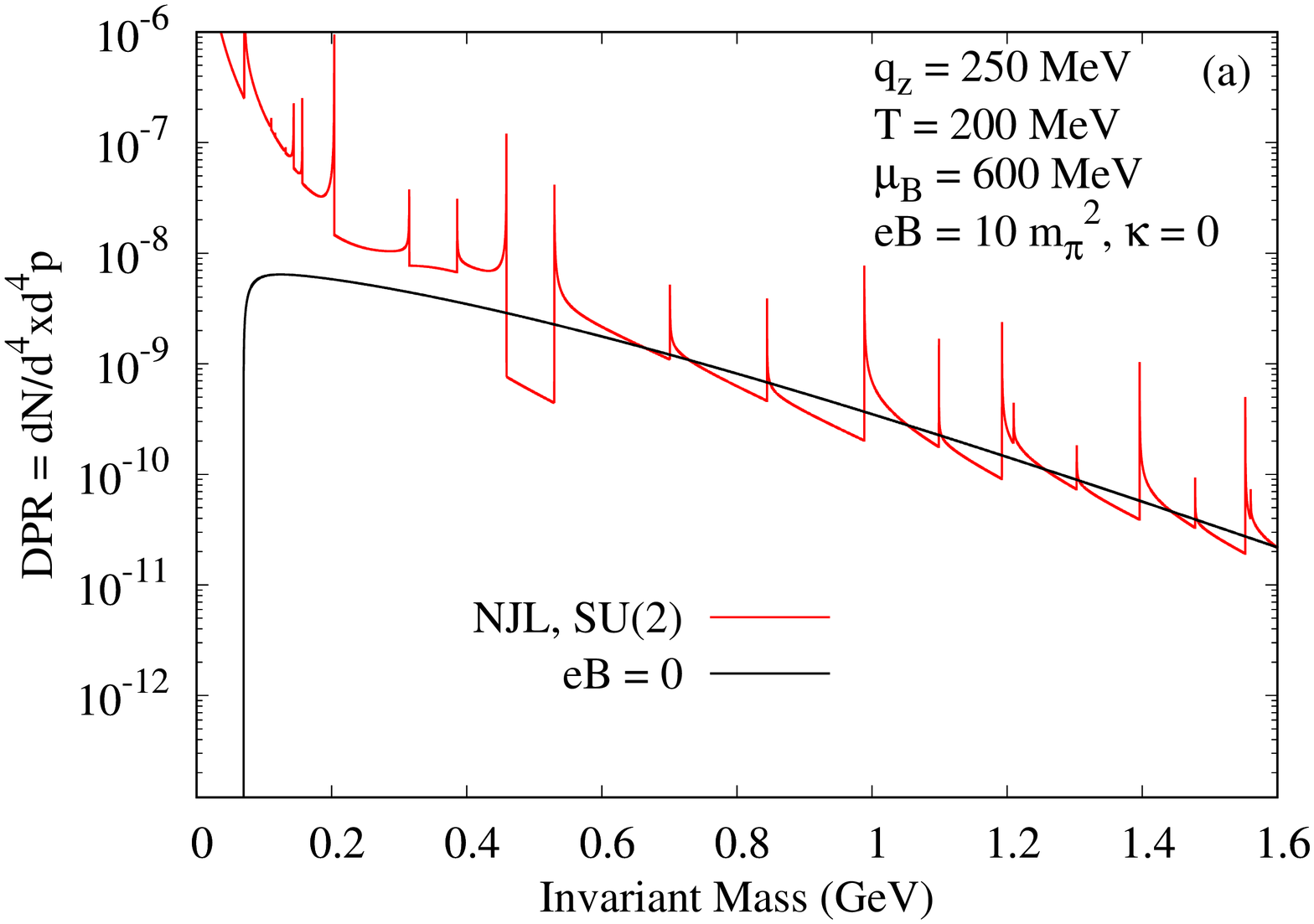}  \includegraphics[angle=-90,scale=0.32]{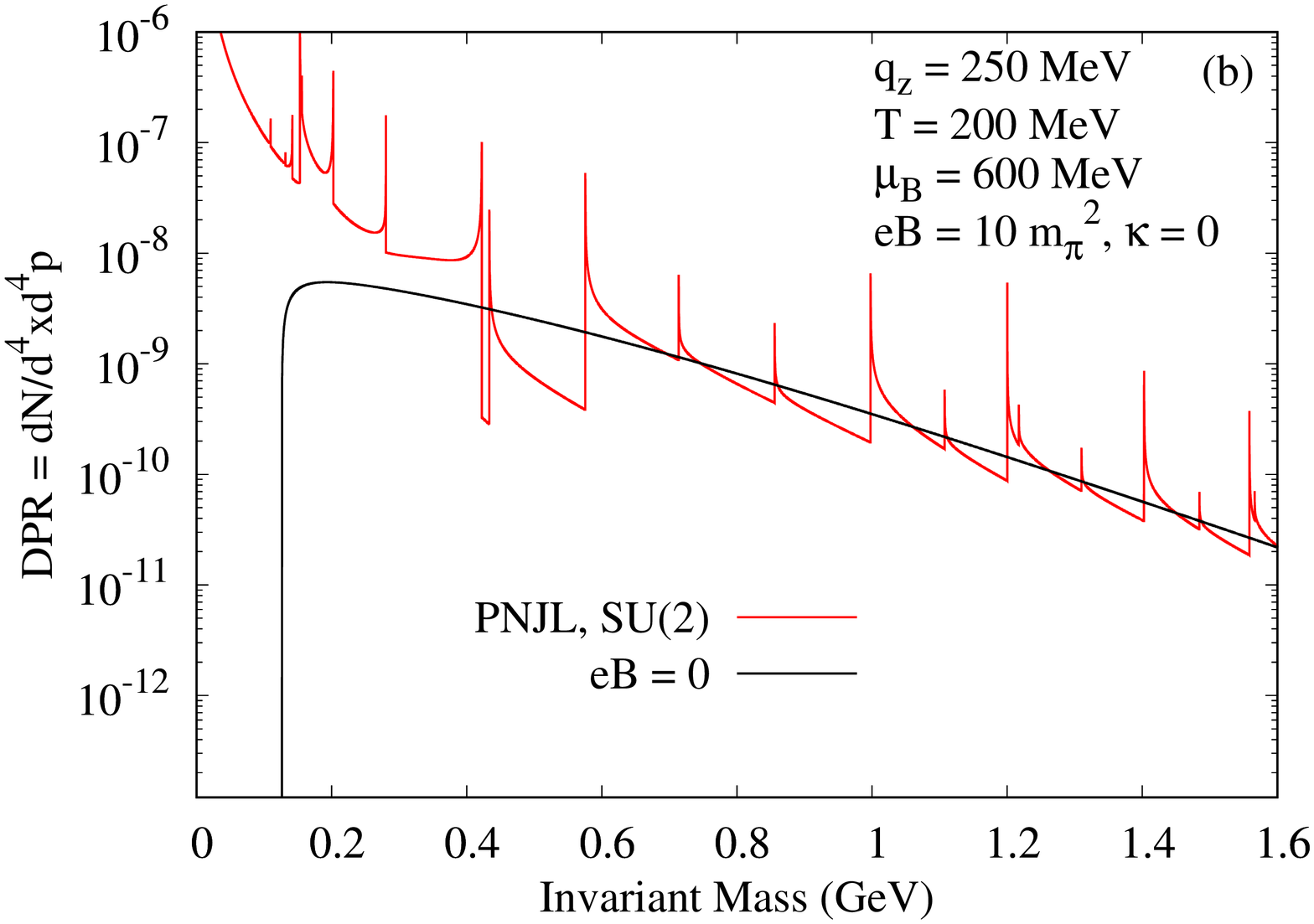}\\
		\includegraphics[angle=-90,scale=0.32]{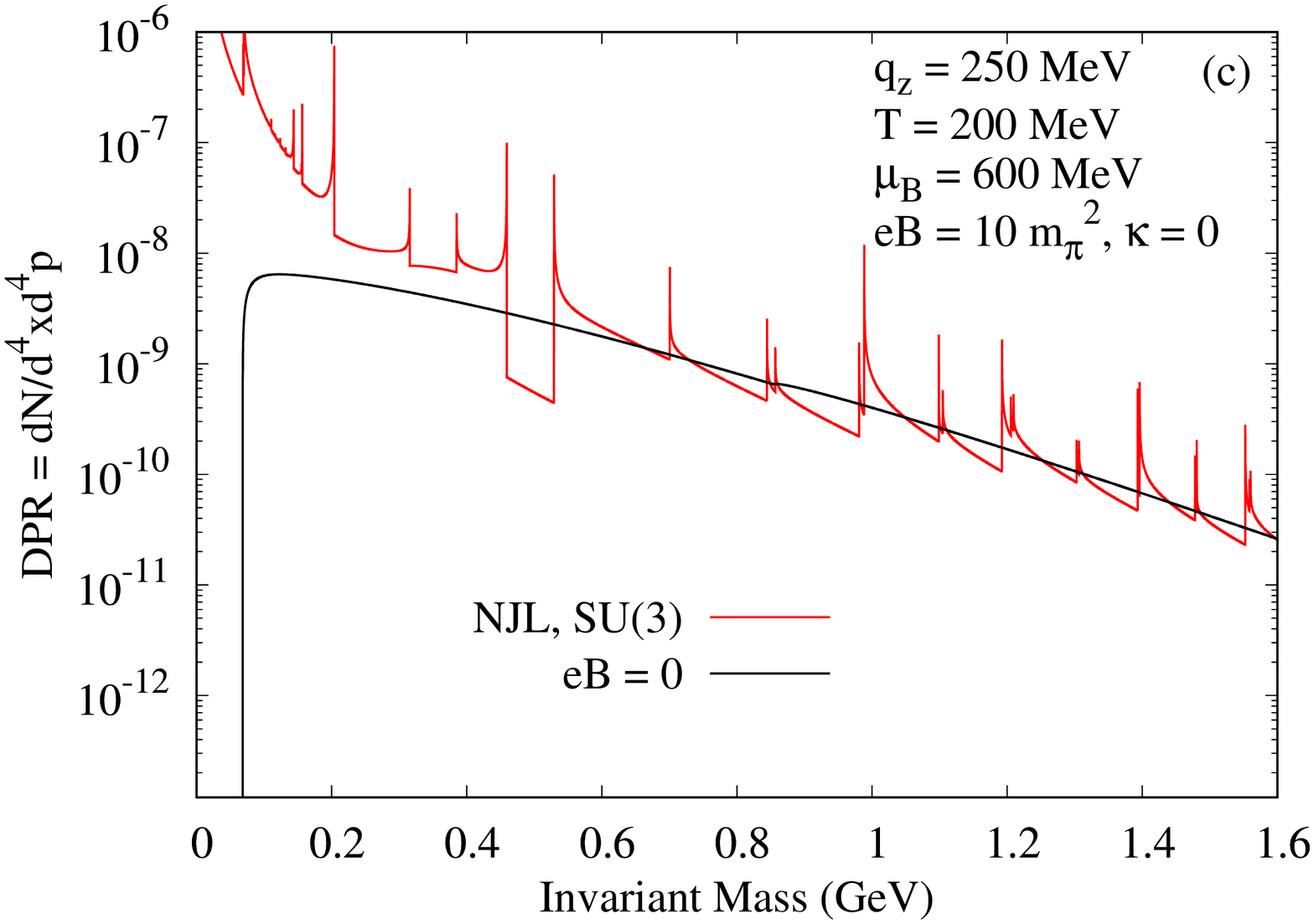}  \includegraphics[angle=-90,scale=0.32]{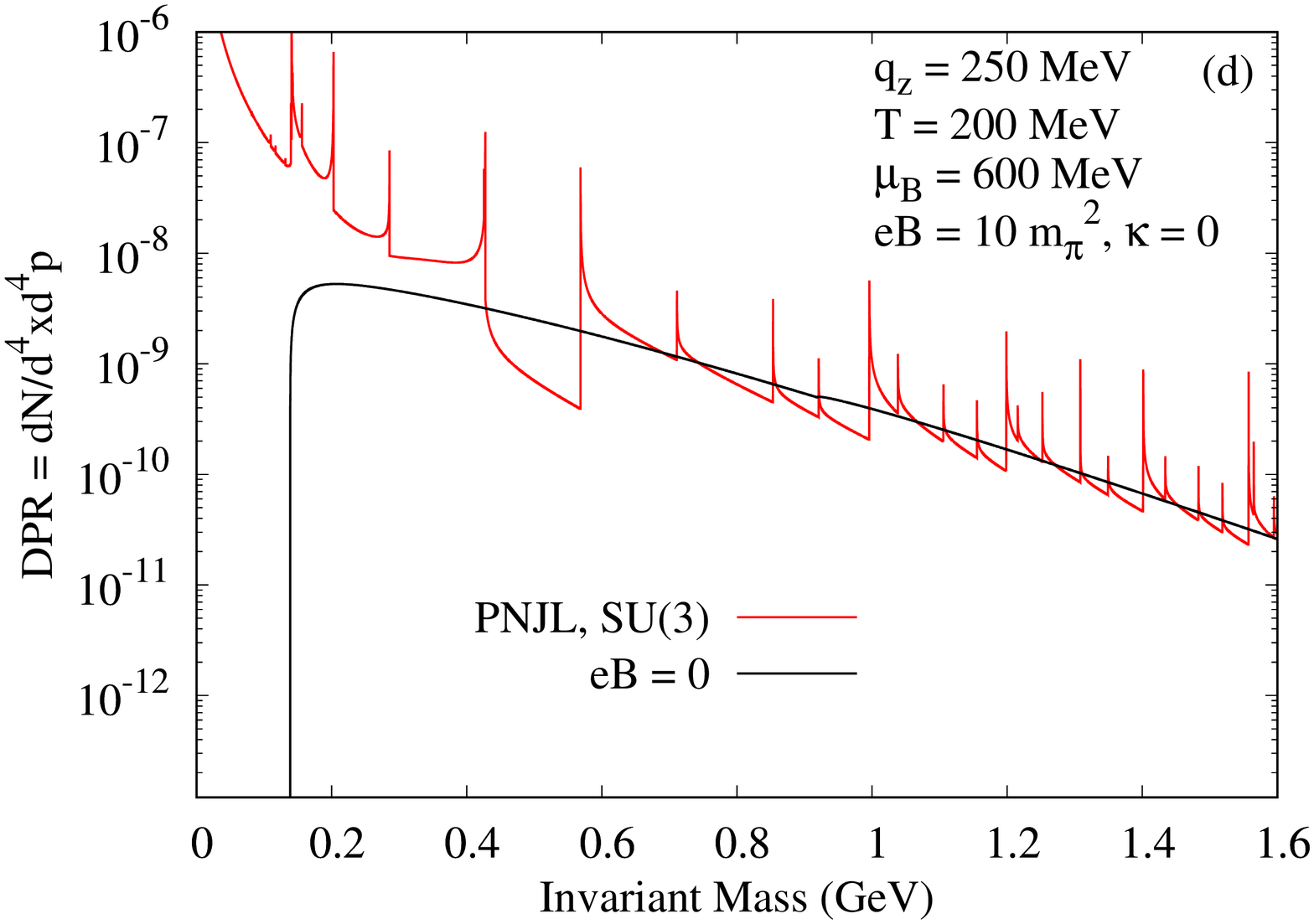}
	\end{center}
	\caption{(Color Online) Dilepton production rate at $q_z=250$ MeV, $T=125$ MeV, $\mu_B=600$ MeV at $ eB = 10~m_\pi^2 $ using (a) two-flavor NJL model  (b) two-flavor PNJL model (c) three-flavor NJL model and (d) three-flavor PNJL model without considering the finite values of AMM of the quarks. The  DPR at $ eB = 0 $ (black solid line) is also shown for comparison. }
	\label{fig.dpr4}
\end{figure}

Now, in Fig.~\ref{fig.dpr4}, we consider a situation where both temperature and chemical potential are sufficiently high i.e. $ T = 200 $~MeV and $ \mu_B = 600$~MeV. From Fig.~\ref{fig.M2} it can be seen that in all the cases at $ T =200 $~MeV the $ u $ and $ d $ quarks are in bare mass limit. As a result, irrespective of the type of models used, unitary-I and Landau cuts have merged with each other and there is a considerable increment in DPRs in all four cases compared to Fig.~\ref{fig.dpr1}. However, it is noteworthy that, since both $ T $ and $ \mu_B $ considered in this case are higher than all the previous cases, the dilepton yield is highest due to availability of larger thermal phase space.

\begin{figure}[h]
	\begin{center}
		\includegraphics[angle=-90,scale=0.32]{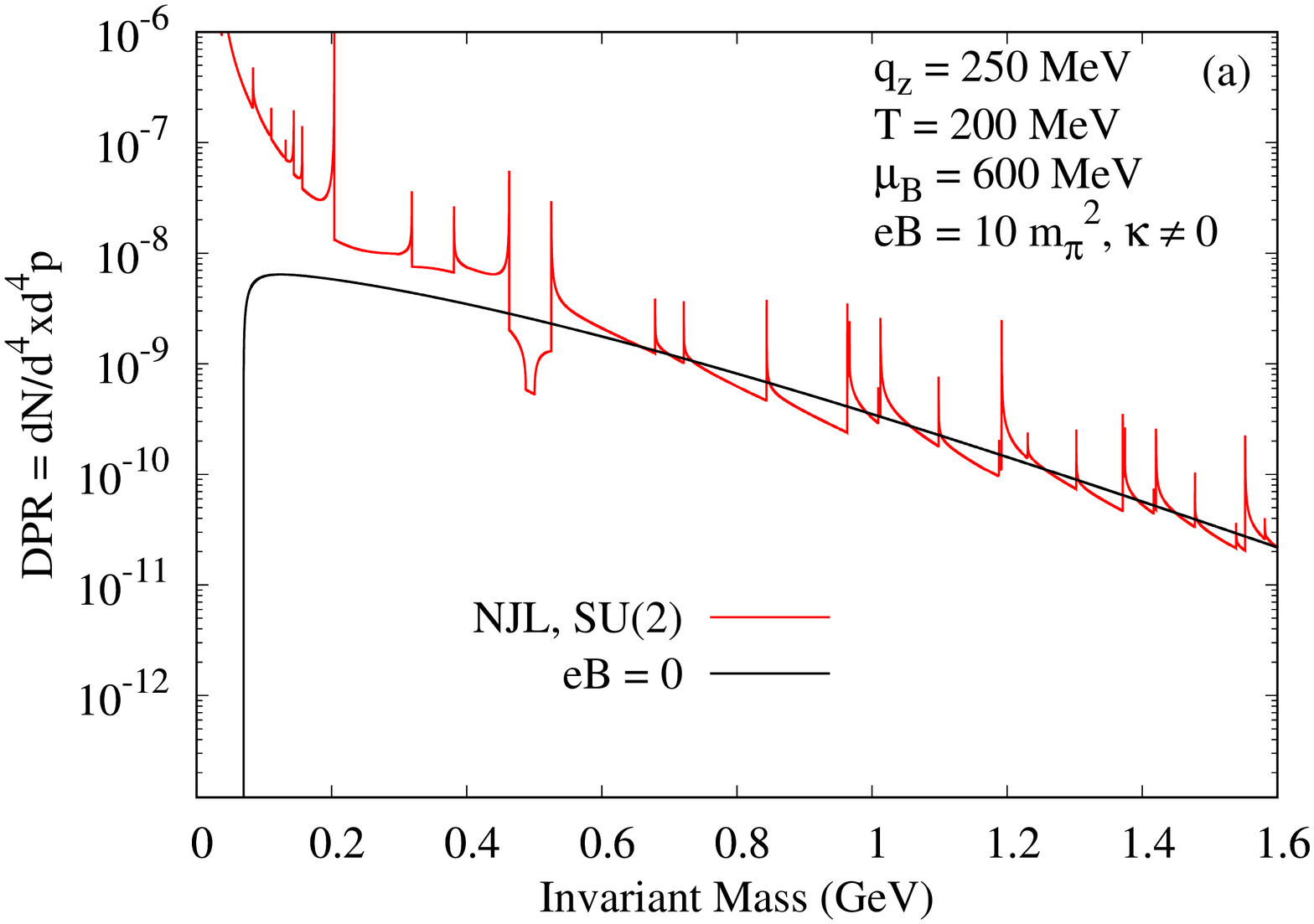}  \includegraphics[angle=-90,scale=0.32]{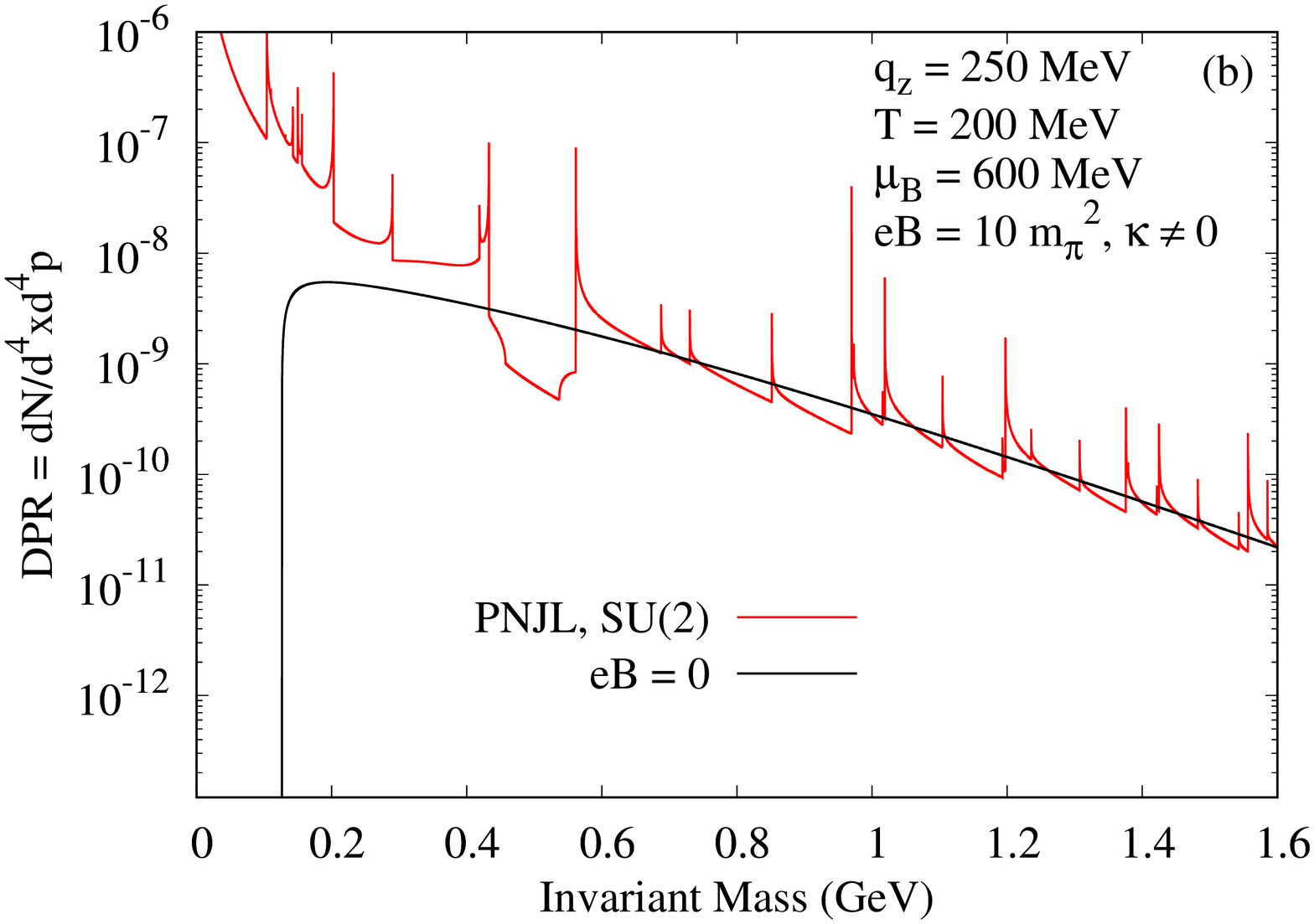}\\
		\includegraphics[angle=-90,scale=0.32]{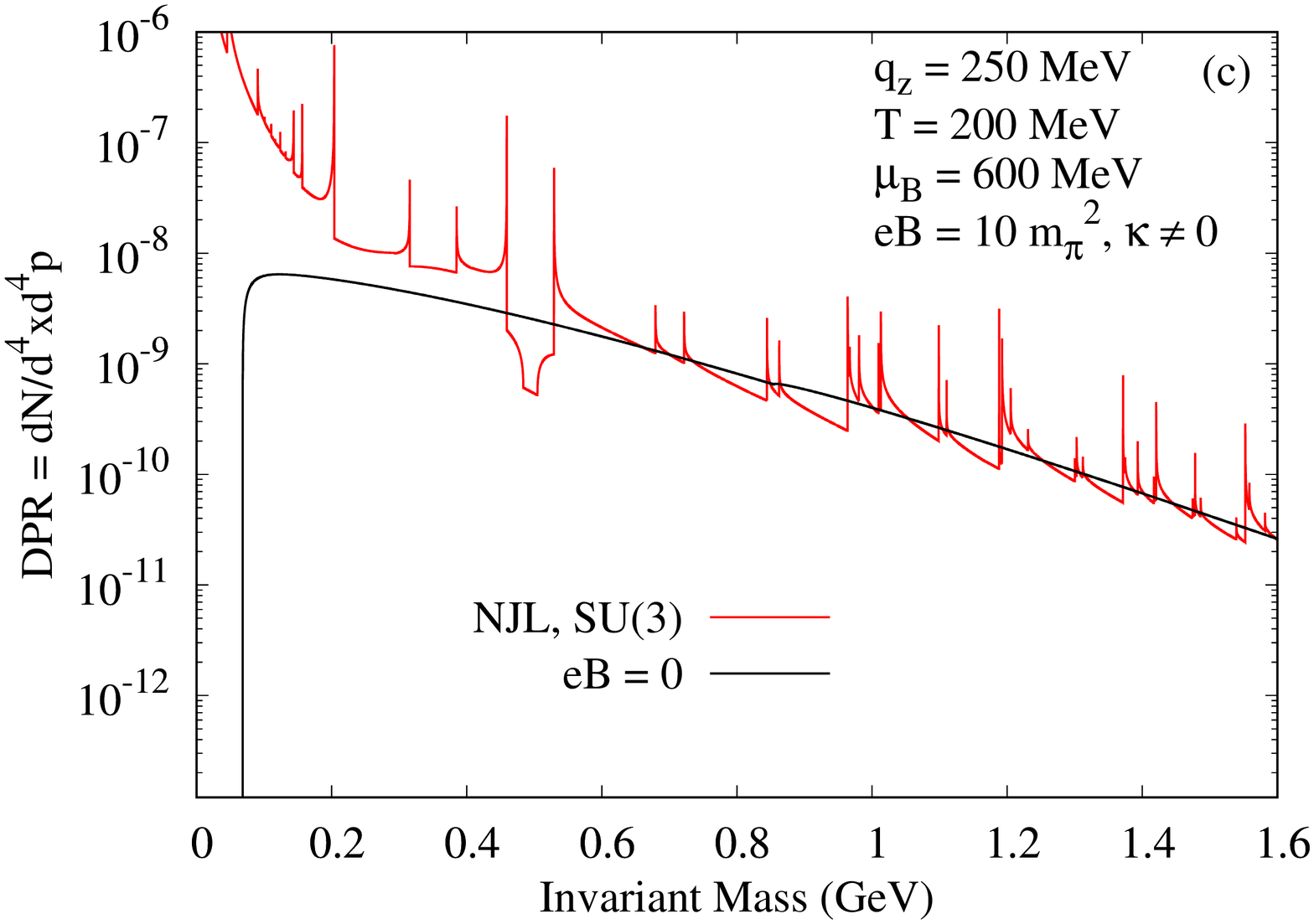}  \includegraphics[angle=-90,scale=0.32]{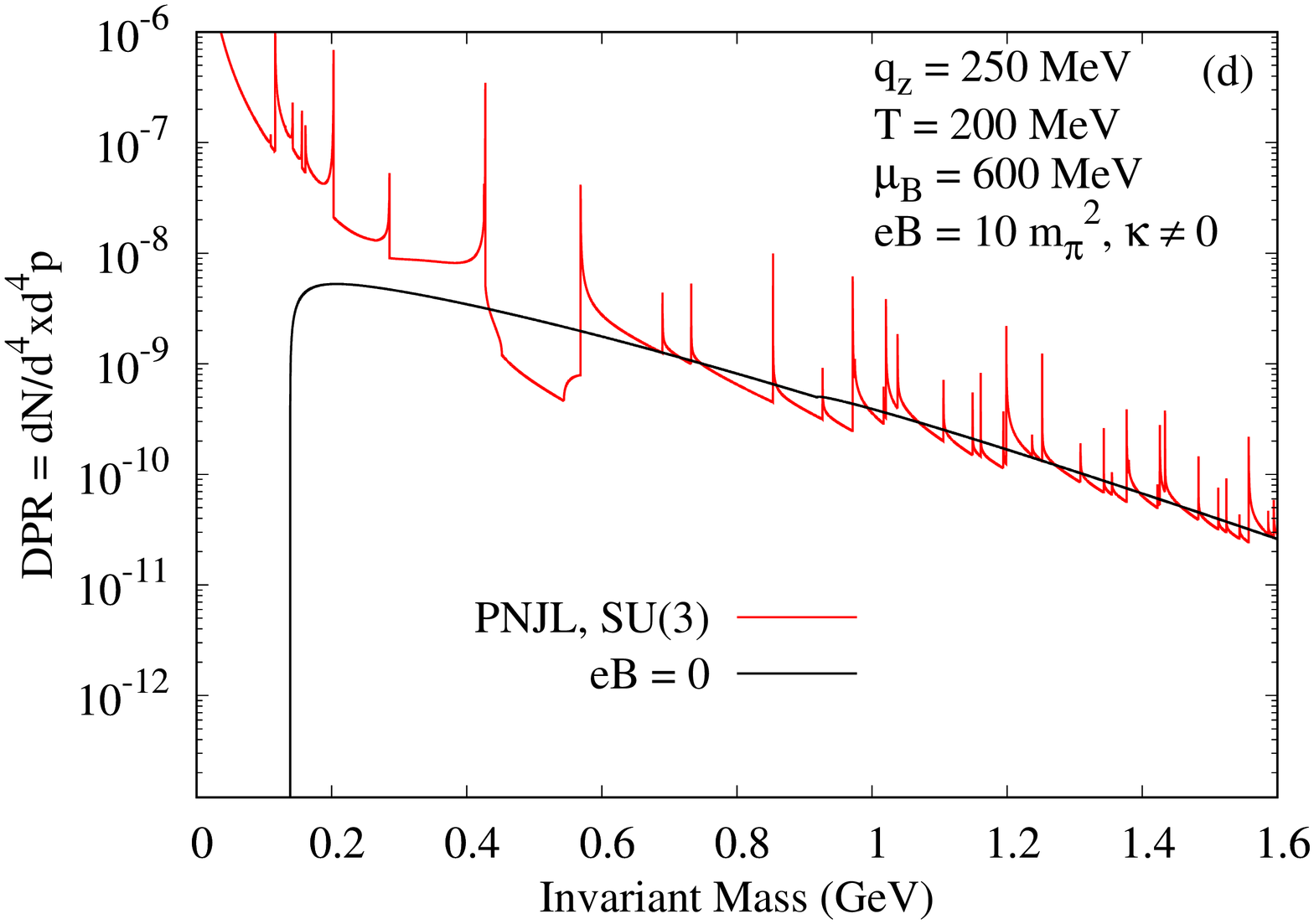}
	\end{center}
	\caption{(Color Online) Dilepton production rate at $q_z=250$ MeV, $T=125$ MeV, $\mu_B=600$ MeV at $ eB = 10~m_\pi^2 $ using (a) two-flavor NJL model  (b) two-flavor PNJL model (c) three-flavor NJL model and (d) three-flavor PNJL model with considering the finite values of AMM of the quarks. The  DPR at $ eB = 0 $ (black solid line) is also shown for comparison. }
	\label{fig.dpr5}
\end{figure}

So far we have observed that, the $ T  $ and/or $ \mu_B $ dependence of constituent quark mass of low lying quarks plays a dominant role in the modifications of DPRs. From Figs.~\ref{fig.M1} and \ref{fig.M2}, it can be seen that, inclusion of finite values of the AMM of the quarks does not yield any significant modifications in the qualitative behaviour of constituent mass. As a result the changes in DPRs are also not appreciable. For example, in Fig.~\ref{fig.dpr5} we have depicted DPRs at $ T = 200 $~MeV and $ \mu_B = 600 $~MeV in presence of a uniform background magnetic field i.e. $ eB = 10~ m_\pi^2  $ as a function of the dilepton invariant mass $ \sqrt{q^2} $ considering the finite values of the AMM of the quarks using (a) two-flavor NJL model, (b) two-flavor PNJL model, (c) three-flavor NJL model and (d) three-flavor PNJL model respectively. As expected, the results are qualitatively same with the scenario when the AMM of the quarks is turned off. However, quantitative differences exists since the expressions for determining the thresholds of unitary-I and Landau cuts will be modified following Eq.~\eqref{eq.LandauCut}.  

\begin{figure}[h]
	\begin{center}
		\includegraphics[angle=-90,scale=0.32]{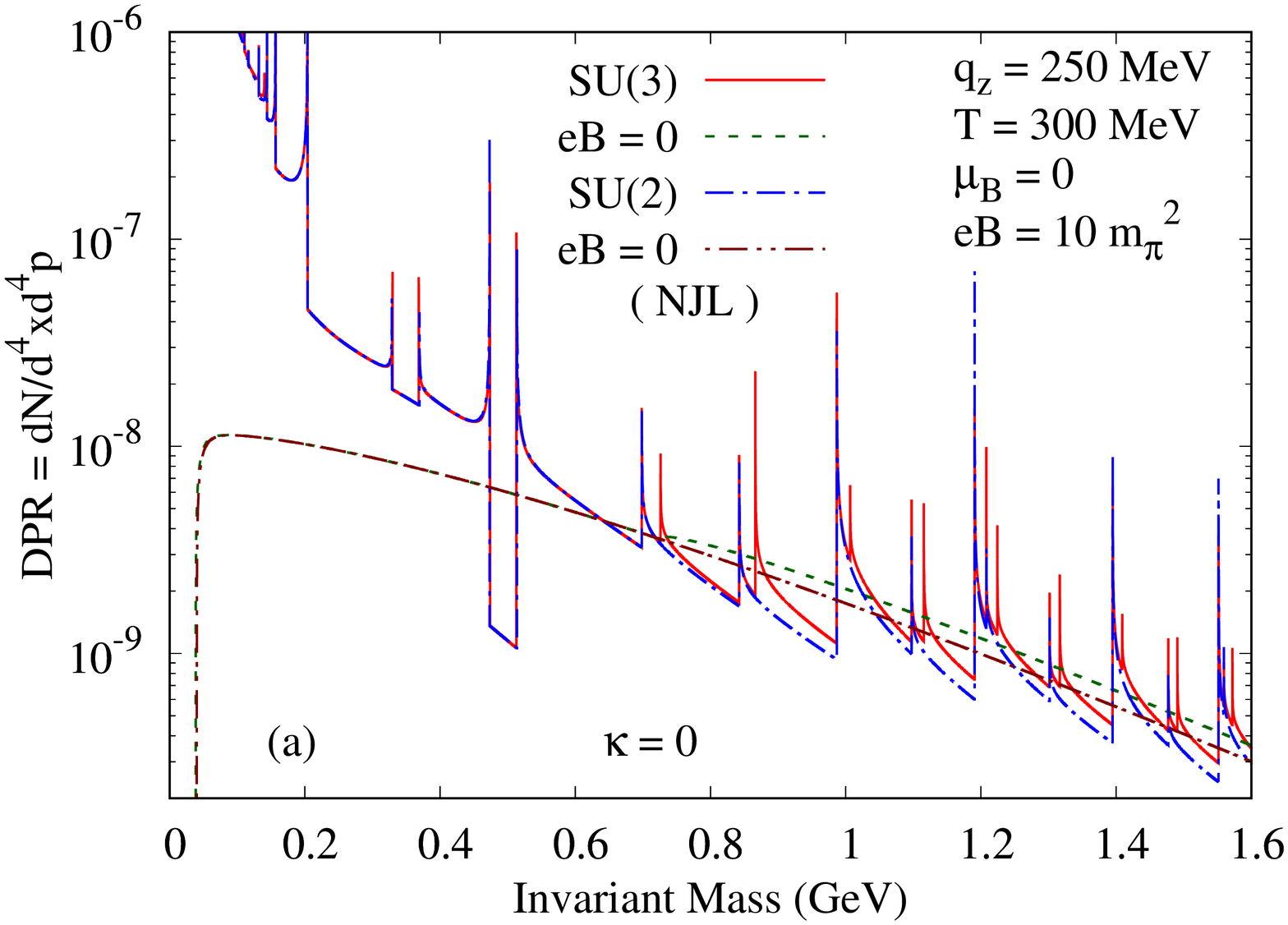}  \includegraphics[angle=-90,scale=0.32]{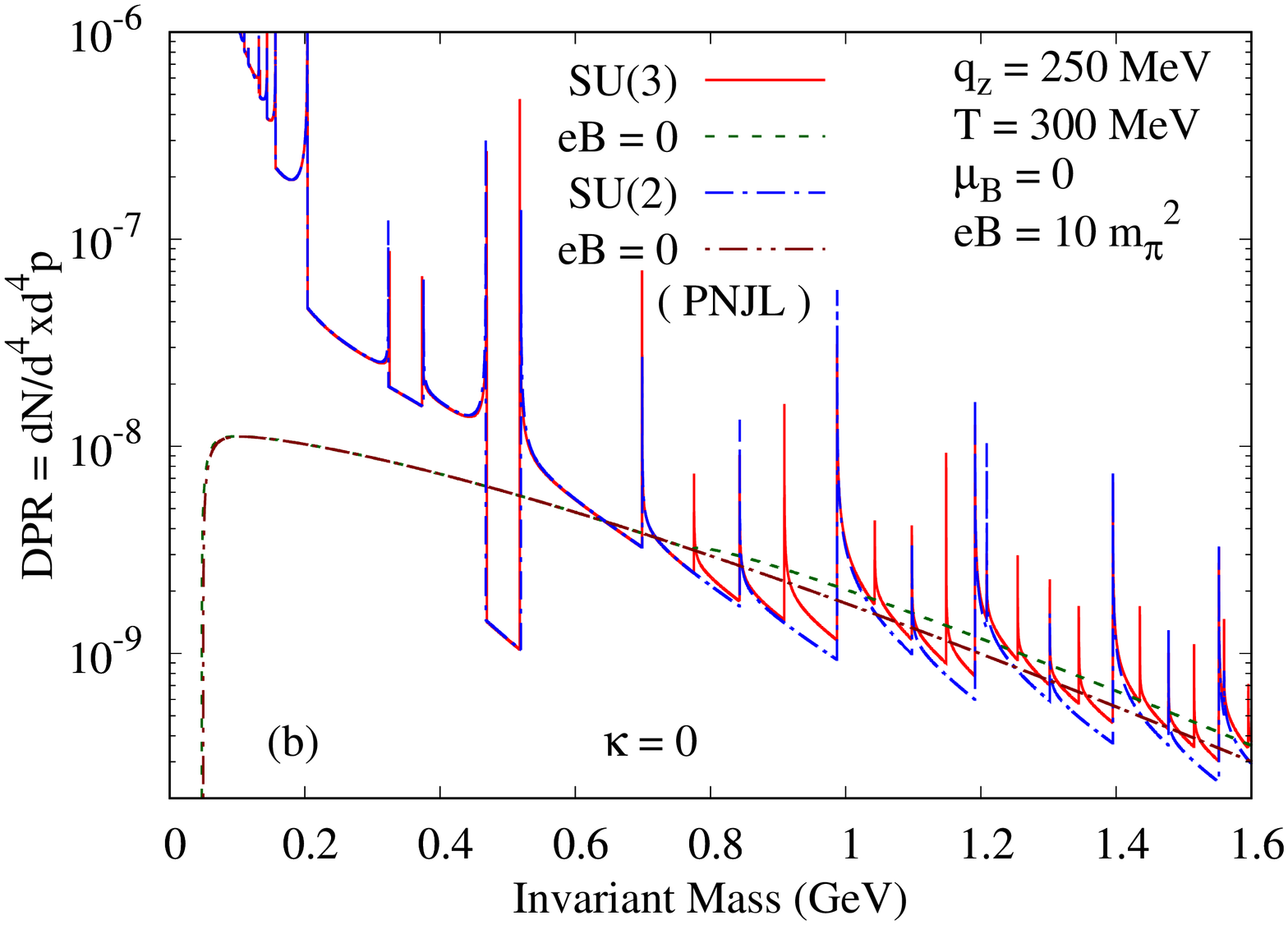}\\
		\includegraphics[angle=-90,scale=0.32]{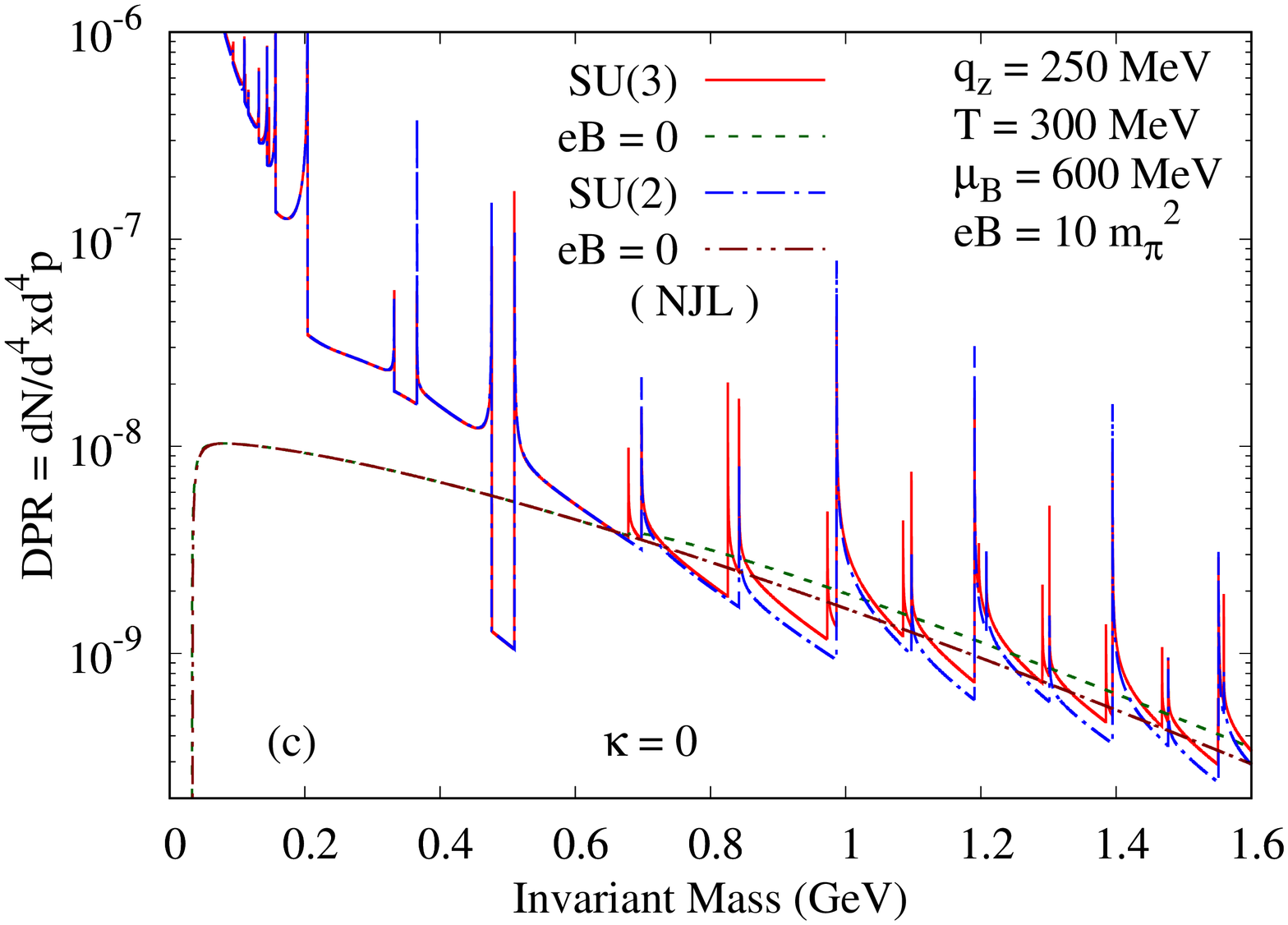}  \includegraphics[angle=-90,scale=0.32]{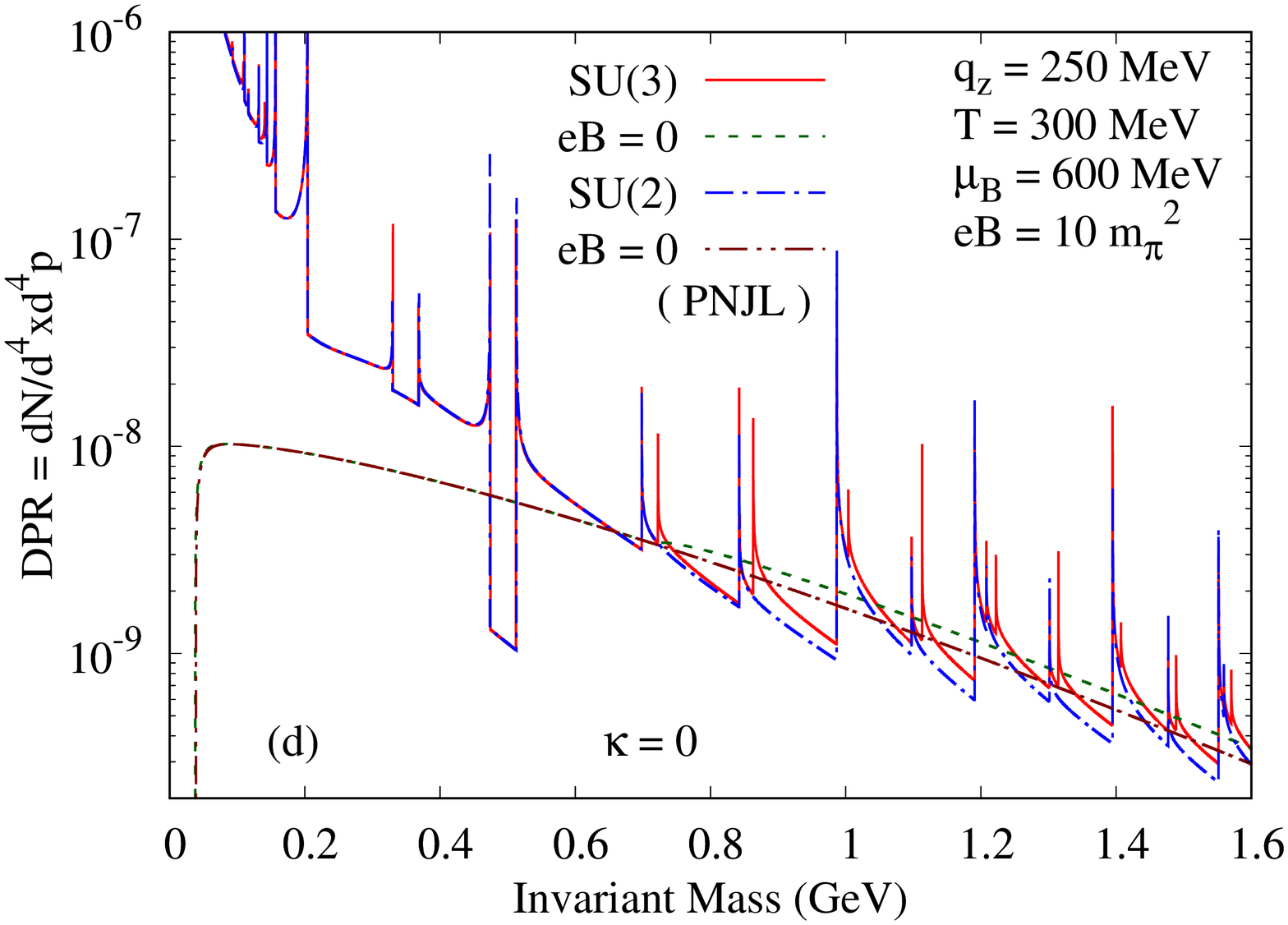}
	\end{center}
	\caption{(Color Online) Comparison of Dilepton production rate at $q_z=250$ MeV, $T=300$ MeV,  at $ eB = 10~m_\pi^2 $ using two and three-flavor NJL and PNJL models at	  $ \mu_B = 0 $ (in figures (a) and (b))  and  $ \mu_B = 600  $~MeV (in figures (c) and (d))   without considering the finite values of AMM of the quarks. The  DPR at $ eB = 0 $ are also shown for comparison. }
	\label{fig.dpr6}
\end{figure}

As the temperature of quark matter produced in HIC is expected to be of the order of $ \sim 300-400 $~MeV, we have evaluated DPR for both two and three-flavor system at higher temperature compared to the previous cases. In Figs.~\ref{fig.dpr6}~(a) and (b)  we have compared  dilepton production rate in presence of nonzero background field at  $T=300$ MeV in two and three-flavor NJL and PNJL models at $ \mu_B = 0 $ without considering the finite values of AMM of the quarks.   At this high value of $ T $, the constituent mass of both  $ u $ and $ d $ quarks, independent of type of models, is in bare quark mass limit indicating partial restoration of chiral symmetry. However the constituent mass of  $ s $ quarks are still larger than its bare mass as can be seen from Figs.~\ref{fig.M1}~(a) and (b).  As a consequence of this, in Fig.~\ref{fig.dpr6}~(a), one can observe the enhancement in the DPRs   for three-flavor model for both $ eB $ zero and nonzero case at higher values of invariant mass  when compared to the two-flavor case.  In Fig.~\ref{fig.dpr6}~(b), Polyakov loop extended NJL model is used. It can be noticed that, the results are almost identical with Fig.~\ref{fig.dpr6}~(a). This is because at high values of $ T $ considered here, $ \Phi, \bar{\Phi } \rightarrow 1$ and it is well known that in this limit NJL and PNJL models lead to similar results~\cite{Hansen:2006ee}. In Figs.~\ref{fig.dpr6}~(c) and (d) we have examined the same system at $ \mu_B = 600$~MeV. From Figs.~\ref{fig.M1} and \ref{fig.M2}, it can be found that, at high $ T $ values, the modifications in the constituent mass of quarks due to inclusion of finite values of $ \mu_B $ are insignificant. So again the results shown in Figs.~\ref{fig.dpr6}~(c) and (d) are almost similar to that in Figs.~\ref{fig.dpr6}~(a) and (b).  It should be noted that, one can also study the effects of strange degrees of freedom in the DPRs considering the finite values of AMM of the quarks in the similar manner. We expect the behaviour will be similar to what we have shown in Fig.~\ref{fig.dpr6}. 

Finally, we would like to discuss how DPR from a hot and dense magentized medium may be modified due to presence of Inverse Magnetic Catalysis (IMC) effect. In the absence of quark chemical potential, results from Latiice QCD simulations~\cite{Bali:2011qj,Bali:2012av,Preis:2010cq,Preis:2012fh} show that  at vanishing temperature, finite magnetic  field has strong tendency to strengthen the spin-zero fermion-antifermion  condensate, resulting in Magnetic Catalysis. However, at finite temperature, an opposite effect is observed which leads to the decrease of pseudo-critical  temperature for chiral symmetry breaking with increase in $ eB $, indicating Inverse Magnetic Catalysis (IMC). It is well-known that NJL-type models are in agreement with the MC effect at $ T = 0 $, as we have already seen in Fig.~\ref{fig.M1}, although, they are  unable to predict the IMC at finite temperature. However, there are several attempts to obtain IMC effect with NJL-type models by introducing $ T $ and $ eB $ dependent couplings or going beyond mean-field and so on (see Ref.~\cite{Bandyopadhyay:2020zte} for a review). Using these different strategies, if one ensures that, at finite values of temperature  IMC is present, then it will modify the $ T $-dependence of constituent mass of quarks which will approach the bare mass limit at comparatively lower values of $ T $ owing to the partial restoration of chiral symmetry. While discussing Figs.~\ref{fig.dpr1} and \ref{fig.dpr2}, we have observed that, the threshold of unitary-I and Landau cuts have strong dependence on the constituent mass of the low lying quarks. It is found that, with decreasing values of constituent mass of quarks the threshold of unitary-I cut moves towards lower values of invariant mass, whereas,  the Landau cut contributions survive up to large values of  invariant mass as $ M $ decreases. So in  presence of the IMC effect, it is expected that the merging of the unitary-I and Landau cuts ( which enables the production of continuous spectrum of dileptons through out the whole invariant mass region ) will occur at smaller values of temperature compared to the scenario when the IMC is absent.
%
\section{SUMMARY \& CONCLUSIONS} \label{sec.summary}
In summary, considering the finite AMM of the quarks, we have calculated the DPR from hot and dense quark matter in the presence of an arbitrary external magnetic field.  Employing the real time formalism of finite temperature field theory and the Schwinger proper time formalism, we have evaluated the electromagnetic spectral function of the vector current correlator at finite temperature, chemical potential, external magnetic field and AMM of the quarks in which the constituent quark mass $M=M(T,\mu,B,\kappa)$ has entered as an input. The three-flavor PNJL model with the gauge invariant Pauli-Villiars scheme has been used to obtain the constituent quark mass in the mean field approximation. The constituent quark mass is found to depend strongly on the external parameters like temperature ($T$), chemical potential ($\mu$), magnetic field ($B$) and AMM ($\kappa$) of the quarks $M=M(T,\mu,B,\kappa)$ and it essentially captures the effect of `strong' interactions specifically around the (pseudo) chiral and confinement-deconfinement phase transition regions. The constituent quark mass $M=M(T,\mu,B,\kappa)$ of different flavors is found to play a primary role in the determination of the thresholds and intensity of dilepton emission. We have also analyzed the complete analytic structure of the in-medium spectral function in the complex energy plane and have found a non-trivial Landau cut in the physical kinematic region resulting from the scattering of the Landau quantized quark/antiquark with the photon; a purely finite magnetic field effect. We have also compared the whole study with the three-flavor NJL and the two-flavor NJL and PNJL models to observe the effects of strangeness and confinement-deconfinement. In all the calculations, we have considered an infinite number of quark Landau levels so that no approximation has been made on the strength of the background magnetic field.

We observe that, in the presence of background magnetic field, independent of the types of model used, DPR acquires contributions from both the unitary-I and Landau cuts owing to the kinematic restrictions for production of physical dileptons with positive energy and time-like four momentum. As a result, the DPR obtained is found to be largely enhanced in the low invariant mass region due to the emergence of the Landau cut contributions, which is purely the effect of finite background field. The detailed analysis of the analytic structure of the two point correlator indicates the non-trivial dependence of the thresholds of unitary-I and Landau cut contributions on magnetic field and AMM of the quarks which in turn affects the constituent quark mass. For a particular value of $ T $ and/or $ \mu_B $, we have found that, the width of the forbidden gap (between the unitary and Landau cuts) is large for both two and three-flavor PNJL models when compared to the scenarios where simple NJL model is used. However, as the values of temperature and/or baryon chemical potential increases the forbidden gap between unitary-I and Landau cut contributions keep on decreasing and finally merges with each other enabling the production of a continuous spectrum of dileptons through out the whole range of invariant mass. We find that, the presence of Polyakov loop can shift this merging of unitary-I and Landau cut contributions towards higher values of temperature.  At these high values of $ T $ and/or $ \mu_B $, a substantial increase in overall magnitude of DPRs is observed as a consequence of enhancement in the availability of the thermal phase space. Finally, at very high values of $ T $  and $ \mu_B $, the DPRs for both NJL and PNJL models are found to be slightly  enhanced when three-flavor model is used due to contributions from strange quarks. With Pauli-Villiars regularization, which is used through out this article, the inclusion of finite values of the AMM of the quarks do not lead to significant change in values of the constituent mass of quarks.


\section*{Acknowledgments}
We acknowledge Dr. Shijun Mao for useful discussions related to the Pauli-Villiars regularization. NC, SS and PR are funded by the Department of Atomic Energy (DAE), Government of India. SG is funded by the Department of Higher Education, Government of West Bengal. 


\bibliography{Nilanjan}

\begin{thebibliography}{86}%
\makeatletter
\providecommand \@ifxundefined [1]{%
 \@ifx{#1\undefined}
}%
\providecommand \@ifnum [1]{%
 \ifnum #1\expandafter \@firstoftwo
 \else \expandafter \@secondoftwo
 \fi
}%
\providecommand \@ifx [1]{%
 \ifx #1\expandafter \@firstoftwo
 \else \expandafter \@secondoftwo
 \fi
}%
\providecommand \natexlab [1]{#1}%
\providecommand \enquote  [1]{``#1''}%
\providecommand \bibnamefont  [1]{#1}%
\providecommand \bibfnamefont [1]{#1}%
\providecommand \citenamefont [1]{#1}%
\providecommand \href@noop [0]{\@secondoftwo}%
\providecommand \href [0]{\begingroup \@sanitize@url \@href}%
\providecommand \@href[1]{\@@startlink{#1}\@@href}%
\providecommand \@@href[1]{\endgroup#1\@@endlink}%
\providecommand \@sanitize@url [0]{\catcode `\\12\catcode `\$12\catcode
  `\&12\catcode `\#12\catcode `\^12\catcode `\_12\catcode `\%12\relax}%
\providecommand \@@startlink[1]{}%
\providecommand \@@endlink[0]{}%
\providecommand \url  [0]{\begingroup\@sanitize@url \@url }%
\providecommand \@url [1]{\endgroup\@href {#1}{\urlprefix }}%
\providecommand \urlprefix  [0]{URL }%
\providecommand \Eprint [0]{\href }%
\providecommand \doibase [0]{http://dx.doi.org/}%
\providecommand \selectlanguage [0]{\@gobble}%
\providecommand \bibinfo  [0]{\@secondoftwo}%
\providecommand \bibfield  [0]{\@secondoftwo}%
\providecommand \translation [1]{[#1]}%
\providecommand \BibitemOpen [0]{}%
\providecommand \bibitemStop [0]{}%
\providecommand \bibitemNoStop [0]{.\EOS\space}%
\providecommand \EOS [0]{\spacefactor3000\relax}%
\providecommand \BibitemShut  [1]{\csname bibitem#1\endcsname}%
\let\auto@bib@innerbib\@empty
\bibitem [{\citenamefont {Wong}(1995)}]{Wong:1995jf}%
  \BibitemOpen
  \bibfield  {author} {\bibinfo {author} {\bibfnamefont {C.~Y.}\ \bibnamefont
  {Wong}},\ }\href@noop {} {\emph {\bibinfo {title} {{Introduction to
  high-energy heavy ion collisions}}}}\ (\bibinfo {year} {1995})\BibitemShut
  {NoStop}%
\bibitem [{\citenamefont {McLerran}\ and\ \citenamefont
  {Toimela}(1985)}]{McLerran:1984ay}%
  \BibitemOpen
  \bibfield  {author} {\bibinfo {author} {\bibfnamefont {L.~D.}\ \bibnamefont
  {McLerran}}\ and\ \bibinfo {author} {\bibfnamefont {T.}~\bibnamefont
  {Toimela}},\ }\href {\doibase 10.1103/PhysRevD.31.545} {\bibfield  {journal}
  {\bibinfo  {journal} {Phys. Rev.}\ }\textbf {\bibinfo {volume} {D31}},\
  \bibinfo {pages} {545} (\bibinfo {year} {1985})}\BibitemShut {NoStop}%
\bibitem [{\citenamefont {Kajantie}\ \emph {et~al.}(1986)\citenamefont
  {Kajantie}, \citenamefont {Kapusta}, \citenamefont {McLerran},\ and\
  \citenamefont {Mekjian}}]{Kajantie:1986dh}%
  \BibitemOpen
  \bibfield  {author} {\bibinfo {author} {\bibfnamefont {K.}~\bibnamefont
  {Kajantie}}, \bibinfo {author} {\bibfnamefont {J.~I.}\ \bibnamefont
  {Kapusta}}, \bibinfo {author} {\bibfnamefont {L.~D.}\ \bibnamefont
  {McLerran}}, \ and\ \bibinfo {author} {\bibfnamefont {A.}~\bibnamefont
  {Mekjian}},\ }\href {\doibase 10.1103/PhysRevD.34.2746} {\bibfield  {journal}
  {\bibinfo  {journal} {Phys. Rev.}\ }\textbf {\bibinfo {volume} {D34}},\
  \bibinfo {pages} {2746} (\bibinfo {year} {1986})}\BibitemShut {NoStop}%
\bibitem [{\citenamefont {Weldon}(1990)}]{Weldon:1990iw}%
  \BibitemOpen
  \bibfield  {author} {\bibinfo {author} {\bibfnamefont {H.~A.}\ \bibnamefont
  {Weldon}},\ }\href {\doibase 10.1103/PhysRevD.42.2384} {\bibfield  {journal}
  {\bibinfo  {journal} {Phys. Rev.}\ }\textbf {\bibinfo {volume} {D42}},\
  \bibinfo {pages} {2384} (\bibinfo {year} {1990})}\BibitemShut {NoStop}%
\bibitem [{\citenamefont {Alam}\ \emph {et~al.}(1996)\citenamefont {Alam},
  \citenamefont {Sinha},\ and\ \citenamefont {Raha}}]{Alam:1996fd}%
  \BibitemOpen
  \bibfield  {author} {\bibinfo {author} {\bibfnamefont {J.}~\bibnamefont
  {Alam}}, \bibinfo {author} {\bibfnamefont {B.}~\bibnamefont {Sinha}}, \ and\
  \bibinfo {author} {\bibfnamefont {S.}~\bibnamefont {Raha}},\ }\href {\doibase
  10.1016/0370-1573(95)00084-4} {\bibfield  {journal} {\bibinfo  {journal}
  {Phys. Rept.}\ }\textbf {\bibinfo {volume} {273}},\ \bibinfo {pages} {243}
  (\bibinfo {year} {1996})}\BibitemShut {NoStop}%
\bibitem [{\citenamefont {Alam}\ \emph {et~al.}(2001)\citenamefont {Alam},
  \citenamefont {Sarkar}, \citenamefont {Roy}, \citenamefont {Hatsuda},\ and\
  \citenamefont {Sinha}}]{Alam:1999sc}%
  \BibitemOpen
  \bibfield  {author} {\bibinfo {author} {\bibfnamefont {J.}~\bibnamefont
  {Alam}}, \bibinfo {author} {\bibfnamefont {S.}~\bibnamefont {Sarkar}},
  \bibinfo {author} {\bibfnamefont {P.}~\bibnamefont {Roy}}, \bibinfo {author}
  {\bibfnamefont {T.}~\bibnamefont {Hatsuda}}, \ and\ \bibinfo {author}
  {\bibfnamefont {B.}~\bibnamefont {Sinha}},\ }\href {\doibase
  10.1006/aphy.2000.6091} {\bibfield  {journal} {\bibinfo  {journal} {Annals
  Phys.}\ }\textbf {\bibinfo {volume} {286}},\ \bibinfo {pages} {159} (\bibinfo
  {year} {2001})},\ \Eprint {http://arxiv.org/abs/hep-ph/9909267}
  {arXiv:hep-ph/9909267 [hep-ph]} \BibitemShut {NoStop}%
\bibitem [{\citenamefont {Rapp}\ and\ \citenamefont
  {Wambach}(2000)}]{Rapp:1999ej}%
  \BibitemOpen
  \bibfield  {author} {\bibinfo {author} {\bibfnamefont {R.}~\bibnamefont
  {Rapp}}\ and\ \bibinfo {author} {\bibfnamefont {J.}~\bibnamefont {Wambach}},\
  }\href {\doibase 10.1007/0-306-47101-9_1} {\bibfield  {journal} {\bibinfo
  {journal} {Adv. Nucl. Phys.}\ }\textbf {\bibinfo {volume} {25}},\ \bibinfo
  {pages} {1} (\bibinfo {year} {2000})},\ \Eprint
  {http://arxiv.org/abs/hep-ph/9909229} {arXiv:hep-ph/9909229 [hep-ph]}
  \BibitemShut {NoStop}%
\bibitem [{\citenamefont {Aurenche}\ \emph {et~al.}(2000)\citenamefont
  {Aurenche}, \citenamefont {Gelis},\ and\ \citenamefont
  {Zaraket}}]{Aurenche:2000gf}%
  \BibitemOpen
  \bibfield  {author} {\bibinfo {author} {\bibfnamefont {P.}~\bibnamefont
  {Aurenche}}, \bibinfo {author} {\bibfnamefont {F.}~\bibnamefont {Gelis}}, \
  and\ \bibinfo {author} {\bibfnamefont {H.}~\bibnamefont {Zaraket}},\ }\href
  {\doibase 10.1103/PhysRevD.62.096012} {\bibfield  {journal} {\bibinfo
  {journal} {Phys. Rev.}\ }\textbf {\bibinfo {volume} {D62}},\ \bibinfo {pages}
  {096012} (\bibinfo {year} {2000})},\ \Eprint
  {http://arxiv.org/abs/hep-ph/0003326} {arXiv:hep-ph/0003326 [hep-ph]}
  \BibitemShut {NoStop}%
\bibitem [{\citenamefont {Arnold}\ \emph {et~al.}(2001)\citenamefont {Arnold},
  \citenamefont {Moore},\ and\ \citenamefont {Yaffe}}]{Arnold:2001ms}%
  \BibitemOpen
  \bibfield  {author} {\bibinfo {author} {\bibfnamefont {P.~B.}\ \bibnamefont
  {Arnold}}, \bibinfo {author} {\bibfnamefont {G.~D.}\ \bibnamefont {Moore}}, \
  and\ \bibinfo {author} {\bibfnamefont {L.~G.}\ \bibnamefont {Yaffe}},\ }\href
  {\doibase 10.1088/1126-6708/2001/12/009} {\bibfield  {journal} {\bibinfo
  {journal} {JHEP}\ }\textbf {\bibinfo {volume} {12}},\ \bibinfo {pages} {009}
  (\bibinfo {year} {2001})},\ \Eprint {http://arxiv.org/abs/hep-ph/0111107}
  {arXiv:hep-ph/0111107 [hep-ph]} \BibitemShut {NoStop}%
\bibitem [{\citenamefont {Rapp}\ and\ \citenamefont {van
  Hees}(2010)}]{Rapp:2009my}%
  \BibitemOpen
  \bibfield  {author} {\bibinfo {author} {\bibfnamefont {R.}~\bibnamefont
  {Rapp}}\ and\ \bibinfo {author} {\bibfnamefont {H.}~\bibnamefont {van
  Hees}},\ }in\ \href {\doibase 10.1142/9789814293297_0003} {\emph {\bibinfo
  {booktitle} {{Quark-gluon plasma 4}}}}\ (\bibinfo {year} {2010})\ pp.\
  \bibinfo {pages} {111--206},\ \Eprint {http://arxiv.org/abs/0903.1096}
  {arXiv:0903.1096 [hep-ph]} \BibitemShut {NoStop}%
\bibitem [{\citenamefont {Chatterjee}\ \emph {et~al.}(2010)\citenamefont
  {Chatterjee}, \citenamefont {Bhattacharya},\ and\ \citenamefont
  {Srivastava}}]{Chatterjee:2009rs}%
  \BibitemOpen
  \bibfield  {author} {\bibinfo {author} {\bibfnamefont {R.}~\bibnamefont
  {Chatterjee}}, \bibinfo {author} {\bibfnamefont {L.}~\bibnamefont
  {Bhattacharya}}, \ and\ \bibinfo {author} {\bibfnamefont {D.~K.}\
  \bibnamefont {Srivastava}},\ }\bibfield  {booktitle} {\emph {\bibinfo
  {booktitle} {{QGP Winter School 2008 Jaipur, India, February 1-3, 2008}}},\
  }\href {\doibase 10.1007/978-3-642-02286-9_7} {\bibfield  {journal} {\bibinfo
   {journal} {Lect. Notes Phys.}\ }\textbf {\bibinfo {volume} {785}},\ \bibinfo
  {pages} {219} (\bibinfo {year} {2010})},\ \Eprint
  {http://arxiv.org/abs/0901.3610} {arXiv:0901.3610 [nucl-th]} \BibitemShut
  {NoStop}%
\bibitem [{\citenamefont {Mallik}\ and\ \citenamefont
  {Sarkar}(2016)}]{Mallik:2016anp}%
  \BibitemOpen
  \bibfield  {author} {\bibinfo {author} {\bibfnamefont {S.}~\bibnamefont
  {Mallik}}\ and\ \bibinfo {author} {\bibfnamefont {S.}~\bibnamefont
  {Sarkar}},\ }\href {\doibase 10.1017/9781316535585} {\emph {\bibinfo {title}
  {{Hadrons at Finite Temperature}}}}\ (\bibinfo  {publisher} {Cambridge
  University Press},\ \bibinfo {address} {Cambridge},\ \bibinfo {year}
  {2016})\BibitemShut {NoStop}%
\bibitem [{\citenamefont {Sarkar}\ and\ \citenamefont
  {Ghosh}(2012)}]{Sarkar:2012ty}%
  \BibitemOpen
  \bibfield  {author} {\bibinfo {author} {\bibfnamefont {S.}~\bibnamefont
  {Sarkar}}\ and\ \bibinfo {author} {\bibfnamefont {S.}~\bibnamefont {Ghosh}},\
  }\href {\doibase 10.1088/1742-6596/374/1/012010} {\bibfield  {journal}
  {\bibinfo  {journal} {J. Phys. Conf. Ser.}\ }\textbf {\bibinfo {volume}
  {374}},\ \bibinfo {pages} {012010} (\bibinfo {year} {2012})},\ \Eprint
  {http://arxiv.org/abs/1204.0893} {arXiv:1204.0893 [nucl-th]} \BibitemShut
  {NoStop}%
\bibitem [{\citenamefont {Kharzeev}\ \emph {et~al.}(2008)\citenamefont
  {Kharzeev}, \citenamefont {McLerran},\ and\ \citenamefont
  {Warringa}}]{Kharzeev:2007jp}%
  \BibitemOpen
  \bibfield  {author} {\bibinfo {author} {\bibfnamefont {D.~E.}\ \bibnamefont
  {Kharzeev}}, \bibinfo {author} {\bibfnamefont {L.~D.}\ \bibnamefont
  {McLerran}}, \ and\ \bibinfo {author} {\bibfnamefont {H.~J.}\ \bibnamefont
  {Warringa}},\ }\href {\doibase 10.1016/j.nuclphysa.2008.02.298} {\bibfield
  {journal} {\bibinfo  {journal} {Nucl. Phys.}\ }\textbf {\bibinfo {volume}
  {A803}},\ \bibinfo {pages} {227} (\bibinfo {year} {2008})},\ \Eprint
  {http://arxiv.org/abs/0711.0950} {arXiv:0711.0950 [hep-ph]} \BibitemShut
  {NoStop}%
\bibitem [{\citenamefont {Skokov}\ \emph {et~al.}(2009)\citenamefont {Skokov},
  \citenamefont {Illarionov},\ and\ \citenamefont {Toneev}}]{Skokov:2009qp}%
  \BibitemOpen
  \bibfield  {author} {\bibinfo {author} {\bibfnamefont {V.}~\bibnamefont
  {Skokov}}, \bibinfo {author} {\bibfnamefont {A.~{\relax Yu}.}\ \bibnamefont
  {Illarionov}}, \ and\ \bibinfo {author} {\bibfnamefont {V.}~\bibnamefont
  {Toneev}},\ }\href {\doibase 10.1142/S0217751X09047570} {\bibfield  {journal}
  {\bibinfo  {journal} {Int. J. Mod. Phys.}\ }\textbf {\bibinfo {volume}
  {A24}},\ \bibinfo {pages} {5925} (\bibinfo {year} {2009})},\ \Eprint
  {http://arxiv.org/abs/0907.1396} {arXiv:0907.1396 [nucl-th]} \BibitemShut
  {NoStop}%
\bibitem [{\citenamefont {Tuchin}(2013{\natexlab{a}})}]{Tuchin:2013apa}%
  \BibitemOpen
  \bibfield  {author} {\bibinfo {author} {\bibfnamefont {K.}~\bibnamefont
  {Tuchin}},\ }\href {\doibase 10.1103/PhysRevC.88.024911} {\bibfield
  {journal} {\bibinfo  {journal} {Phys. Rev. C}\ }\textbf {\bibinfo {volume}
  {88}},\ \bibinfo {pages} {024911} (\bibinfo {year} {2013}{\natexlab{a}})},\
  \Eprint {http://arxiv.org/abs/1305.5806} {arXiv:1305.5806 [hep-ph]}
  \BibitemShut {NoStop}%
\bibitem [{\citenamefont {Tuchin}(2016)}]{Tuchin:2015oka}%
  \BibitemOpen
  \bibfield  {author} {\bibinfo {author} {\bibfnamefont {K.}~\bibnamefont
  {Tuchin}},\ }\href {\doibase 10.1103/PhysRevC.93.014905} {\bibfield
  {journal} {\bibinfo  {journal} {Phys. Rev. C}\ }\textbf {\bibinfo {volume}
  {93}},\ \bibinfo {pages} {014905} (\bibinfo {year} {2016})},\ \Eprint
  {http://arxiv.org/abs/1508.06925} {arXiv:1508.06925 [hep-ph]} \BibitemShut
  {NoStop}%
\bibitem [{\citenamefont {Tuchin}(2013{\natexlab{b}})}]{Tuchin:2013ie}%
  \BibitemOpen
  \bibfield  {author} {\bibinfo {author} {\bibfnamefont {K.}~\bibnamefont
  {Tuchin}},\ }\href {\doibase 10.1155/2013/490495} {\bibfield  {journal}
  {\bibinfo  {journal} {Adv. High Energy Phys.}\ }\textbf {\bibinfo {volume}
  {2013}},\ \bibinfo {pages} {490495} (\bibinfo {year} {2013}{\natexlab{b}})},\
  \Eprint {http://arxiv.org/abs/1301.0099} {arXiv:1301.0099 [hep-ph]}
  \BibitemShut {NoStop}%
\bibitem [{\citenamefont {Gursoy}\ \emph {et~al.}(2014)\citenamefont {Gursoy},
  \citenamefont {Kharzeev},\ and\ \citenamefont {Rajagopal}}]{Gursoy:2014aka}%
  \BibitemOpen
  \bibfield  {author} {\bibinfo {author} {\bibfnamefont {U.}~\bibnamefont
  {Gursoy}}, \bibinfo {author} {\bibfnamefont {D.}~\bibnamefont {Kharzeev}}, \
  and\ \bibinfo {author} {\bibfnamefont {K.}~\bibnamefont {Rajagopal}},\ }\href
  {\doibase 10.1103/PhysRevC.89.054905} {\bibfield  {journal} {\bibinfo
  {journal} {Phys. Rev.}\ }\textbf {\bibinfo {volume} {C89}},\ \bibinfo {pages}
  {054905} (\bibinfo {year} {2014})},\ \Eprint {http://arxiv.org/abs/1401.3805}
  {arXiv:1401.3805 [hep-ph]} \BibitemShut {NoStop}%
\bibitem [{\citenamefont {Duncan}\ and\ \citenamefont
  {Thompson}(1992)}]{Duncan:1992hi}%
  \BibitemOpen
  \bibfield  {author} {\bibinfo {author} {\bibfnamefont {R.~C.}\ \bibnamefont
  {Duncan}}\ and\ \bibinfo {author} {\bibfnamefont {C.}~\bibnamefont
  {Thompson}},\ }\href {\doibase 10.1086/186413} {\bibfield  {journal}
  {\bibinfo  {journal} {Astrophys. J.}\ }\textbf {\bibinfo {volume} {392}},\
  \bibinfo {pages} {L9} (\bibinfo {year} {1992})}\BibitemShut {NoStop}%
\bibitem [{\citenamefont {Thompson}\ and\ \citenamefont
  {Duncan}(1993)}]{Thompson:1993hn}%
  \BibitemOpen
  \bibfield  {author} {\bibinfo {author} {\bibfnamefont {C.}~\bibnamefont
  {Thompson}}\ and\ \bibinfo {author} {\bibfnamefont {R.~C.}\ \bibnamefont
  {Duncan}},\ }\href {\doibase 10.1086/172580} {\bibfield  {journal} {\bibinfo
  {journal} {Astrophys. J.}\ }\textbf {\bibinfo {volume} {408}},\ \bibinfo
  {pages} {194} (\bibinfo {year} {1993})}\BibitemShut {NoStop}%
\bibitem [{\citenamefont {Vachaspati}(1991)}]{Vachaspati:1991nm}%
  \BibitemOpen
  \bibfield  {author} {\bibinfo {author} {\bibfnamefont {T.}~\bibnamefont
  {Vachaspati}},\ }\href {\doibase 10.1016/0370-2693(91)90051-Q} {\bibfield
  {journal} {\bibinfo  {journal} {Phys. Lett.}\ }\textbf {\bibinfo {volume}
  {B265}},\ \bibinfo {pages} {258} (\bibinfo {year} {1991})}\BibitemShut
  {NoStop}%
\bibitem [{\citenamefont {Campanelli}(2013)}]{Campanelli:2013mea}%
  \BibitemOpen
  \bibfield  {author} {\bibinfo {author} {\bibfnamefont {L.}~\bibnamefont
  {Campanelli}},\ }\href {\doibase 10.1103/PhysRevLett.111.061301} {\bibfield
  {journal} {\bibinfo  {journal} {Phys. Rev. Lett.}\ }\textbf {\bibinfo
  {volume} {111}},\ \bibinfo {pages} {061301} (\bibinfo {year} {2013})},\
  \Eprint {http://arxiv.org/abs/1304.6534} {arXiv:1304.6534 [astro-ph.CO]}
  \BibitemShut {NoStop}%
\bibitem [{\citenamefont {Miransky}\ and\ \citenamefont
  {Shovkovy}(2015)}]{Miransky:2015ava}%
  \BibitemOpen
  \bibfield  {author} {\bibinfo {author} {\bibfnamefont {V.~A.}\ \bibnamefont
  {Miransky}}\ and\ \bibinfo {author} {\bibfnamefont {I.~A.}\ \bibnamefont
  {Shovkovy}},\ }\href {\doibase 10.1016/j.physrep.2015.02.003} {\bibfield
  {journal} {\bibinfo  {journal} {Phys. Rept.}\ }\textbf {\bibinfo {volume}
  {576}},\ \bibinfo {pages} {1} (\bibinfo {year} {2015})},\ \Eprint
  {http://arxiv.org/abs/1503.00732} {arXiv:1503.00732 [hep-ph]} \BibitemShut
  {NoStop}%
\bibitem [{\citenamefont {Kharzeev}\ \emph
  {et~al.}(2013{\natexlab{a}})\citenamefont {Kharzeev}, \citenamefont
  {Landsteiner}, \citenamefont {Schmitt},\ and\ \citenamefont
  {Yee}}]{Kharzeev:2013jha}%
  \BibitemOpen
  \bibinfo {editor} {\bibfnamefont {D.}~\bibnamefont {Kharzeev}}, \bibinfo
  {editor} {\bibfnamefont {K.}~\bibnamefont {Landsteiner}}, \bibinfo {editor}
  {\bibfnamefont {A.}~\bibnamefont {Schmitt}}, \ and\ \bibinfo {editor}
  {\bibfnamefont {H.-U.}\ \bibnamefont {Yee}},\ eds.,\ \href {\doibase
  10.1007/978-3-642-37305-3} {\emph {\bibinfo {title} {{Strongly Interacting
  Matter in Magnetic Fields}}}},\ Vol.\ \bibinfo {volume} {871}\ (\bibinfo
  {year} {2013})\BibitemShut {NoStop}%
\bibitem [{\citenamefont {Kharzeev}\ \emph
  {et~al.}(2013{\natexlab{b}})\citenamefont {Kharzeev}, \citenamefont
  {Landsteiner}, \citenamefont {Schmitt},\ and\ \citenamefont
  {Yee}}]{Kharzeev:2012ph}%
  \BibitemOpen
  \bibfield  {author} {\bibinfo {author} {\bibfnamefont {D.~E.}\ \bibnamefont
  {Kharzeev}}, \bibinfo {author} {\bibfnamefont {K.}~\bibnamefont
  {Landsteiner}}, \bibinfo {author} {\bibfnamefont {A.}~\bibnamefont
  {Schmitt}}, \ and\ \bibinfo {author} {\bibfnamefont {H.-U.}\ \bibnamefont
  {Yee}},\ }\href {\doibase 10.1007/978-3-642-37305-3_1} {\bibfield  {journal}
  {\bibinfo  {journal} {Lect. Notes Phys.}\ }\textbf {\bibinfo {volume}
  {871}},\ \bibinfo {pages} {1} (\bibinfo {year} {2013}{\natexlab{b}})},\
  \Eprint {http://arxiv.org/abs/1211.6245} {arXiv:1211.6245 [hep-ph]}
  \BibitemShut {NoStop}%
\bibitem [{\citenamefont {Kharzeev}\ and\ \citenamefont
  {Zhitnitsky}(2007)}]{Kharzeev:2007tn}%
  \BibitemOpen
  \bibfield  {author} {\bibinfo {author} {\bibfnamefont {D.}~\bibnamefont
  {Kharzeev}}\ and\ \bibinfo {author} {\bibfnamefont {A.}~\bibnamefont
  {Zhitnitsky}},\ }\href {\doibase 10.1016/j.nuclphysa.2007.10.001} {\bibfield
  {journal} {\bibinfo  {journal} {Nucl. Phys.}\ }\textbf {\bibinfo {volume}
  {A797}},\ \bibinfo {pages} {67} (\bibinfo {year} {2007})},\ \Eprint
  {http://arxiv.org/abs/0706.1026} {arXiv:0706.1026 [hep-ph]} \BibitemShut
  {NoStop}%
\bibitem [{\citenamefont {Chernodub}(2010)}]{Chernodub:2010qx}%
  \BibitemOpen
  \bibfield  {author} {\bibinfo {author} {\bibfnamefont {M.~N.}\ \bibnamefont
  {Chernodub}},\ }\href {\doibase 10.1103/PhysRevD.82.085011} {\bibfield
  {journal} {\bibinfo  {journal} {Phys. Rev.}\ }\textbf {\bibinfo {volume}
  {D82}},\ \bibinfo {pages} {085011} (\bibinfo {year} {2010})},\ \Eprint
  {http://arxiv.org/abs/1008.1055} {arXiv:1008.1055 [hep-ph]} \BibitemShut
  {NoStop}%
\bibitem [{\citenamefont {Chernodub}(2013)}]{Chernodub:2012tf}%
  \BibitemOpen
  \bibfield  {author} {\bibinfo {author} {\bibfnamefont {M.~N.}\ \bibnamefont
  {Chernodub}},\ }\href {\doibase 10.1007/978-3-642-37305-3_6} {\bibfield
  {journal} {\bibinfo  {journal} {Lect. Notes Phys.}\ }\textbf {\bibinfo
  {volume} {871}},\ \bibinfo {pages} {143} (\bibinfo {year} {2013})},\ \Eprint
  {http://arxiv.org/abs/1208.5025} {arXiv:1208.5025 [hep-ph]} \BibitemShut
  {NoStop}%
\bibitem [{\citenamefont {Fukushima}\ \emph {et~al.}(2008)\citenamefont
  {Fukushima}, \citenamefont {Kharzeev},\ and\ \citenamefont
  {Warringa}}]{Fukushima:2008xe}%
  \BibitemOpen
  \bibfield  {author} {\bibinfo {author} {\bibfnamefont {K.}~\bibnamefont
  {Fukushima}}, \bibinfo {author} {\bibfnamefont {D.~E.}\ \bibnamefont
  {Kharzeev}}, \ and\ \bibinfo {author} {\bibfnamefont {H.~J.}\ \bibnamefont
  {Warringa}},\ }\href {\doibase 10.1103/PhysRevD.78.074033} {\bibfield
  {journal} {\bibinfo  {journal} {Phys. Rev.}\ }\textbf {\bibinfo {volume}
  {D78}},\ \bibinfo {pages} {074033} (\bibinfo {year} {2008})},\ \Eprint
  {http://arxiv.org/abs/0808.3382} {arXiv:0808.3382 [hep-ph]} \BibitemShut
  {NoStop}%
\bibitem [{\citenamefont {Kharzeev}\ and\ \citenamefont
  {Warringa}(2009)}]{Kharzeev:2009pj}%
  \BibitemOpen
  \bibfield  {author} {\bibinfo {author} {\bibfnamefont {D.~E.}\ \bibnamefont
  {Kharzeev}}\ and\ \bibinfo {author} {\bibfnamefont {H.~J.}\ \bibnamefont
  {Warringa}},\ }\href {\doibase 10.1103/PhysRevD.80.034028} {\bibfield
  {journal} {\bibinfo  {journal} {Phys. Rev.}\ }\textbf {\bibinfo {volume}
  {D80}},\ \bibinfo {pages} {034028} (\bibinfo {year} {2009})},\ \Eprint
  {http://arxiv.org/abs/0907.5007} {arXiv:0907.5007 [hep-ph]} \BibitemShut
  {NoStop}%
\bibitem [{\citenamefont {Bali}\ \emph
  {et~al.}(2012{\natexlab{a}})\citenamefont {Bali}, \citenamefont {Bruckmann},
  \citenamefont {Endrodi}, \citenamefont {Fodor}, \citenamefont {Katz},
  \citenamefont {Krieg}, \citenamefont {Schafer},\ and\ \citenamefont
  {Szabo}}]{Bali:2011qj}%
  \BibitemOpen
  \bibfield  {author} {\bibinfo {author} {\bibfnamefont {G.~S.}\ \bibnamefont
  {Bali}}, \bibinfo {author} {\bibfnamefont {F.}~\bibnamefont {Bruckmann}},
  \bibinfo {author} {\bibfnamefont {G.}~\bibnamefont {Endrodi}}, \bibinfo
  {author} {\bibfnamefont {Z.}~\bibnamefont {Fodor}}, \bibinfo {author}
  {\bibfnamefont {S.~D.}\ \bibnamefont {Katz}}, \bibinfo {author}
  {\bibfnamefont {S.}~\bibnamefont {Krieg}}, \bibinfo {author} {\bibfnamefont
  {A.}~\bibnamefont {Schafer}}, \ and\ \bibinfo {author} {\bibfnamefont
  {K.~K.}\ \bibnamefont {Szabo}},\ }\href {\doibase 10.1007/JHEP02(2012)044}
  {\bibfield  {journal} {\bibinfo  {journal} {JHEP}\ }\textbf {\bibinfo
  {volume} {02}},\ \bibinfo {pages} {044} (\bibinfo {year}
  {2012}{\natexlab{a}})},\ \Eprint {http://arxiv.org/abs/1111.4956}
  {arXiv:1111.4956 [hep-lat]} \BibitemShut {NoStop}%
\bibitem [{\citenamefont {Shovkovy}(2013)}]{Shovkovy:2012zn}%
  \BibitemOpen
  \bibfield  {author} {\bibinfo {author} {\bibfnamefont {I.~A.}\ \bibnamefont
  {Shovkovy}},\ }\href {\doibase 10.1007/978-3-642-37305-3_2} {\bibfield
  {journal} {\bibinfo  {journal} {Lect. Notes Phys.}\ }\textbf {\bibinfo
  {volume} {871}},\ \bibinfo {pages} {13} (\bibinfo {year} {2013})},\ \Eprint
  {http://arxiv.org/abs/1207.5081} {arXiv:1207.5081 [hep-ph]} \BibitemShut
  {NoStop}%
\bibitem [{\citenamefont {Gusynin}\ \emph {et~al.}(1994)\citenamefont
  {Gusynin}, \citenamefont {Miransky},\ and\ \citenamefont
  {Shovkovy}}]{Gusynin:1994re}%
  \BibitemOpen
  \bibfield  {author} {\bibinfo {author} {\bibfnamefont {V.~P.}\ \bibnamefont
  {Gusynin}}, \bibinfo {author} {\bibfnamefont {V.~A.}\ \bibnamefont
  {Miransky}}, \ and\ \bibinfo {author} {\bibfnamefont {I.~A.}\ \bibnamefont
  {Shovkovy}},\ }\href {\doibase 10.1103/PhysRevLett.76.1005,
  10.1103/PhysRevLett.73.3499} {\bibfield  {journal} {\bibinfo  {journal}
  {Phys. Rev. Lett.}\ }\textbf {\bibinfo {volume} {73}},\ \bibinfo {pages}
  {3499} (\bibinfo {year} {1994})},\ \bibinfo {note} {[Erratum: Phys. Rev.
  Lett.76,1005(1996)]},\ \Eprint {http://arxiv.org/abs/hep-ph/9405262}
  {arXiv:hep-ph/9405262 [hep-ph]} \BibitemShut {NoStop}%
\bibitem [{\citenamefont {Gusynin}\ \emph {et~al.}(1996)\citenamefont
  {Gusynin}, \citenamefont {Miransky},\ and\ \citenamefont
  {Shovkovy}}]{Gusynin:1995nb}%
  \BibitemOpen
  \bibfield  {author} {\bibinfo {author} {\bibfnamefont {V.~P.}\ \bibnamefont
  {Gusynin}}, \bibinfo {author} {\bibfnamefont {V.~A.}\ \bibnamefont
  {Miransky}}, \ and\ \bibinfo {author} {\bibfnamefont {I.~A.}\ \bibnamefont
  {Shovkovy}},\ }\href {\doibase 10.1016/0550-3213(96)00021-1} {\bibfield
  {journal} {\bibinfo  {journal} {Nucl. Phys.}\ }\textbf {\bibinfo {volume}
  {B462}},\ \bibinfo {pages} {249} (\bibinfo {year} {1996})},\ \Eprint
  {http://arxiv.org/abs/hep-ph/9509320} {arXiv:hep-ph/9509320 [hep-ph]}
  \BibitemShut {NoStop}%
\bibitem [{\citenamefont {Gusynin}\ \emph {et~al.}(1999)\citenamefont
  {Gusynin}, \citenamefont {Miransky},\ and\ \citenamefont
  {Shovkovy}}]{Gusynin:1999pq}%
  \BibitemOpen
  \bibfield  {author} {\bibinfo {author} {\bibfnamefont {V.~P.}\ \bibnamefont
  {Gusynin}}, \bibinfo {author} {\bibfnamefont {V.~A.}\ \bibnamefont
  {Miransky}}, \ and\ \bibinfo {author} {\bibfnamefont {I.~A.}\ \bibnamefont
  {Shovkovy}},\ }\href {\doibase 10.1016/S0550-3213(99)00573-8} {\bibfield
  {journal} {\bibinfo  {journal} {Nucl. Phys.}\ }\textbf {\bibinfo {volume}
  {B563}},\ \bibinfo {pages} {361} (\bibinfo {year} {1999})},\ \Eprint
  {http://arxiv.org/abs/hep-ph/9908320} {arXiv:hep-ph/9908320 [hep-ph]}
  \BibitemShut {NoStop}%
\bibitem [{\citenamefont {Preis}\ \emph {et~al.}(2011)\citenamefont {Preis},
  \citenamefont {Rebhan},\ and\ \citenamefont {Schmitt}}]{Preis:2010cq}%
  \BibitemOpen
  \bibfield  {author} {\bibinfo {author} {\bibfnamefont {F.}~\bibnamefont
  {Preis}}, \bibinfo {author} {\bibfnamefont {A.}~\bibnamefont {Rebhan}}, \
  and\ \bibinfo {author} {\bibfnamefont {A.}~\bibnamefont {Schmitt}},\ }\href
  {\doibase 10.1007/JHEP03(2011)033} {\bibfield  {journal} {\bibinfo  {journal}
  {JHEP}\ }\textbf {\bibinfo {volume} {03}},\ \bibinfo {pages} {033} (\bibinfo
  {year} {2011})},\ \Eprint {http://arxiv.org/abs/1012.4785} {arXiv:1012.4785
  [hep-th]} \BibitemShut {NoStop}%
\bibitem [{\citenamefont {Preis}\ \emph {et~al.}(2013)\citenamefont {Preis},
  \citenamefont {Rebhan},\ and\ \citenamefont {Schmitt}}]{Preis:2012fh}%
  \BibitemOpen
  \bibfield  {author} {\bibinfo {author} {\bibfnamefont {F.}~\bibnamefont
  {Preis}}, \bibinfo {author} {\bibfnamefont {A.}~\bibnamefont {Rebhan}}, \
  and\ \bibinfo {author} {\bibfnamefont {A.}~\bibnamefont {Schmitt}},\ }\href
  {\doibase 10.1007/978-3-642-37305-3_3} {\bibfield  {journal} {\bibinfo
  {journal} {Lect. Notes Phys.}\ }\textbf {\bibinfo {volume} {871}},\ \bibinfo
  {pages} {51} (\bibinfo {year} {2013})},\ \Eprint
  {http://arxiv.org/abs/1208.0536} {arXiv:1208.0536 [hep-ph]} \BibitemShut
  {NoStop}%
\bibitem [{\citenamefont {Chernodub}\ \emph {et~al.}(2012)\citenamefont
  {Chernodub}, \citenamefont {Van~Doorsselaere},\ and\ \citenamefont
  {Verschelde}}]{Chernodub:2011gs}%
  \BibitemOpen
  \bibfield  {author} {\bibinfo {author} {\bibfnamefont {M.~N.}\ \bibnamefont
  {Chernodub}}, \bibinfo {author} {\bibfnamefont {J.}~\bibnamefont
  {Van~Doorsselaere}}, \ and\ \bibinfo {author} {\bibfnamefont
  {H.}~\bibnamefont {Verschelde}},\ }\href {\doibase
  10.1103/PhysRevD.85.045002} {\bibfield  {journal} {\bibinfo  {journal} {Phys.
  Rev.}\ }\textbf {\bibinfo {volume} {D85}},\ \bibinfo {pages} {045002}
  (\bibinfo {year} {2012})},\ \Eprint {http://arxiv.org/abs/1111.4401}
  {arXiv:1111.4401 [hep-ph]} \BibitemShut {NoStop}%
\bibitem [{\citenamefont {Chernodub}(2011)}]{Chernodub:2011mc}%
  \BibitemOpen
  \bibfield  {author} {\bibinfo {author} {\bibfnamefont {M.~N.}\ \bibnamefont
  {Chernodub}},\ }\href {\doibase 10.1103/PhysRevLett.106.142003} {\bibfield
  {journal} {\bibinfo  {journal} {Phys. Rev. Lett.}\ }\textbf {\bibinfo
  {volume} {106}},\ \bibinfo {pages} {142003} (\bibinfo {year} {2011})},\
  \Eprint {http://arxiv.org/abs/1101.0117} {arXiv:1101.0117 [hep-ph]}
  \BibitemShut {NoStop}%
\bibitem [{\citenamefont {Tuchin}(2013{\natexlab{c}})}]{Tuchin:2012mf}%
  \BibitemOpen
  \bibfield  {author} {\bibinfo {author} {\bibfnamefont {K.}~\bibnamefont
  {Tuchin}},\ }\href {\doibase 10.1103/PhysRevC.87.024912} {\bibfield
  {journal} {\bibinfo  {journal} {Phys. Rev.}\ }\textbf {\bibinfo {volume}
  {C87}},\ \bibinfo {pages} {024912} (\bibinfo {year} {2013}{\natexlab{c}})},\
  \Eprint {http://arxiv.org/abs/1206.0485} {arXiv:1206.0485 [hep-ph]}
  \BibitemShut {NoStop}%
\bibitem [{\citenamefont {Tuchin}(2013{\natexlab{d}})}]{Tuchin:2013bda}%
  \BibitemOpen
  \bibfield  {author} {\bibinfo {author} {\bibfnamefont {K.}~\bibnamefont
  {Tuchin}},\ }\href {\doibase 10.1103/PhysRevC.88.024910} {\bibfield
  {journal} {\bibinfo  {journal} {Phys. Rev.}\ }\textbf {\bibinfo {volume}
  {C88}},\ \bibinfo {pages} {024910} (\bibinfo {year} {2013}{\natexlab{d}})},\
  \Eprint {http://arxiv.org/abs/1305.0545} {arXiv:1305.0545 [nucl-th]}
  \BibitemShut {NoStop}%
\bibitem [{\citenamefont {Sadooghi}\ and\ \citenamefont
  {Taghinavaz}(2017)}]{Sadooghi:2016jyf}%
  \BibitemOpen
  \bibfield  {author} {\bibinfo {author} {\bibfnamefont {N.}~\bibnamefont
  {Sadooghi}}\ and\ \bibinfo {author} {\bibfnamefont {F.}~\bibnamefont
  {Taghinavaz}},\ }\href {\doibase 10.1016/j.aop.2016.11.008} {\bibfield
  {journal} {\bibinfo  {journal} {Annals Phys.}\ }\textbf {\bibinfo {volume}
  {376}},\ \bibinfo {pages} {218} (\bibinfo {year} {2017})},\ \Eprint
  {http://arxiv.org/abs/1601.04887} {arXiv:1601.04887 [hep-ph]} \BibitemShut
  {NoStop}%
\bibitem [{\citenamefont {Bandyopadhyay}\ \emph {et~al.}(2016)\citenamefont
  {Bandyopadhyay}, \citenamefont {Islam},\ and\ \citenamefont
  {Mustafa}}]{Bandyopadhyay:2016fyd}%
  \BibitemOpen
  \bibfield  {author} {\bibinfo {author} {\bibfnamefont {A.}~\bibnamefont
  {Bandyopadhyay}}, \bibinfo {author} {\bibfnamefont {C.~A.}\ \bibnamefont
  {Islam}}, \ and\ \bibinfo {author} {\bibfnamefont {M.~G.}\ \bibnamefont
  {Mustafa}},\ }\href {\doibase 10.1103/PhysRevD.94.114034} {\bibfield
  {journal} {\bibinfo  {journal} {Phys. Rev.}\ }\textbf {\bibinfo {volume}
  {D94}},\ \bibinfo {pages} {114034} (\bibinfo {year} {2016})},\ \Eprint
  {http://arxiv.org/abs/1602.06769} {arXiv:1602.06769 [hep-ph]} \BibitemShut
  {NoStop}%
\bibitem [{\citenamefont {Bandyopadhyay}\ and\ \citenamefont
  {Mallik}(2017)}]{Bandyopadhyay:2017raf}%
  \BibitemOpen
  \bibfield  {author} {\bibinfo {author} {\bibfnamefont {A.}~\bibnamefont
  {Bandyopadhyay}}\ and\ \bibinfo {author} {\bibfnamefont {S.}~\bibnamefont
  {Mallik}},\ }\href {\doibase 10.1103/PhysRevD.95.074019} {\bibfield
  {journal} {\bibinfo  {journal} {Phys. Rev.}\ }\textbf {\bibinfo {volume}
  {D95}},\ \bibinfo {pages} {074019} (\bibinfo {year} {2017})},\ \Eprint
  {http://arxiv.org/abs/1704.01364} {arXiv:1704.01364 [hep-ph]} \BibitemShut
  {NoStop}%
\bibitem [{\citenamefont {Ghosh}\ and\ \citenamefont
  {Chandra}(2018)}]{Ghosh:2018xhh}%
  \BibitemOpen
  \bibfield  {author} {\bibinfo {author} {\bibfnamefont {S.}~\bibnamefont
  {Ghosh}}\ and\ \bibinfo {author} {\bibfnamefont {V.}~\bibnamefont
  {Chandra}},\ }\href {\doibase 10.1103/PhysRevD.98.076006} {\bibfield
  {journal} {\bibinfo  {journal} {Phys. Rev.}\ }\textbf {\bibinfo {volume}
  {D98}},\ \bibinfo {pages} {076006} (\bibinfo {year} {2018})},\ \Eprint
  {http://arxiv.org/abs/1808.05176} {arXiv:1808.05176 [hep-ph]} \BibitemShut
  {NoStop}%
\bibitem [{\citenamefont {Islam}\ \emph {et~al.}(2019)\citenamefont {Islam},
  \citenamefont {Bandyopadhyay}, \citenamefont {Roy},\ and\ \citenamefont
  {Sarkar}}]{Islam:2018sog}%
  \BibitemOpen
  \bibfield  {author} {\bibinfo {author} {\bibfnamefont {C.~A.}\ \bibnamefont
  {Islam}}, \bibinfo {author} {\bibfnamefont {A.}~\bibnamefont
  {Bandyopadhyay}}, \bibinfo {author} {\bibfnamefont {P.~K.}\ \bibnamefont
  {Roy}}, \ and\ \bibinfo {author} {\bibfnamefont {S.}~\bibnamefont {Sarkar}},\
  }\href {\doibase 10.1103/PhysRevD.99.094028} {\bibfield  {journal} {\bibinfo
  {journal} {Phys. Rev.}\ }\textbf {\bibinfo {volume} {D99}},\ \bibinfo {pages}
  {094028} (\bibinfo {year} {2019})},\ \Eprint
  {http://arxiv.org/abs/1812.10380} {arXiv:1812.10380 [hep-ph]} \BibitemShut
  {NoStop}%
\bibitem [{\citenamefont {Ghosh}\ \emph
  {et~al.}(2020{\natexlab{a}})\citenamefont {Ghosh}, \citenamefont {Chaudhuri},
  \citenamefont {Sarkar},\ and\ \citenamefont {Roy}}]{Ghosh:2020xwp}%
  \BibitemOpen
  \bibfield  {author} {\bibinfo {author} {\bibfnamefont {S.}~\bibnamefont
  {Ghosh}}, \bibinfo {author} {\bibfnamefont {N.}~\bibnamefont {Chaudhuri}},
  \bibinfo {author} {\bibfnamefont {S.}~\bibnamefont {Sarkar}}, \ and\ \bibinfo
  {author} {\bibfnamefont {P.}~\bibnamefont {Roy}},\ }\href {\doibase
  10.1103/PhysRevD.101.096002} {\bibfield  {journal} {\bibinfo  {journal}
  {Phys. Rev. D}\ }\textbf {\bibinfo {volume} {101}},\ \bibinfo {pages}
  {096002} (\bibinfo {year} {2020}{\natexlab{a}})},\ \Eprint
  {http://arxiv.org/abs/2004.09203} {arXiv:2004.09203 [nucl-th]} \BibitemShut
  {NoStop}%
\bibitem [{\citenamefont {Nambu}\ and\ \citenamefont
  {Jona-Lasinio}(1961{\natexlab{a}})}]{Nambu1}%
  \BibitemOpen
  \bibfield  {author} {\bibinfo {author} {\bibfnamefont {Y.}~\bibnamefont
  {Nambu}}\ and\ \bibinfo {author} {\bibfnamefont {G.}~\bibnamefont
  {Jona-Lasinio}},\ }\href {\doibase 10.1103/PhysRev.124.246} {\bibfield
  {journal} {\bibinfo  {journal} {Phys. Rev.}\ }\textbf {\bibinfo {volume}
  {124}},\ \bibinfo {pages} {246} (\bibinfo {year} {1961}{\natexlab{a}})},\
  \bibinfo {note} {[,141(1961)]}\BibitemShut {NoStop}%
\bibitem [{\citenamefont {Nambu}\ and\ \citenamefont
  {Jona-Lasinio}(1961{\natexlab{b}})}]{Nambu2}%
  \BibitemOpen
  \bibfield  {author} {\bibinfo {author} {\bibfnamefont {Y.}~\bibnamefont
  {Nambu}}\ and\ \bibinfo {author} {\bibfnamefont {G.}~\bibnamefont
  {Jona-Lasinio}},\ }\href {\doibase 10.1103/PhysRev.122.345} {\bibfield
  {journal} {\bibinfo  {journal} {Phys. Rev.}\ }\textbf {\bibinfo {volume}
  {122}},\ \bibinfo {pages} {345} (\bibinfo {year} {1961}{\natexlab{b}})},\
  \bibinfo {note} {[,127(1961)]}\BibitemShut {NoStop}%
\bibitem [{\citenamefont {Klevansky}(1992)}]{Klevansky}%
  \BibitemOpen
  \bibfield  {author} {\bibinfo {author} {\bibfnamefont {S.~P.}\ \bibnamefont
  {Klevansky}},\ }\href {\doibase 10.1103/RevModPhys.64.649} {\bibfield
  {journal} {\bibinfo  {journal} {Rev. Mod. Phys.}\ }\textbf {\bibinfo {volume}
  {64}},\ \bibinfo {pages} {649} (\bibinfo {year} {1992})}\BibitemShut
  {NoStop}%
\bibitem [{\citenamefont {Vogl}\ and\ \citenamefont {Weise}(1991)}]{Vogl}%
  \BibitemOpen
  \bibfield  {author} {\bibinfo {author} {\bibfnamefont {U.}~\bibnamefont
  {Vogl}}\ and\ \bibinfo {author} {\bibfnamefont {W.}~\bibnamefont {Weise}},\
  }\href {\doibase 10.1016/0146-6410(91)90005-9} {\bibfield  {journal}
  {\bibinfo  {journal} {Prog. Part. Nucl. Phys.}\ }\textbf {\bibinfo {volume}
  {27}},\ \bibinfo {pages} {195} (\bibinfo {year} {1991})}\BibitemShut
  {NoStop}%
\bibitem [{\citenamefont {Buballa}(2005)}]{Buballa}%
  \BibitemOpen
  \bibfield  {author} {\bibinfo {author} {\bibfnamefont {M.}~\bibnamefont
  {Buballa}},\ }\href {\doibase 10.1016/j.physrep.2004.11.004} {\bibfield
  {journal} {\bibinfo  {journal} {Phys. Rept.}\ }\textbf {\bibinfo {volume}
  {407}},\ \bibinfo {pages} {205} (\bibinfo {year} {2005})},\ \Eprint
  {http://arxiv.org/abs/hep-ph/0402234} {arXiv:hep-ph/0402234 [hep-ph]}
  \BibitemShut {NoStop}%
\bibitem [{\citenamefont {McLerran}\ and\ \citenamefont
  {Svetitsky}(1981)}]{McLerran:1981pb}%
  \BibitemOpen
  \bibfield  {author} {\bibinfo {author} {\bibfnamefont {L.~D.}\ \bibnamefont
  {McLerran}}\ and\ \bibinfo {author} {\bibfnamefont {B.}~\bibnamefont
  {Svetitsky}},\ }\href {\doibase 10.1103/PhysRevD.24.450} {\bibfield
  {journal} {\bibinfo  {journal} {Phys. Rev. D}\ }\textbf {\bibinfo {volume}
  {24}},\ \bibinfo {pages} {450} (\bibinfo {year} {1981})}\BibitemShut
  {NoStop}%
\bibitem [{\citenamefont {Cheng}\ \emph {et~al.}(2008)\citenamefont {Cheng}
  \emph {et~al.}}]{Cheng:2007jq}%
  \BibitemOpen
  \bibfield  {author} {\bibinfo {author} {\bibfnamefont {M.}~\bibnamefont
  {Cheng}} \emph {et~al.},\ }\href {\doibase 10.1103/PhysRevD.77.014511}
  {\bibfield  {journal} {\bibinfo  {journal} {Phys. Rev. D}\ }\textbf {\bibinfo
  {volume} {77}},\ \bibinfo {pages} {014511} (\bibinfo {year} {2008})},\
  \Eprint {http://arxiv.org/abs/0710.0354} {arXiv:0710.0354 [hep-lat]}
  \BibitemShut {NoStop}%
\bibitem [{\citenamefont {Ratti}\ \emph {et~al.}(2006)\citenamefont {Ratti},
  \citenamefont {Thaler},\ and\ \citenamefont {Weise}}]{Ratti:2005jh}%
  \BibitemOpen
  \bibfield  {author} {\bibinfo {author} {\bibfnamefont {C.}~\bibnamefont
  {Ratti}}, \bibinfo {author} {\bibfnamefont {M.~A.}\ \bibnamefont {Thaler}}, \
  and\ \bibinfo {author} {\bibfnamefont {W.}~\bibnamefont {Weise}},\ }\href
  {\doibase 10.1103/PhysRevD.73.014019} {\bibfield  {journal} {\bibinfo
  {journal} {Phys. Rev. D}\ }\textbf {\bibinfo {volume} {73}},\ \bibinfo
  {pages} {014019} (\bibinfo {year} {2006})},\ \Eprint
  {http://arxiv.org/abs/hep-ph/0506234} {arXiv:hep-ph/0506234} \BibitemShut
  {NoStop}%
\bibitem [{\citenamefont {Ratti}\ \emph {et~al.}(2007)\citenamefont {Ratti},
  \citenamefont {Roessner}, \citenamefont {Thaler},\ and\ \citenamefont
  {Weise}}]{Ratti:2006wg}%
  \BibitemOpen
  \bibfield  {author} {\bibinfo {author} {\bibfnamefont {C.}~\bibnamefont
  {Ratti}}, \bibinfo {author} {\bibfnamefont {S.}~\bibnamefont {Roessner}},
  \bibinfo {author} {\bibfnamefont {M.}~\bibnamefont {Thaler}}, \ and\ \bibinfo
  {author} {\bibfnamefont {W.}~\bibnamefont {Weise}},\ }\href {\doibase
  10.1140/epjc/s10052-006-0065-x} {\bibfield  {journal} {\bibinfo  {journal}
  {Eur. Phys. J. C}\ }\textbf {\bibinfo {volume} {49}},\ \bibinfo {pages} {213}
  (\bibinfo {year} {2007})},\ \Eprint {http://arxiv.org/abs/hep-ph/0609218}
  {arXiv:hep-ph/0609218} \BibitemShut {NoStop}%
\bibitem [{\citenamefont {Klevansky}\ and\ \citenamefont
  {Lemmer}(1989)}]{Klevansky:1989vi}%
  \BibitemOpen
  \bibfield  {author} {\bibinfo {author} {\bibfnamefont {S.}~\bibnamefont
  {Klevansky}}\ and\ \bibinfo {author} {\bibfnamefont {R.~H.}\ \bibnamefont
  {Lemmer}},\ }\href {\doibase 10.1103/PhysRevD.39.3478} {\bibfield  {journal}
  {\bibinfo  {journal} {Phys. Rev. D}\ }\textbf {\bibinfo {volume} {39}},\
  \bibinfo {pages} {3478} (\bibinfo {year} {1989})}\BibitemShut {NoStop}%
\bibitem [{\citenamefont {Gatto}\ and\ \citenamefont
  {Ruggieri}(2010)}]{Gatto:2010qs}%
  \BibitemOpen
  \bibfield  {author} {\bibinfo {author} {\bibfnamefont {R.}~\bibnamefont
  {Gatto}}\ and\ \bibinfo {author} {\bibfnamefont {M.}~\bibnamefont
  {Ruggieri}},\ }\href {\doibase 10.1103/PhysRevD.82.054027} {\bibfield
  {journal} {\bibinfo  {journal} {Phys. Rev. D}\ }\textbf {\bibinfo {volume}
  {82}},\ \bibinfo {pages} {054027} (\bibinfo {year} {2010})},\ \Eprint
  {http://arxiv.org/abs/1007.0790} {arXiv:1007.0790 [hep-ph]} \BibitemShut
  {NoStop}%
\bibitem [{\citenamefont {Fayazbakhsh}\ \emph {et~al.}(2012)\citenamefont
  {Fayazbakhsh}, \citenamefont {Sadeghian},\ and\ \citenamefont
  {Sadooghi}}]{Fayazbakhsh:2012vr}%
  \BibitemOpen
  \bibfield  {author} {\bibinfo {author} {\bibfnamefont {S.}~\bibnamefont
  {Fayazbakhsh}}, \bibinfo {author} {\bibfnamefont {S.}~\bibnamefont
  {Sadeghian}}, \ and\ \bibinfo {author} {\bibfnamefont {N.}~\bibnamefont
  {Sadooghi}},\ }\href {\doibase 10.1103/PhysRevD.86.085042} {\bibfield
  {journal} {\bibinfo  {journal} {Phys. Rev. D}\ }\textbf {\bibinfo {volume}
  {86}},\ \bibinfo {pages} {085042} (\bibinfo {year} {2012})},\ \Eprint
  {http://arxiv.org/abs/1206.6051} {arXiv:1206.6051 [hep-ph]} \BibitemShut
  {NoStop}%
\bibitem [{\citenamefont {Ruggieri}\ \emph {et~al.}(2013)\citenamefont
  {Ruggieri}, \citenamefont {Tachibana},\ and\ \citenamefont
  {Greco}}]{Ruggieri:2013cya}%
  \BibitemOpen
  \bibfield  {author} {\bibinfo {author} {\bibfnamefont {M.}~\bibnamefont
  {Ruggieri}}, \bibinfo {author} {\bibfnamefont {M.}~\bibnamefont {Tachibana}},
  \ and\ \bibinfo {author} {\bibfnamefont {V.}~\bibnamefont {Greco}},\ }\href
  {\doibase 10.1007/JHEP07(2013)165} {\bibfield  {journal} {\bibinfo  {journal}
  {JHEP}\ }\textbf {\bibinfo {volume} {07}},\ \bibinfo {pages} {165} (\bibinfo
  {year} {2013})},\ \Eprint {http://arxiv.org/abs/1305.0137} {arXiv:1305.0137
  [hep-ph]} \BibitemShut {NoStop}%
\bibitem [{\citenamefont {Andersen}\ \emph {et~al.}(2016)\citenamefont
  {Andersen}, \citenamefont {Naylor},\ and\ \citenamefont
  {Tranberg}}]{Andersen:2014xxa}%
  \BibitemOpen
  \bibfield  {author} {\bibinfo {author} {\bibfnamefont {J.~O.}\ \bibnamefont
  {Andersen}}, \bibinfo {author} {\bibfnamefont {W.~R.}\ \bibnamefont
  {Naylor}}, \ and\ \bibinfo {author} {\bibfnamefont {A.}~\bibnamefont
  {Tranberg}},\ }\href {\doibase 10.1103/RevModPhys.88.025001} {\bibfield
  {journal} {\bibinfo  {journal} {Rev. Mod. Phys.}\ }\textbf {\bibinfo {volume}
  {88}},\ \bibinfo {pages} {025001} (\bibinfo {year} {2016})},\ \Eprint
  {http://arxiv.org/abs/1411.7176} {arXiv:1411.7176 [hep-ph]} \BibitemShut
  {NoStop}%
\bibitem [{\citenamefont {Ferreira}\ \emph
  {et~al.}(2014{\natexlab{a}})\citenamefont {Ferreira}, \citenamefont {Costa},\
  and\ \citenamefont {Provid\^encia}}]{Ferreira:2013oda}%
  \BibitemOpen
  \bibfield  {author} {\bibinfo {author} {\bibfnamefont {M.}~\bibnamefont
  {Ferreira}}, \bibinfo {author} {\bibfnamefont {P.}~\bibnamefont {Costa}}, \
  and\ \bibinfo {author} {\bibfnamefont {C.}~\bibnamefont {Provid\^encia}},\
  }\href {\doibase 10.1103/PhysRevD.89.036006} {\bibfield  {journal} {\bibinfo
  {journal} {Phys. Rev. D}\ }\textbf {\bibinfo {volume} {89}},\ \bibinfo
  {pages} {036006} (\bibinfo {year} {2014}{\natexlab{a}})},\ \Eprint
  {http://arxiv.org/abs/1312.6733} {arXiv:1312.6733 [hep-ph]} \BibitemShut
  {NoStop}%
\bibitem [{\citenamefont {Ferreira}\ \emph
  {et~al.}(2014{\natexlab{b}})\citenamefont {Ferreira}, \citenamefont {Costa},\
  and\ \citenamefont {Provid\^encia}}]{Ferreira:2014exa}%
  \BibitemOpen
  \bibfield  {author} {\bibinfo {author} {\bibfnamefont {M.}~\bibnamefont
  {Ferreira}}, \bibinfo {author} {\bibfnamefont {P.}~\bibnamefont {Costa}}, \
  and\ \bibinfo {author} {\bibfnamefont {C.}~\bibnamefont {Provid\^encia}},\
  }\href {\doibase 10.1103/PhysRevD.90.016012} {\bibfield  {journal} {\bibinfo
  {journal} {Phys. Rev. D}\ }\textbf {\bibinfo {volume} {90}},\ \bibinfo
  {pages} {016012} (\bibinfo {year} {2014}{\natexlab{b}})},\ \Eprint
  {http://arxiv.org/abs/1406.3608} {arXiv:1406.3608 [hep-ph]} \BibitemShut
  {NoStop}%
\bibitem [{\citenamefont {Ferreira}\ \emph
  {et~al.}(2014{\natexlab{c}})\citenamefont {Ferreira}, \citenamefont {Costa},
  \citenamefont {Louren\c{c}o}, \citenamefont {Frederico},\ and\ \citenamefont
  {Provid\^encia}}]{Ferreira:2014kpa}%
  \BibitemOpen
  \bibfield  {author} {\bibinfo {author} {\bibfnamefont {M.}~\bibnamefont
  {Ferreira}}, \bibinfo {author} {\bibfnamefont {P.}~\bibnamefont {Costa}},
  \bibinfo {author} {\bibfnamefont {O.}~\bibnamefont {Louren\c{c}o}}, \bibinfo
  {author} {\bibfnamefont {T.}~\bibnamefont {Frederico}}, \ and\ \bibinfo
  {author} {\bibfnamefont {C.}~\bibnamefont {Provid\^encia}},\ }\href {\doibase
  10.1103/PhysRevD.89.116011} {\bibfield  {journal} {\bibinfo  {journal} {Phys.
  Rev. D}\ }\textbf {\bibinfo {volume} {89}},\ \bibinfo {pages} {116011}
  (\bibinfo {year} {2014}{\natexlab{c}})},\ \Eprint
  {http://arxiv.org/abs/1404.5577} {arXiv:1404.5577 [hep-ph]} \BibitemShut
  {NoStop}%
\bibitem [{\citenamefont {Mao}(2016{\natexlab{a}})}]{Mao:2016fha}%
  \BibitemOpen
  \bibfield  {author} {\bibinfo {author} {\bibfnamefont {S.}~\bibnamefont
  {Mao}},\ }\href {\doibase 10.1016/j.physletb.2016.05.018} {\bibfield
  {journal} {\bibinfo  {journal} {Phys. Lett. B}\ }\textbf {\bibinfo {volume}
  {758}},\ \bibinfo {pages} {195} (\bibinfo {year} {2016}{\natexlab{a}})},\
  \Eprint {http://arxiv.org/abs/1602.06503} {arXiv:1602.06503 [hep-ph]}
  \BibitemShut {NoStop}%
\bibitem [{\citenamefont {Mao}(2016{\natexlab{b}})}]{Mao:2016lsr}%
  \BibitemOpen
  \bibfield  {author} {\bibinfo {author} {\bibfnamefont {S.}~\bibnamefont
  {Mao}},\ }\href {\doibase 10.1103/PhysRevD.94.036007} {\bibfield  {journal}
  {\bibinfo  {journal} {Phys. Rev. D}\ }\textbf {\bibinfo {volume} {94}},\
  \bibinfo {pages} {036007} (\bibinfo {year} {2016}{\natexlab{b}})},\ \Eprint
  {http://arxiv.org/abs/1605.04526} {arXiv:1605.04526 [hep-th]} \BibitemShut
  {NoStop}%
\bibitem [{\citenamefont {Fayazbakhsh}\ and\ \citenamefont
  {Sadooghi}(2014)}]{Fayazbakhsh:2014mca}%
  \BibitemOpen
  \bibfield  {author} {\bibinfo {author} {\bibfnamefont {S.}~\bibnamefont
  {Fayazbakhsh}}\ and\ \bibinfo {author} {\bibfnamefont {N.}~\bibnamefont
  {Sadooghi}},\ }\href {\doibase 10.1103/PhysRevD.90.105030} {\bibfield
  {journal} {\bibinfo  {journal} {Phys. Rev. D}\ }\textbf {\bibinfo {volume}
  {90}},\ \bibinfo {pages} {105030} (\bibinfo {year} {2014})},\ \Eprint
  {http://arxiv.org/abs/1408.5457} {arXiv:1408.5457 [hep-ph]} \BibitemShut
  {NoStop}%
\bibitem [{\citenamefont {Avancini}\ \emph {et~al.}(2019)\citenamefont
  {Avancini}, \citenamefont {Farias},\ and\ \citenamefont
  {Tavares}}]{Avancini:2018svs}%
  \BibitemOpen
  \bibfield  {author} {\bibinfo {author} {\bibfnamefont {S.~S.}\ \bibnamefont
  {Avancini}}, \bibinfo {author} {\bibfnamefont {R.~L.}\ \bibnamefont
  {Farias}}, \ and\ \bibinfo {author} {\bibfnamefont {W.~R.}\ \bibnamefont
  {Tavares}},\ }\href {\doibase 10.1103/PhysRevD.99.056009} {\bibfield
  {journal} {\bibinfo  {journal} {Phys. Rev. D}\ }\textbf {\bibinfo {volume}
  {99}},\ \bibinfo {pages} {056009} (\bibinfo {year} {2019})},\ \Eprint
  {http://arxiv.org/abs/1812.00945} {arXiv:1812.00945 [hep-ph]} \BibitemShut
  {NoStop}%
\bibitem [{\citenamefont {Chaudhuri}\ \emph {et~al.}(2019)\citenamefont
  {Chaudhuri}, \citenamefont {Ghosh}, \citenamefont {Sarkar},\ and\
  \citenamefont {Roy}}]{Chaudhuri:2019lbw}%
  \BibitemOpen
  \bibfield  {author} {\bibinfo {author} {\bibfnamefont {N.}~\bibnamefont
  {Chaudhuri}}, \bibinfo {author} {\bibfnamefont {S.}~\bibnamefont {Ghosh}},
  \bibinfo {author} {\bibfnamefont {S.}~\bibnamefont {Sarkar}}, \ and\ \bibinfo
  {author} {\bibfnamefont {P.}~\bibnamefont {Roy}},\ }\href {\doibase
  10.1103/PhysRevD.99.116025} {\bibfield  {journal} {\bibinfo  {journal} {Phys.
  Rev.}\ }\textbf {\bibinfo {volume} {D99}},\ \bibinfo {pages} {116025}
  (\bibinfo {year} {2019})},\ \Eprint {http://arxiv.org/abs/1907.03990}
  {arXiv:1907.03990 [nucl-th]} \BibitemShut {NoStop}%
\bibitem [{\citenamefont {Chaudhuri}\ \emph {et~al.}(2020)\citenamefont
  {Chaudhuri}, \citenamefont {Ghosh}, \citenamefont {Sarkar},\ and\
  \citenamefont {Roy}}]{Chaudhuri:2020lga}%
  \BibitemOpen
  \bibfield  {author} {\bibinfo {author} {\bibfnamefont {N.}~\bibnamefont
  {Chaudhuri}}, \bibinfo {author} {\bibfnamefont {S.}~\bibnamefont {Ghosh}},
  \bibinfo {author} {\bibfnamefont {S.}~\bibnamefont {Sarkar}}, \ and\ \bibinfo
  {author} {\bibfnamefont {P.}~\bibnamefont {Roy}},\ }\href {\doibase
  10.1140/epja/s10050-020-00222-9} {\bibfield  {journal} {\bibinfo  {journal}
  {Eur. Phys. J. A}\ }\textbf {\bibinfo {volume} {56}},\ \bibinfo {pages} {213}
  (\bibinfo {year} {2020})},\ \Eprint {http://arxiv.org/abs/2003.05692}
  {arXiv:2003.05692 [nucl-th]} \BibitemShut {NoStop}%
\bibitem [{\citenamefont {Ghosh}\ \emph
  {et~al.}(2020{\natexlab{b}})\citenamefont {Ghosh}, \citenamefont {Mukherjee},
  \citenamefont {Chaudhuri}, \citenamefont {Roy},\ and\ \citenamefont
  {Sarkar}}]{Ghosh:2020qvg}%
  \BibitemOpen
  \bibfield  {author} {\bibinfo {author} {\bibfnamefont {S.}~\bibnamefont
  {Ghosh}}, \bibinfo {author} {\bibfnamefont {A.}~\bibnamefont {Mukherjee}},
  \bibinfo {author} {\bibfnamefont {N.}~\bibnamefont {Chaudhuri}}, \bibinfo
  {author} {\bibfnamefont {P.}~\bibnamefont {Roy}}, \ and\ \bibinfo {author}
  {\bibfnamefont {S.}~\bibnamefont {Sarkar}},\ }\href {\doibase
  10.1103/PhysRevD.101.056023} {\bibfield  {journal} {\bibinfo  {journal}
  {Phys. Rev. D}\ }\textbf {\bibinfo {volume} {101}},\ \bibinfo {pages}
  {056023} (\bibinfo {year} {2020}{\natexlab{b}})},\ \Eprint
  {http://arxiv.org/abs/2003.02024} {arXiv:2003.02024 [hep-ph]} \BibitemShut
  {NoStop}%
\bibitem [{\citenamefont {Mei}\ and\ \citenamefont {Mao}(2020)}]{Mei:2020jzn}%
  \BibitemOpen
  \bibfield  {author} {\bibinfo {author} {\bibfnamefont {J.}~\bibnamefont
  {Mei}}\ and\ \bibinfo {author} {\bibfnamefont {S.}~\bibnamefont {Mao}},\
  }\href {\doibase 10.1103/PhysRevD.102.114035} {\bibfield  {journal} {\bibinfo
   {journal} {Phys. Rev. D}\ }\textbf {\bibinfo {volume} {102}},\ \bibinfo
  {pages} {114035} (\bibinfo {year} {2020})},\ \Eprint
  {http://arxiv.org/abs/2008.12123} {arXiv:2008.12123 [hep-ph]} \BibitemShut
  {NoStop}%
\bibitem [{\citenamefont {Greiner}\ \emph {et~al.}(2011)\citenamefont
  {Greiner}, \citenamefont {Haque}, \citenamefont {Mustafa},\ and\
  \citenamefont {Thoma}}]{Greiner:2010zg}%
  \BibitemOpen
  \bibfield  {author} {\bibinfo {author} {\bibfnamefont {C.}~\bibnamefont
  {Greiner}}, \bibinfo {author} {\bibfnamefont {N.}~\bibnamefont {Haque}},
  \bibinfo {author} {\bibfnamefont {M.~G.}\ \bibnamefont {Mustafa}}, \ and\
  \bibinfo {author} {\bibfnamefont {M.~H.}\ \bibnamefont {Thoma}},\ }\href
  {\doibase 10.1103/PhysRevC.83.014908} {\bibfield  {journal} {\bibinfo
  {journal} {Phys. Rev.}\ }\textbf {\bibinfo {volume} {C83}},\ \bibinfo {pages}
  {014908} (\bibinfo {year} {2011})},\ \Eprint {http://arxiv.org/abs/1010.2169}
  {arXiv:1010.2169 [hep-ph]} \BibitemShut {NoStop}%
\bibitem [{\citenamefont {Ghosh}\ \emph {et~al.}(2017)\citenamefont {Ghosh},
  \citenamefont {Mukherjee}, \citenamefont {Mandal}, \citenamefont {Sarkar},\
  and\ \citenamefont {Roy}}]{Ghosh:2017rjo}%
  \BibitemOpen
  \bibfield  {author} {\bibinfo {author} {\bibfnamefont {S.}~\bibnamefont
  {Ghosh}}, \bibinfo {author} {\bibfnamefont {A.}~\bibnamefont {Mukherjee}},
  \bibinfo {author} {\bibfnamefont {M.}~\bibnamefont {Mandal}}, \bibinfo
  {author} {\bibfnamefont {S.}~\bibnamefont {Sarkar}}, \ and\ \bibinfo {author}
  {\bibfnamefont {P.}~\bibnamefont {Roy}},\ }\href {\doibase
  10.1103/PhysRevD.96.116020} {\bibfield  {journal} {\bibinfo  {journal} {Phys.
  Rev.}\ }\textbf {\bibinfo {volume} {D96}},\ \bibinfo {pages} {116020}
  (\bibinfo {year} {2017})},\ \Eprint {http://arxiv.org/abs/1704.05319}
  {arXiv:1704.05319 [hep-ph]} \BibitemShut {NoStop}%
\bibitem [{\citenamefont {Ghosh}\ \emph {et~al.}(2019)\citenamefont {Ghosh},
  \citenamefont {Mukherjee}, \citenamefont {Roy},\ and\ \citenamefont
  {Sarkar}}]{Ghosh:2019fet}%
  \BibitemOpen
  \bibfield  {author} {\bibinfo {author} {\bibfnamefont {S.}~\bibnamefont
  {Ghosh}}, \bibinfo {author} {\bibfnamefont {A.}~\bibnamefont {Mukherjee}},
  \bibinfo {author} {\bibfnamefont {P.}~\bibnamefont {Roy}}, \ and\ \bibinfo
  {author} {\bibfnamefont {S.}~\bibnamefont {Sarkar}},\ }\href {\doibase
  10.1103/PhysRevD.99.096004} {\bibfield  {journal} {\bibinfo  {journal} {Phys.
  Rev.}\ }\textbf {\bibinfo {volume} {D99}},\ \bibinfo {pages} {096004}
  (\bibinfo {year} {2019})},\ \Eprint {http://arxiv.org/abs/1901.02290}
  {arXiv:1901.02290 [hep-ph]} \BibitemShut {NoStop}%
\bibitem [{\citenamefont {Hansen}\ \emph {et~al.}(2007)\citenamefont {Hansen},
  \citenamefont {Alberico}, \citenamefont {Beraudo}, \citenamefont {Molinari},
  \citenamefont {Nardi},\ and\ \citenamefont {Ratti}}]{Hansen:2006ee}%
  \BibitemOpen
  \bibfield  {author} {\bibinfo {author} {\bibfnamefont {H.}~\bibnamefont
  {Hansen}}, \bibinfo {author} {\bibfnamefont {W.}~\bibnamefont {Alberico}},
  \bibinfo {author} {\bibfnamefont {A.}~\bibnamefont {Beraudo}}, \bibinfo
  {author} {\bibfnamefont {A.}~\bibnamefont {Molinari}}, \bibinfo {author}
  {\bibfnamefont {M.}~\bibnamefont {Nardi}}, \ and\ \bibinfo {author}
  {\bibfnamefont {C.}~\bibnamefont {Ratti}},\ }\href {\doibase
  10.1103/PhysRevD.75.065004} {\bibfield  {journal} {\bibinfo  {journal} {Phys.
  Rev. D}\ }\textbf {\bibinfo {volume} {75}},\ \bibinfo {pages} {065004}
  (\bibinfo {year} {2007})},\ \Eprint {http://arxiv.org/abs/hep-ph/0609116}
  {arXiv:hep-ph/0609116} \BibitemShut {NoStop}%
\bibitem [{\citenamefont {Roessner}\ \emph {et~al.}(2007)\citenamefont
  {Roessner}, \citenamefont {Ratti},\ and\ \citenamefont
  {Weise}}]{Roessner:2006xn}%
  \BibitemOpen
  \bibfield  {author} {\bibinfo {author} {\bibfnamefont {S.}~\bibnamefont
  {Roessner}}, \bibinfo {author} {\bibfnamefont {C.}~\bibnamefont {Ratti}}, \
  and\ \bibinfo {author} {\bibfnamefont {W.}~\bibnamefont {Weise}},\ }\href
  {\doibase 10.1103/PhysRevD.75.034007} {\bibfield  {journal} {\bibinfo
  {journal} {Phys. Rev. D}\ }\textbf {\bibinfo {volume} {75}},\ \bibinfo
  {pages} {034007} (\bibinfo {year} {2007})},\ \Eprint
  {http://arxiv.org/abs/hep-ph/0609281} {arXiv:hep-ph/0609281} \BibitemShut
  {NoStop}%
\bibitem [{\citenamefont {Wang}\ and\ \citenamefont
  {Zhuang}(2018)}]{Wang:2017vtn}%
  \BibitemOpen
  \bibfield  {author} {\bibinfo {author} {\bibfnamefont {Z.}~\bibnamefont
  {Wang}}\ and\ \bibinfo {author} {\bibfnamefont {P.}~\bibnamefont {Zhuang}},\
  }\href {\doibase 10.1103/PhysRevD.97.034026} {\bibfield  {journal} {\bibinfo
  {journal} {Phys. Rev. D}\ }\textbf {\bibinfo {volume} {97}},\ \bibinfo
  {pages} {034026} (\bibinfo {year} {2018})},\ \Eprint
  {http://arxiv.org/abs/1712.00554} {arXiv:1712.00554 [hep-ph]} \BibitemShut
  {NoStop}%
\bibitem [{\citenamefont {Carignano}\ and\ \citenamefont
  {Buballa}(2012)}]{Carignano:2011gr}%
  \BibitemOpen
  \bibfield  {author} {\bibinfo {author} {\bibfnamefont {S.}~\bibnamefont
  {Carignano}}\ and\ \bibinfo {author} {\bibfnamefont {M.}~\bibnamefont
  {Buballa}},\ }\href {\doibase 10.5506/APhysPolBSupp.5.641} {\bibfield
  {journal} {\bibinfo  {journal} {Acta Phys. Polon. Supp.}\ }\textbf {\bibinfo
  {volume} {5}},\ \bibinfo {pages} {641} (\bibinfo {year} {2012})},\ \Eprint
  {http://arxiv.org/abs/1111.4400} {arXiv:1111.4400 [hep-ph]} \BibitemShut
  {NoStop}%
\bibitem [{\citenamefont {Mao}(2014)}]{Mao:2014hga}%
  \BibitemOpen
  \bibfield  {author} {\bibinfo {author} {\bibfnamefont {S.}~\bibnamefont
  {Mao}},\ }\href {\doibase 10.1103/PhysRevD.89.116006} {\bibfield  {journal}
  {\bibinfo  {journal} {Phys. Rev. D}\ }\textbf {\bibinfo {volume} {89}},\
  \bibinfo {pages} {116006} (\bibinfo {year} {2014})},\ \Eprint
  {http://arxiv.org/abs/1402.4564} {arXiv:1402.4564 [nucl-th]} \BibitemShut
  {NoStop}%
\bibitem [{\citenamefont {Carignano}\ and\ \citenamefont
  {Buballa}(2020)}]{Carignano:2019ivp}%
  \BibitemOpen
  \bibfield  {author} {\bibinfo {author} {\bibfnamefont {S.}~\bibnamefont
  {Carignano}}\ and\ \bibinfo {author} {\bibfnamefont {M.}~\bibnamefont
  {Buballa}},\ }\href {\doibase 10.1103/PhysRevD.101.014026} {\bibfield
  {journal} {\bibinfo  {journal} {Phys. Rev. D}\ }\textbf {\bibinfo {volume}
  {101}},\ \bibinfo {pages} {014026} (\bibinfo {year} {2020})},\ \Eprint
  {http://arxiv.org/abs/1910.03604} {arXiv:1910.03604 [hep-ph]} \BibitemShut
  {NoStop}%
\bibitem [{\citenamefont {Aguirre}(2020)}]{Aguirre:2020tiy}%
  \BibitemOpen
  \bibfield  {author} {\bibinfo {author} {\bibfnamefont {R.}~\bibnamefont
  {Aguirre}},\ }\href {\doibase 10.1103/PhysRevD.102.096025} {\bibfield
  {journal} {\bibinfo  {journal} {Phys. Rev. D}\ }\textbf {\bibinfo {volume}
  {102}},\ \bibinfo {pages} {096025} (\bibinfo {year} {2020})},\ \Eprint
  {http://arxiv.org/abs/2009.01828} {arXiv:2009.01828 [hep-ph]} \BibitemShut
  {NoStop}%
\bibitem [{\citenamefont {Hattori}\ \emph {et~al.}(2021)\citenamefont
  {Hattori}, \citenamefont {Taya},\ and\ \citenamefont
  {Yoshida}}]{Hattori:2020htm}%
  \BibitemOpen
  \bibfield  {author} {\bibinfo {author} {\bibfnamefont {K.}~\bibnamefont
  {Hattori}}, \bibinfo {author} {\bibfnamefont {H.}~\bibnamefont {Taya}}, \
  and\ \bibinfo {author} {\bibfnamefont {S.}~\bibnamefont {Yoshida}},\ }\href
  {\doibase 10.1007/JHEP01(2021)093} {\bibfield  {journal} {\bibinfo  {journal}
  {JHEP}\ }\textbf {\bibinfo {volume} {01}},\ \bibinfo {pages} {093} (\bibinfo
  {year} {2021})},\ \Eprint {http://arxiv.org/abs/2010.13492} {arXiv:2010.13492
  [hep-ph]} \BibitemShut {NoStop}%
\bibitem [{\citenamefont {Bali}\ \emph
  {et~al.}(2012{\natexlab{b}})\citenamefont {Bali}, \citenamefont {Collins},
  \citenamefont {Deka}, \citenamefont {Glassle}, \citenamefont {Gockeler},
  \citenamefont {Najjar}, \citenamefont {Nobile}, \citenamefont {Pleiter},
  \citenamefont {Schafer},\ and\ \citenamefont {Sternbeck}}]{Bali:2012av}%
  \BibitemOpen
  \bibfield  {author} {\bibinfo {author} {\bibfnamefont {G.~S.}\ \bibnamefont
  {Bali}}, \bibinfo {author} {\bibfnamefont {S.}~\bibnamefont {Collins}},
  \bibinfo {author} {\bibfnamefont {M.}~\bibnamefont {Deka}}, \bibinfo {author}
  {\bibfnamefont {B.}~\bibnamefont {Glassle}}, \bibinfo {author} {\bibfnamefont
  {M.}~\bibnamefont {Gockeler}}, \bibinfo {author} {\bibfnamefont
  {J.}~\bibnamefont {Najjar}}, \bibinfo {author} {\bibfnamefont
  {A.}~\bibnamefont {Nobile}}, \bibinfo {author} {\bibfnamefont
  {D.}~\bibnamefont {Pleiter}}, \bibinfo {author} {\bibfnamefont
  {A.}~\bibnamefont {Schafer}}, \ and\ \bibinfo {author} {\bibfnamefont
  {A.}~\bibnamefont {Sternbeck}},\ }\href {\doibase 10.1103/PhysRevD.86.054504}
  {\bibfield  {journal} {\bibinfo  {journal} {Phys. Rev. D}\ }\textbf {\bibinfo
  {volume} {86}},\ \bibinfo {pages} {054504} (\bibinfo {year}
  {2012}{\natexlab{b}})},\ \Eprint {http://arxiv.org/abs/1207.1110}
  {arXiv:1207.1110 [hep-lat]} \BibitemShut {NoStop}%
\bibitem [{\citenamefont {Bandyopadhyay}\ and\ \citenamefont
  {Farias}(2020)}]{Bandyopadhyay:2020zte}%
  \BibitemOpen
  \bibfield  {author} {\bibinfo {author} {\bibfnamefont {A.}~\bibnamefont
  {Bandyopadhyay}}\ and\ \bibinfo {author} {\bibfnamefont {R.~L.~S.}\
  \bibnamefont {Farias}},\ }\href@noop {} {\  (\bibinfo {year} {2020})},\
  \Eprint {http://arxiv.org/abs/2003.11054} {arXiv:2003.11054 [hep-ph]}
  \BibitemShut {NoStop}%
\end{thebibliography}%

\end{document}